\documentclass[a4paper,12pt]{article}
\pdfoutput=1 

\usepackage{jheppub} 

\usepackage[T1]{fontenc} 

\usepackage{amssymb,amsmath}
\usepackage{slashed,bbold}
\usepackage{multirow}
\usepackage{enumitem}
\setenumerate[1]{label={\alph*)}, leftmargin=18pt}
\setitemize[1]{leftmargin=18pt}
\usepackage[table,dvipsnames]{xcolor}
\usepackage{hyperref}
\hypersetup{
	colorlinks = true,
	citecolor  = NavyBlue,
	linkcolor  = NavyBlue,
	urlcolor = NavyBlue
}

\usepackage{cleveref}
\crefname{table}{Table}{Tables}
\crefname{equation}{Eq.}{Eqs.}
\crefname{appendix}{App.}{Apps.}
\crefname{section}{Sec.}{Secs.}
\crefname{figure}{Fig.}{Figs.}

\usepackage[compat=1.1.0]{tikz-feynman}

\newcommand{\s}{\hspace{0.8pt}}
\DeclareMathSymbol{\shortminus}{\mathbin}{AMSa}{"39}

\newcommand{\dd}{\text{d}}
\newcommand{\A}{\mathcal{A}}
\newcommand{\D}{\mathcal{D}}

\newcommand{\J}{J}
\renewcommand{\L}{\mathcal{L}}
\newcommand{\M}{\mathcal{M}}
\newcommand{\Mfun}{\mathcal{F}}
\renewcommand{\O}{\mathcal{O}}
\newcommand{\R}{\mathcal{R}}
\newcommand{\T}{\mathcal{T}}
\newcommand{\V}{\mathcal{V}}

\newcommand{\DDelta}{\boldsymbol{\Delta}}
\newcommand{\VV}{V^{(\xi)}}

\newcommand{\dphi}[1]{(\partial_{#1}\phi)}
\newcommand{\Dphi}[1]{(\partial^{#1}\phi)}
\newcommand{\dphit}[1]{(\partial_{#1}\widetilde\phi)}
\newcommand{\Dphit}[1]{(\partial^{#1}\widetilde\phi)}
\newcommand{\ddphi}[2]{(\D_{#2}\partial_{#1}\phi)}

\newcommand{\dG}{\Theta}

\newcommand{\oseq}{\overset{\text{OS}}{=}}

\preprint{
\vspace{-8pt}
\begin{flushright}
CERN-TH-2025-188
\end{flushright}
}

\title{Geometric Building Blocks of\\ Effective Field Theory Amplitudes}

\author[a,b,c]{Timothy~Cohen,}
\author[d]{Xu-Xiang~Li,}
\author[d]{and Zhengkang~Zhang\s}

\affiliation[a]{Theoretical Physics Department, CERN, 1211 Geneva, Switzerland}
\affiliation[b]{Theoretical Particle Physics Laboratory, EPFL, 1015 Lausanne, Switzerland}
\affiliation[c]{Institute for Fundamental Science, University of Oregon, Eugene, OR 97403, USA}
\affiliation[d]{Department of Physics \& Astronomy, University of Utah, Salt Lake City, UT 84112, USA}

\emailAdd{tim.cohen@cern.ch}
\emailAdd{xuxiang.li@utah.edu}
\emailAdd{z.k.zhang@utah.edu}

\abstract{On-shell amplitudes are invariant under field redefinitions. Nonderivative field redefinitions have a natural interpretation as coordinate transformations on the target manifold. General field redefinitions, which may involve derivatives, can be viewed as coordinate transformations on the field configuration manifold. We present a unified perspective for the geometry of both the target manifold and the field configuration manifold for scalar effective field theories. In both cases, we identify vertices that can be used to build the tree-level amplitudes, with the property that they transform covariantly in the vacuum and on-shell limits. We identify a choice of metric on the field configuration manifold, for which amplitude expressions on the target manifold can be easily reproduced from their counterparts on the field configuration manifold. This clarifies the relation between the well-established framework of field space geometry and recent proposals for functional geometry.} 

\begin{document}
\maketitle
\flushbottom
\setcounter{page}{2}

\section{Introduction}
\label{sec:Introduction}

The laws of physics do not depend on the coordinate system chosen by physicists. In an Effective Field Theory (EFT), the dynamical degrees of freedom are fields, and the coordinate system corresponds to a field basis. Given an EFT action $S[\phi]$, we are free to redefine the fields:
\begin{equation}
\phi = f \bigl(\widetilde\phi\,,\, \partial_\mu\widetilde\phi\,,\, \partial_\mu\partial_\nu\widetilde\phi\,,\, \dots\bigr) 
\label{eq:redef}
\end{equation}
for some function $f$, and express the same theory using a redefined action $\widetilde S[\widetilde\phi]$ in terms of the new set of fields. The two actions before and after the field redefinition, $S[\phi]$ and $\widetilde S[\widetilde\phi]$, must contain the same physics (e.g., they must predict the same on-shell amplitudes) because all we have done is a change of basis.  

This seemingly innocuous statement has been a source of complication and confusion across the EFT literature. For example, the Standard Model EFT (SMEFT) features thousands of effective operators already at the dimension-six level \cite{Grzadkowski:2010es}. Many field bases have been proposed for SMEFT \cite{Hagiwara:1993ck,Giudice:2007fh,Grzadkowski:2010es,Elias-Miro:2013mua,Elias-Miro:2013eta} and, because of the sheer number of operators, it is typically not straightforward to compare different analyses if they do not use the same basis. Another crucial question is what theories of new physics have their low-energy imprints captured by SMEFT, and what theories require the more general framework of the Higgs EFT (HEFT). Delineating the boundary between different classes of UV completions of the Standard Model requires going beyond the classification in terms of linearly vs.\ nonlinearly realized electroweak symmetry, in part because this distinction becomes ambiguous under basis change.

To appreciate the problem in a concrete setting, consider the standard algorithm for computing scattering amplitudes in an EFT. Starting from the action, we derive the Feynman rules and draw diagrams to connect vertices with propagators. These conventional building blocks of EFT amplitudes are field basis dependent. For example, a set of irrelevant operators that appear to modify the propagators in one basis might have their effects encoded in vertex corrections in a different basis. Only when the building blocks are assembled into on-shell amplitudes do we see that the final results are the same across all bases. 

In this paper, we develop an approach where the individual off-shell building blocks of EFT amplitudes have well-defined transformation properties under field redefinitions at every step of the calculation. This makes the ultimate field redefinition invariance of the on-shell amplitudes manifest. This approach has the benefit that the physical implications of a given theory are not obscured by change of basis.  

Realizing this goal naturally evokes ideas from differential geometry, which physicists are usually introduced to in the context of general relativity. Indeed, a central goal of general relativity is to express physical laws in terms of tensors which transform covariantly under coordinate changes on the spacetime manifold. In order to adopt the same strategy for EFTs, we would like to interpret fields as charting some manifold. This enables the introduction of geometric quantities, which serve as the elemental objects with which we can construct observables such as scattering amplitudes. This is the underlying logic of the EFT geometry program, which has a long history~\cite{Coleman:1969sm, Callan:1969sn, Honerkamp:1971sh, Volkov:1973vd, Tataru:1975ys, Alvarez-Gaume:1981exa, Alvarez-Gaume:1981exv, Vilkovisky:1984st, DeWitt:1984sjp, Gaillard:1985uh, DeWitt:1985sg, Georgi:1991ch} and has undergone significant new developments in recent years: SMEFT vs.\ HEFT classification of Standard Model extensions~\cite{Alonso:2015fsp, Alonso:2016oah, Cohen:2020xca, Cohen:2021ucp, Alonso:2021rac}, all order in $\frac{v^2}{\Lambda^2}$ expressions for electroweak and Higgs observables~\cite{Helset:2020yio,Hays:2020scx,Corbett:2021eux,Martin:2023fad}, geometric soft theorems~\cite{Cheung:2021yog, Derda:2024jvo, Cohen:2025dex}, EFT matching and renormalization group evolution equations~\cite{Alonso:2017tdy, Alonso:2022ffe, Helset:2022pde, Assi:2023zid, Jenkins:2023rtg, Jenkins:2023bls, Li:2024ciy, Aigner:2025xyt, Assi:2025fsm}, and incorporation of fermions~\cite{Finn:2020nvn,Gattus:2023gep,Gattus:2024ird,Assi:2023zid,Assi:2025fsm,Craig:2025uoc}, to name a few. 

In the simplest version of EFT geometry, the manifold of interest is the target manifold of the field map (also commonly referred to as the field space manifold). To be concrete, consider $N$ real scalar fields $\phi^i(x)$, with $i=1,\dots,N$. We say that the values of $\phi^i(x)$ at each spacetime point $x$ provide a coordinate chart on an $N$-dimensional manifold, which is the target manifold of the map $\phi$. The associated geometric framework is known as {\it field space geometry}. It is built on two key observations:
\begin{itemize}
    \item field redefinitions that {\it do not involve derivatives}, $\phi = f(\widetilde\phi)$, can be viewed as coordinate transformations on the target manifold, and
    \item for an EFT {\it truncated at $\O(\partial^2)$}, defined by $\L= \frac{1}{2}\,g_{ij}(\phi)\,(\partial_\mu\phi^i) (\partial^\mu\phi^j) -U(\phi)$, the coefficient function $g_{ij}(\phi)$ of the two-derivative term naturally defines a metric on the target manifold.
\end{itemize}
With a metric in hand, it becomes straightforward to apply the machinery familiar from general relativity to introduce other geometric notions like the Levi-Civita connection, covariant derivative, Riemann curvature tensor, etc. One can then derive on-shell amplitudes in terms of the geometric objects, which results in expressions that are manifestly invariant under {\it nonderivative} field redefinitions~\cite{Cohen:2021ucp,Cheung:2021yog,Finn:2019aip,Finn:2020nvn,Gattus:2023gep,Gattus:2024ird}. 

Accommodating {\it derivative} field redefinitions, on the other hand, is more challenging. When derivatives of $\phi$ are present, \cref{eq:redef} is not a coordinate transformation on the target manifold. Moreover, it is often necessary in phenomenological applications to go beyond $\O(\partial^2)$ in the EFT Lagrangian. Progress toward a more general geometric construction has been made in two complementary directions, by including derivatives of fields as additional coordinates~\cite{Craig:2023wni,Craig:2023hhp,Alminawi:2023qtf,Lee:2024xqa} or by defining geometric objects on the manifold of field configurations~\cite{Cohen:2022uuw,Cheung:2022vnd,Cohen:2023ekv,Cohen:2024bml}. The former approach is naturally formulated in the jet bundle formalism, while the latter approach draws on the language of functional methods in field theory. Our present work is in line with the second direction and, following Ref.~\cite{Cohen:2022uuw}, we refer to this approach as {\it functional geometry}. Two key observations underlying the initial proposal of functional geometry are:
\begin{itemize}
	\item general field redefinitions of the form \cref{eq:redef}, which replace $\phi$ by a functional of $\widetilde\phi$, can be viewed as coordinate transformations on the field configuration manifold (referred to as the functional manifold in Refs.~\cite{Cohen:2023ekv,Cohen:2024bml}), and
	\item for a general EFT, amplitudes can be constructed recursively by taking functional derivatives of correlation functions.
\end{itemize}
The second point follows from the standard path integral formulation of field theory: the second functional derivative of $S[\phi]$ gives the inverse propagator, and higher functional derivatives give the standard Feynman vertices. Furthermore, replacing the classical action $S[\phi]$ by the one-particle-irreducible (1PI) effective action, one obtains the loop-corrected propagator and vertices. 

Importantly, although on-shell amplitudes can only be measured at the physical vacuum, in order to take functional derivatives we must work with objects defined on the entire field configuration manifold. The focal point of the analyses in Refs.~\cite{Cohen:2022uuw,Cohen:2023ekv,Cohen:2024bml} is a set of objects $\M_{1\dots n}$, which are a natural generalization of on-shell amplitudes away from the vacuum and continued to off-shell momenta. Away from the vacuum and on-shell limits, however, $\M_{1\dots n}$ generally do not transform as tensors. This leads to the notion of {\it on-shell covariance}, meaning the inhomogeneous pieces that appear upon performing a field redefinition vanish at the vacuum for on-shell external momenta~\cite{Cohen:2022uuw,Cohen:2023ekv}. Ref.~\cite{Cohen:2024bml} further elaborated this notion by showing that not only are $\M_{1\dots n}$ on-shell covariant, for general scalar EFTs at tree level, they can also be recursively constructed from on-shell covariant building blocks. The main finding of Ref.~\cite{Cohen:2024bml} can be summarized as:
\begin{equation}
\M_{1\dots n} = \Mfun_{1\dots n}\bigl(\Delta^{ab}, \{ V_{a_1\dots a_m}\}\bigr)= \Mfun_{1\dots n}\bigl(\Delta^{ab}, \{ \V_{a_1\dots a_m}\}\bigr) \,,
\label{eq:M}
\end{equation}
where $\Delta^{ab}$ and $V_{a_1\dots a_m}$ are the standard Feynman propagator and vertices, and $\V_{a_1\dots a_m}$ represents a new set of vertices that are on-shell covariant.\footnote{As a technical note, the $\V_{a_1\dots a_m}$ in this work are the fully symmetric components of the $\V_{a_1\dots a_m}$ introduced in Ref.~\cite{Cohen:2024bml}. \Cref{eq:M} holds for both versions of $\V_{a_1\dots a_m}$.} The statement of \cref{eq:M} is that one can take the standard tree-level expression of $\M_{1\dots n}$, and ``upgrade'' $V_{a_1\dots a_m}$ to $\V_{a_1\dots a_m}$, in order to obtain a manifestly on-shell covariant expression for the off-shell, off-vacuum generalization of the amplitude. Note that $\Delta^{ab}$ is already on-shell covariant, so nothing needs to be done to the propagators. See \cref{fig:amp} for a graphic illustration.

\begin{figure}[t]
	\begin{align*}
	\begin{tikzpicture}[baseline = (v)]
	\begin{feynman}
	\vertex[blob, minimum size = 20pt] (v) {};
	\vertex[above = 25pt of v] (t) {};
	\vertex[left = 25pt of t] (phi1) {};
	\vertex[right = 25pt of t] (phi4) {};
	\vertex[below = 25pt of v] (b) {};
	\vertex[left = 25pt of b] (phi2) {};
	\vertex[right = 25pt of b] (phi3) {};
	\diagram*{ (phi1) -- (v) -- (phi2), (phi4) -- (v) -- (phi3) };
	\end{feynman}
	\end{tikzpicture}
	\;\;&=\;\;
	\begin{tikzpicture}[baseline = (v)]
	\begin{feynman}
	\vertex[dot, minimum size = 0pt] (v) {};
	\vertex[above = 25pt of v] (t) {};
	\vertex[left = 25pt of t] (phi1) {};
	\vertex[right = 25pt of t] (phi4) {};
	\vertex[below = 25pt of v] (b) {};
	\vertex[left = 25pt of b] (phi2) {};
	\vertex[right = 25pt of b] (phi3) {};
	\diagram*{ (phi1) -- (v) -- (phi2), (phi4) -- (v) -- (phi3) };
	\end{feynman}
	\end{tikzpicture}
	\;\;+\;\;
	\begin{tikzpicture}[baseline = (v12)]
	\begin{feynman}
	\vertex[dot, minimum size = 0pt] (v12) {};
	\vertex[left = 15pt of v12] (l) {};
	\vertex[above = 25pt of l] (phi1) {};
	\vertex[below = 25pt of l] (phi2) {};
	\vertex[right = 25pt of v12, dot, minimum size = 0pt] (v34) {};
	\vertex[right = 15pt of v34] (r) {};
	\vertex[above = 25pt of r] (phi4) {};
	\vertex[below = 25pt of r] (phi3) {};
	\diagram*{ (phi1) -- (v12) -- (phi2), (v12) -- (v34), (phi3) -- (v34) -- (phi4) };
	\end{feynman}
	\end{tikzpicture}
	\;\;+ \;\; \text{(crossings)} \\[5pt]
	&=\;\;
	\begin{tikzpicture}[baseline = (v)]
	\begin{feynman}
	\vertex[empty dot, minimum size = 15pt, line width = 0.6pt] (v) {$\mathcal{V}$};
	\vertex[above = 25pt of v] (t) {};
	\vertex[left = 25pt of t] (phi1) {};
	\vertex[right = 25pt of t] (phi4) {};
	\vertex[below = 25pt of v] (b) {};
	\vertex[left = 25pt of b] (phi2) {};
	\vertex[right = 25pt of b] (phi3) {};
	\diagram*{ (phi1) -- (v) -- (phi2), (phi4) -- (v) -- (phi3) };
	\end{feynman}
	\end{tikzpicture}
	\;\;+\;\;
	\begin{tikzpicture}[baseline = (v12)]
	\begin{feynman}
	\vertex[empty dot, minimum size = 15pt, line width = 0.6pt] (v12) {$\mathcal{V}$};
	\vertex[left = 15pt of v12] (l) {};
	\vertex[above = 25pt of l] (phi1) {};
	\vertex[below = 25pt of l] (phi2) {};
	\vertex[right = 25pt of v12, empty dot, minimum size = 15pt, line width = 0.6pt] (v34) {$\mathcal{V}$};
	\vertex[right = 15pt of v34] (r) {};
	\vertex[above = 25pt of r] (phi4) {};
	\vertex[below = 25pt of r] (phi3) {};
	\diagram*{ (phi1) -- (v12) -- (phi2), (v12) -- (v34), (phi3) -- (v34) -- (phi4) };
	\end{feynman}
	\end{tikzpicture}
	\;\;+ \;\; \text{(crossings)}
	\end{align*}
	\caption{\label{fig:amp}
		Graphic illustration of \cref{eq:M} for $n=4$. $\V$ represents a new type of vertices built from geometric objects on the field configuration manifold, which transform covariantly in the vacuum and on-shell limits. The geometric expression in the second line makes on-shell covariance of the amplitude manifest. 
	}
\end{figure}

A primary goal of the present paper is to elucidate the relation between recent developments in functional geometry~\cite{Cohen:2022uuw,Cohen:2023ekv,Cohen:2024bml} and existing results in field space geometry, focusing on scalar EFTs. In particular, we will show that \cref{eq:M} is not unique to functional geometry; there is also a field space geometry version of this equation which we derive for arbitrary-point tree-level amplitudes in scalar EFTs truncated at $\O(\partial^2)$. This will result in a unified perspective on the notion of on-shell covariance of EFT amplitudes and their building blocks, both on the target manifold and on the field configuration manifold. Reconciling previous formulations of field space geometry and functional geometry, which appear to have adopted different starting points as we can see from the bullet points above, is an essential step toward refining many of the recent results in field space geometry, e.g., classification of EFTs and soft theorems, to accommodate the more general set of field redefinitions that include derivatives. This paper also serves as a jumping-off point to discover the generalizations of these results to more generic EFTs.

This paper is organized as follows. In \cref{sec:field}, we formulate EFT amplitude calculations in field space geometry in a way that can be easily generalized to functional geometry. Many of the results in this section are already known from previous works, but an important novelty here is that we consider correlation functions defined on the entire target manifold (not just at the vacuum for on-shell momenta), and show that they can be constructed from on-shell covariant building blocks. In \cref{sec:functional}, we follow a similar strategy to derive EFT amplitudes in functional geometry. In Ref.~\cite{Cohen:2024bml}, the main result, \cref{eq:M}, was obtained from an off-shell recursion relation which does not seem to have a field space geometry counterpart. We reproduce this result (more precisely, a slight variant of the result in Ref.~\cite{Cohen:2024bml} which involves symmetric vertices) using a different approach based on normal coordinates, as a natural generalization of the discussion in \cref{sec:field}. The notion of on-shell covariance which unites the developments in both \cref{sec:field,sec:functional} relies on the existence of a connection but not necessarily a metric, so we stay agnostic about the definition of metric throughout \cref{sec:functional}. In \cref{sec:geometry}, we discuss proposals to establish a metric---hence a Riemannian geometry and its associated Levi-Civita connection---on the field configuration manifold. A new development here is that we outline a procedure to define an unambiguous metric from a general EFT action, which allows us to easily reproduce amplitude results in field space geometry from the more general framework of functional geometry. We conclude in \cref{sec:conclusions} and provide additional technical details in the appendices. We consistently work at tree level throughout this paper, and leave a generalization of our results to loop-level amplitudes to future work.

\section{EFT amplitudes on the target manifold}
\label{sec:field}

In this section, we revisit the well-established framework of field space geometry.  We provide a new perspective on the on-shell covariance of EFT amplitudes, as defined on the target manifold.  This sets the stage for a natural generalization to functional geometry as defined on the field configuration manifold.

\subsection{Target manifold}

Consider an EFT of $N$ real scalars $\phi^i$, with $i=1,\dots, N$. The action truncated at $\O(\partial^2)$ can be written as:
\begin{equation}
	S[\phi] = \int_x\, \biggl[ \,\frac{1}{2} \,g_{ij}(\phi)(\partial_\mu\phi^i) (\partial^\mu\phi^j ) -U(\phi) \biggr] \,,
	\label{eq:L_2d}
\end{equation}
where 
\begin{equation}
    \int_x\;\; \equiv\; \int \dd^d x\,,
\end{equation}
and $g_{ij}(\phi) = g_{ji}(\phi)$ is symmetric and positive-definite at the vacuum, defined to be the minimum of $U(\phi)$. The starting point of field space geometry is that $g_{ij}(\phi)$ is a metric on the target manifold, while $U(\phi)$ is a scalar function~\cite{Alonso:2015fsp,Alonso:2016oah}. Indeed, under a coordinate transformation on the target manifold, i.e., a nonderivative field redefinition, $\phi^i = f^i(\widetilde\phi)$ where $f^i$ are a set of analytic functions:
\begin{align}
	S[\phi] &= \int_x\,\biggl[ \, \frac{1}{2} \,g_{ij}\bigl(f(\widetilde{\phi})\bigr)\,\frac{\partial\phi^i}{\partial\widetilde{\phi}^k}\frac{\partial\phi^j}{\partial\widetilde{\phi}^l} \, (\partial_\mu\widetilde{\phi}^k) (\partial^\mu\widetilde\phi^l) -U\bigl(f(\widetilde\phi)\bigr)  \biggr] \notag\\[5pt]
	& \equiv \int_x\,\biggl[ \, \frac{1}{2}\, \widetilde{g}_{kl}(\widetilde{\phi})(\partial_\mu\widetilde{\phi}^k) (\partial^\mu\widetilde\phi^l) -\widetilde{U}(\widetilde{\phi}) \biggr] = \widetilde S[\widetilde{\phi}] \,,
	\label{eq:L_transform}
\end{align}
where
\begin{equation}
    \widetilde{g}_{kl}(\widetilde{\phi}) = \frac{\partial\phi^i}{\partial\widetilde{\phi}^k}\frac{\partial\phi^j}{\partial\widetilde{\phi}^l}\, g_{ij}\bigl(f(\widetilde{\phi})\bigr)
    \qquad\text{and}\qquad
    \widetilde{U}\bigl(\widetilde{\phi}\bigr) = U\bigl(f(\widetilde{\phi})\bigr) \,,
\end{equation}
in agreement with the transformations of a $(0,2)$ tensor and a scalar, respectively.\footnote{As usual, we refer to ``tensor functions'' simply as ``tensors.'' It should be clear from the context whether we are talking about a tensor function on the entire manifold or a tensor at a given point on the manifold.} 

The identification of $g_{ij}(\phi)$ as a metric endows the target manifold with a Riemannian structure. From the metric, we can obtain the Levi-Civita connection and Riemann curvature tensor in the usual way:
\begin{subequations}
\label{eq:Gamma_R}
\begin{align}
    \Gamma^i_{jk} &= \frac{1}{2}\,g^{il}\, \bigl( g_{lj,k} + g_{lk,\s j} - g_{jk,l} \bigr) \,,\label{eq:Gamma_fs}\\[5pt]
    R^i{}_{jkl} &= \Gamma^i_{jl,k} -\Gamma^i_{jk,l} + \Gamma^i_{km} \Gamma^m_{jl} - \Gamma^i_{lm} \Gamma^m_{kl}\,,\\[5pt]
    R_{ijkl} &= g_{im}R^m{}_{jkl} = \frac{1}{2}\, \bigl(g_{il,\s jk} +g_{jk,il} -g_{ik,\s jl} -g_{jl,ik} \bigr) \notag\\[2pt]
    &\hspace{75pt}+ g_{mn}\Gamma^m_{il}\Gamma^n_{jk} -g_{mn} \Gamma^m_{ik} \Gamma^n_{jl}\,,
\end{align}
\end{subequations}
where comma denotes a partial derivative on the target manifold, e.g., $g_{ij,k} = \frac{\partial g_{ij}^{}}{\partial\phi^k}$. Covariant derivatives are denoted by semicolons, e.g., for a tensor $T^{i\dots}{}_{j\dots}$,
\begin{equation}
    T^{i\dots}{}_{j\dots;k} = T^{i\dots}{}_{j\dots,k} + \bigl(\Gamma^i_{kl} \,T^{l\dots}{}_{j\dots} + \,\cdots \bigr) - \bigl(\Gamma^l_{kj}\, T^{i\dots}{}_{l\dots} +\,\cdots\bigr) \,.
\end{equation}
Of particular interest are covariant derivatives of the scalar potential $U(\phi)$:
\begin{subequations}
\label{eq:U_cd}
\begin{align}
    U_{;i_1} &= U_{,i_1} \,,\\[5pt]
	U_{;i_1 i_2} &= U_{,i_1 i_2} - \Gamma^j_{i_1 i_2} \, U_{,\s j} \,,\\[5pt]
	U_{;i_1 i_2 i_3} &= U_{, i_1 i_2 i_3} - (\Gamma^j_{i_1 i_2}\, U_{,\s j i_3})_\text{3 terms} - \Gamma^j_{i_1 i_2 i_3} \, U_{,\s j} \,,\\[5pt]
    U_{;i_1 i_2 i_3 i_4} &= U_{,i_1 i_2 i_3 i_4} -(\Gamma^j_{i_1 i_2} \, U_{,\s j i_3 i_4})_\text{6 terms} + (\Gamma^j_{i_1 i_2}\, \Gamma^k_{i_3 i_4}\, U_{,\s jk})_\text{3 terms} \notag\\[2pt]
	&\quad\; - (\Gamma^j_{i_1 i_2 i_3}\, U_{,\s j i_4})_\text{4 terms} - \Gamma^j_{i_1 i_2 i_3 i_4} \, U_{,\s j} \,,
\end{align}
\end{subequations}
where we have introduced the generalized Christoffel symbols $\Gamma^j_{i_1\dots i_n}$. They are defined recursively via:
\begin{equation}
    \Gamma^j_{i_1\dots i_n i_{n+1}} \equiv \Gamma^j_{i_1\dots i_n, i_{n+1}} - \sum_{a=1}^n \Gamma^k_{i_{n+1} i_a} \, \Gamma^j_{i_1\dots \slashed{i_a} k\dots i_n} \,,
    \label{eq:Gamma_def}
\end{equation}
where $\slashed{i_a} k$ means the index $i_a$ is replaced by $k$. We use $(\dots)_\text{\,$n$ terms}$ as a shorthand for a sum of inequivalent $n$ terms of the form of the expression in the parentheses. For example, $(\Gamma^j_{i_1 i_2}\, U_{,\s j i_3})_\text{3 terms} = \Gamma^j_{i_1 i_2}\, U_{,\s j i_3} + \Gamma^j_{i_1 i_3}\, U_{,\s j i_2} + \Gamma^j_{i_2 i_3}\, U_{,\s j i_1}$ is obtained by summing over permutations of open indices, while $(\Gamma^j_{i_1 i_2 i_3}\, U_{,\s j i_4})_\text{4 terms} = \Gamma^j_{i_1 i_2 i_3}\, U_{,\s j i_4} + \Gamma^j_{i_1 i_2 i_4}\, U_{,\s j i_3} + \Gamma^j_{i_1 i_3 i_4}\, U_{,\s j i_2} + \Gamma^j_{i_2 i_3 i_4}\, U_{,\s j i_1}$ is obtained by summing over terms where each of $i_1, \dots, i_4$ plays the role of $i_4$ in $\Gamma^j_{i_1 i_2 i_3}\, U_{,\s j i_4}$, and the ordering of the remaining three indices is preserved.

\subsection{Amplitudes and correlation functions}

To derive scattering amplitudes from the geometric quantities introduced above, we need to first find the particle states. At tree level, this amounts to diagonalizing the quadratic part of \cref{eq:L_2d} around the vacuum $\overline\phi$. We expand $\phi^i(x) = \overline\phi{}^i + \eta^i(x)$ with $\eta$ the quantum fluctuation, and obtain:
\begin{align}
	S[\overline\phi+\eta] = \;& \int_x \,\biggl[
	\, \frac{1}{2} \biggl( \overline g_{ij} + \overline g_{ij,k} \,\eta^k +\frac{1}{2}\,\overline g_{ij,kl} \,\eta^k \eta^l + \cdots \biggr) \, (\partial_\mu\eta^i) (\partial^\mu\eta^j) \notag\\[5pt]
	& \qquad -\overline U -\frac{1}{2}\,\overline U_{,ij} \,\eta^i\eta^j -\frac{1}{3!}\,\overline U_{,ijk}\,\eta^i\eta^j\eta^k -\cdots \biggr] \,,
	\label{eq:L_2d_expand}
\end{align}
where bar means evaluated at $\phi = \overline\phi$, e.g., $\overline g_{ij}\equiv g_{ij}(\overline\phi)$. The metric can be diagonalized by a vielbein $e^\alpha{}_i$ at any point on the manifold.\footnote{Since the dimension of the manifold is arbitrary, we use the term ``vielbein'' as opposed to ``tetrad'' (which is often used in general relativity since the spacetime manifold is four dimensional).} At the vacuum point $\overline\phi$, we can choose the vielbein such that $\overline U_{,ij}$ is also diagonalized:
\begin{equation}
	\overline g_{ij} = \overline e^\alpha{}_i \,\overline e^\beta{}_j\,\delta_{\alpha\beta} \,,\qquad\text{and}\qquad
	\overline U_{,ij} = \overline e^\alpha{}_i \,\overline e^\beta{}_j\, m^2_{\alpha\beta}\,,
\end{equation}
where $m^2_{\alpha\beta} = m_\alpha^2\,\delta^{\vphantom{2}}_{\alpha\beta}$ is the diagonal mass matrix. We use Greek letters $\alpha,\beta,\dots$ to label mass eigenstates.

It is more convenient to work with momentum space fields, which are related to position space fields via:
\begin{equation}
    \eta^i(p) = \int_x \,e^{ip\cdot x}\, \eta^i(x) \qquad \text{and} \qquad
    \eta^i(x) = \int_p \,e^{-ip\cdot x}\, \eta^i(p) \,,
\end{equation}
where
\begin{equation}
    \int_p\;\; \equiv\; \int\frac{\dd^dp}{(2\pi)^d} \,.
\end{equation}
In terms of $\eta^i(p)$, the action reads:
\begin{align}
	S[\overline\phi+\eta] &= S[\overline\phi] + \frac{1}{2} \int_p \,\eta^i(-p) \,\overline\Delta^{\,-1}_{ij}(p) \,\eta^j(p) +\O(\eta^3) \notag\\[5pt]
	&= S[\overline\phi] + \frac{1}{2} \int_p \,\bigl(\overline e^\alpha{}_i\,\eta^i(-p)\bigr)\, \overline\Delta^{\,-1}_{\alpha\beta}(p)\, \bigl(\overline e^\beta{}_j\,\eta^j(p) \bigr) +\O(\eta^3)\,,
\end{align}
where 
\begin{subequations}
	\begin{align}
	\overline\Delta^{\,-1}_{ij}(p) &\equiv \overline g_{ij}\,p^2 -\overline U_{,ij} = \overline e^\alpha{}_i\,\overline e^\beta{}_j\,\overline\Delta^{\,-1}_{\alpha\beta}(p)\,,\label{eq:Delta_inv}\\[5pt]
	\overline\Delta^{\,-1}_{\alpha\beta}(p) &\equiv \delta_{\alpha\beta} (p^2 - m_\alpha^2) \,.
	\end{align}
\end{subequations}
The inverse of these expressions, $\overline\Delta^{\,ij}(p)$, $\overline\Delta^{\alpha\beta}(p)$, which satisfy $\overline\Delta^{\,ij}(p)\, \overline\Delta^{\,-1}_{jk}(p) = \delta^i_k$, $\overline\Delta^{\alpha\beta}(p)\, \overline\Delta^{\,-1}_{\beta\gamma}(p) = \delta^\alpha_\gamma$, give propagators in the flavor and mass bases, respectively.

By the Lehmann-Symanzik-Zimmermann (LSZ) formula, scattering amplitudes are given by residues on the poles of time-ordered correlation functions of mass eigenstate fields $\overline e^\alpha{}_i\,\eta^i$. At tree level (neglecting loop contributions to wave function renormalization factors), we have:
\begin{align}
	& \bigl\langle \overline e^{\alpha_1}{}_{i_1} \,\eta^{i_1}(p_1) \dots \, \overline e^{\alpha_n}{}_{i_n}\, \eta^{i_n}(p_n)\bigr\rangle_\text{c} \notag\\[5pt]
	&\hspace{12pt}= \frac{i}{p_1^2-m_{\alpha_1}^2}\, \cdots \, \frac{i}{p_n^2-m_{\alpha_n}^2} \,(2\pi)^d\, \delta^d(p_1+\cdots + p_n)
	\, i\A_{\alpha_1\dots\alpha_n}(-p_1,\dots,-p_n) \notag\\[5pt]
	&\hspace{26pt} +\, \text{(non-singular)}\notag\\[8pt]
    &\hspace{12pt}= i^{n+1}\,\overline\Delta^{\alpha_1\beta_1}(p_1) \dots\, \overline\Delta^{\alpha_n\beta_n}(p_n) \, (2\pi)^d\, \delta^d(p_1+\cdots + p_n)
    \, \A_{\beta_1\dots\beta_n}(-p_1,\dots,-p_n) \notag\\[5pt]
    &\hspace{26pt}+\, \text{(non-singular)}\,,
    \label{eq:lsz}
\end{align}
where ``c'' means ``connected,''\footnote{Technically, the LSZ formula applies for full correlation functions. Here we are extracting the connected piece, which is proportional to the overall momentum-conserving $\delta$-function, to define $\A$.} and all momenta in the argument of $\A$ are incoming. Our convention here is such that upper (lower) indices are consistently associated with outgoing (incoming) momenta. This explains the awkward minus signs in \cref{eq:lsz} but will turn out convenient later. Multiplying both sides of \cref{eq:lsz} by inverse propagators and taking the on-shell limit, we obtain:
\begin{align}
    &
    \hspace{-25pt}
    (2\pi)^d\, \delta^d(p_1+\cdots + p_n)  \,\A_{\alpha_1\dots\alpha_n}(-p_1,\dots,-p_n) \notag\\[5pt]
    =\;& \frac{1}{i^{n+1}} \, \lim_{p_a^2\to m_{\alpha_a}^2} \overline e_{\alpha_1}{}^{i_1} \,\dots\, \overline e_{\alpha_n}{}^{i_n} \,\overline\Delta^{\,-1}_{i_1j_1}(p_1)\,\dots\,\overline\Delta^{\,-1}_{i_nj_n}(p_n) \, \bigl\langle \eta^{j_1}(p_1) \dots \, \eta^{j_n}(p_n)\bigr\rangle_\text{c} \,,
\label{eq:lsz_fs}
\end{align}
where $\overline e_\alpha{}^i$ is the inverse of $\overline e^\alpha{}_i$ satisfying $\overline e_\alpha{}^i \,\overline e^\beta{}_i = \delta_\alpha^\beta$, and we have used the relation between $\overline\Delta^{\,-1}_{ij}(p)$ and $\overline\Delta^{\,-1}_{\alpha\beta}(p)$ from \cref{eq:Delta_inv}. 

The object $\bigl\langle \eta^{i_1}(p_1) \dots \, \eta^{i_n}(p_n)\bigr\rangle_\text{c}$ in \cref{eq:lsz_fs} represents a connected correlation function among the fluctuations of the fields around the vacuum point $\overline\phi$ on the target manifold (we leave time ordering implicit throughout the paper). It can be computed from the path integral by taking functional derivatives with respect to a source that couples to $\eta$:
\begin{align}
    &
    \hspace{-25pt}
    \bigl\langle \eta^{i_1}(p_1) \dots \, \eta^{i_n}(p_n)\bigr\rangle_{\text{c}} \notag\\[5pt]
    =\;& \biggl[\frac{(2\pi)^{d}}{i}\biggr]^n\,
    \frac{\delta^n}{\delta\J_{i_1}(p_1) \dots \delta\J_{i_n}(p_n)}
    \log \int\D\eta\, e^{i \bigl(S[\,\overline\phi+\eta] +\int_x\J_i\, \eta^i\bigr)} \biggr|_{\J_i(x) = 0} \,,
     \label{eq:ccf}
\end{align}
where $\J_i(p) = \int_x \, e^{-ip\cdot x} \, \J_i(x)$, consistent with the sign convention discussed below \cref{eq:lsz}. Substituting \cref{eq:L_2d_expand} into \cref{eq:ccf} and perturbatively evaluating the path integral, one can derive the standard Feynman rules, where the propagator is given by $\overline\Delta^{\, ij}(p)$ and vertices are given by functions of $\overline g_{ij,k\dots}$, $\overline U_{,ijk\dots}$ and momenta.

As mentioned in the Introduction, in order to connect to recent developments in functional geometry, we must study correlation functions defined away from the vacuum. Expanding the action around an arbitrary point $\phi$ on the target manifold, we have:
\begin{align}
    S[\phi+\eta]
    = \;& \int_x \,\biggl[ \, \frac{1}{2} \biggl( g_{ij}(\phi) + g_{ij,k}(\phi) \,\eta^k +\frac{1}{2}\,g_{ij,kl}(\phi) \,\eta^k \eta^l + \cdots \biggr) \, (\partial_\mu\eta^i) (\partial^\mu\eta^j)  \notag\\[5pt]
	& -U(\phi) -U_{,i}(\phi)\, \eta^i
	-\frac{1}{2}\,U_{,ij}(\phi) \,\eta^i\eta^j -\frac{1}{3!}\, U_{,ijk}(\phi)\,\eta^i\eta^j\eta^k -\cdots \biggr]\,.
    \label{eq:L_2d_expand_phi}
\end{align}
\cref{eq:L_2d_expand_phi} takes the same form as \cref{eq:L_2d_expand}, but with derivatives of $U$ and $g_{ij}$ evaluated at $\phi$ instead of $\overline\phi$. If we include a constant source $\J_i(x) = U_{,i}(\phi)$ to cancel the linear term, and define:
\begin{align}
    &\bigl\langle \eta^{i_1}(p_1) \dots \, \eta^{i_n}(p_n)\bigr\rangle_{\text{c}, \,\phi} \notag\\[8pt]
    &\hspace{20pt}\equiv 
	 \biggl[\frac{(2\pi)^{d}}{i}\biggr]^n\,
     \frac{\delta^n}{\delta\J_{i_1}(p_1) \dots \delta\J_{i_n}(p_n)}
     \log \int\D\eta\, e^{i \bigl(S[\phi+\eta] +\int_x\J_i\, \eta^i\bigr)} \biggr|_{\J_i(x) = U_{,i}(\phi)} \,,
     \label{eq:ccf_phi}
\end{align}
we would obtain the same set of Feynman rules as we would from \cref{eq:ccf}, but with $\overline\Delta^{\,ij}(p)$, $\overline g_{ij,k\dots}$, $\overline U_{,ijk\dots}$ replaced by $\Delta^{ij}(\phi\,; p)$, $g_{ij,k\dots}(\phi)$, $U_{,ijk\dots}(\phi)$, where $\Delta^{ij}(\phi\,; p)$ is the inverse of
\begin{equation}
    \Delta^{-1}_{ij}(\phi\,; p) = g_{ij}(\phi)\, p^2 -U_{,ij}(\phi) \,.
    \label{eq:inv_prop}
\end{equation}

A more useful set of quantities to consider are {\it amputated correlation functions}, which we denote by $\M_{i_1\dots i_n}$. They are defined at any point $\phi$ on the target manifold by factoring out the external propagators and the overall momentum-conserving $\delta$-function from connected correlation functions:
\begin{align}
     &\bigl\langle \eta^{i_1}(p_1) \dots \, \eta^{i_n}(p_n)\bigr\rangle_{\text{c}, \,\phi} \notag\\[5pt]
     &\hspace{5pt}\equiv i^n \, \Delta^{i_1j_1}(\phi\,;p_1) \dots \Delta^{i_nj_n}(\phi\,;p_n) \, (2\pi)^d\, \delta^d(p_1+\cdots + p_n)
     \, i\M_{j_1\dots j_n}(\phi\,;-p_1, \dots, -p_n) \,.
     \label{eq:M_def}
\end{align}
Combining \cref{eq:lsz_fs,eq:M_def}, we obtain:
\begin{equation}
	\A_{\alpha_1\dots\alpha_n}(p_1,\dots,p_n) = \lim_{p_a^2\to m_{\alpha_a}^2} \overline e_{\alpha_1}{}^{i_1} \dots \, \overline e_{\alpha_n}{}^{i_n}\, \overline\M_{i_1\dots i_n}(p_1, \dots, p_n) \,,
	\label{eq:A_fs}
\end{equation}
where $\overline{\M}_{i_1\dots i_n}(p_1, \dots, p_n)$ denotes $\M_{i_1\dots i_n}(\phi\,;p_1, \dots, p_n)$ evaluated at $\phi =\overline\phi$ (corresponding to $J_i(\phi)=0$).

\subsection{On-shell covariance}
\label{sec:field_osc}

From \cref{eq:A_fs}, we see that on-shell amplitudes $\A_{\alpha_1\dots\alpha_n}(p_1,\dots,p_n)$ can be obtained from amputated correlation functions $\M_{i_1\dots i_n}(\phi\,;p_1, \dots, p_n)$ on the target manifold by
\begin{itemize}
    \item[1)] taking the vacuum limit $\phi\to\overline\phi$,
    \item[2)] contracting with vielbein factors $\overline e_{\alpha_a}{}^{i_a}$ for all external particles, and 
    \item[3)] taking the on-shell limit $p_a^2 \to m_{\alpha_a}^2$.
\end{itemize}
For a fixed set of particles labeled by $\alpha_1\dots \alpha_n$, the vielbein factors transform as vectors under coordinate changes on the target manifold. Since $\A_{\alpha_1\dots\alpha_n}(p_1,\dots,p_n)$ are physical observables, \cref{eq:A_fs} implies that in the vacuum and on-shell limits, amputated correlation functions $\M_{i_1\dots i_n}(\phi\,;p_1, \dots, p_n)$ must transform as tensors. More precisely, under a nonderivative field redefinition $\phi^i = f^i(\widetilde{\phi})$,
\begin{equation}
    \widetilde\M_{i_1\dots i_n} (\widetilde\phi) = \frac{\partial\phi^{j_1}}{\partial\widetilde{\phi}^{i_1}} \cdots \frac{\partial\phi^{j_n}}{\partial\widetilde{\phi}^{i_n}}\, \M_{j_1\dots j_n}\bigl(f(\widetilde{\phi})\bigr) + X_{i_1\dots i_n}(\widetilde{\phi}) \,,
\label{eq:M_transform}
\end{equation}
where $X_{i_1\dots i_n}$ satisfies
\begin{equation}
\lim_{p_a^2\to m_{\alpha_a}^2}\overline e_{\alpha_1}{}^{j_1}\,\frac{\partial\widetilde\phi^{i_1}}{\partial\phi^{j_1}} \,\cdots\; \overline e_{\alpha_n}{}^{j_n}\,\frac{\partial\widetilde\phi^{i_n}}{\partial\phi^{j_n}}\; \overline X_{i_1\dots i_n} =0 \,,
\label{eq:X_long}
\end{equation}
and momentum arguments $p_1, \dots, p_n$ are implicit. We will abbreviate \cref{eq:X_long} as:
\begin{equation}
X_{i_1\dots i_n} \oseq\, 0 \,,
\label{eq:X}
\end{equation}
where ``\,$\oseq$\,'' means equal upon performing the three operations listed at the beginning of this subsection. We call this relation between objects on the target manifold {\it on-shell equivalence} to emphasize the ``taking the on-shell limit'' operation, though it is understood that the vacuum limit is also taken and the open indices are contracted with vielbein factors. \cref{eq:X} says $X_{i_1\dots i_n}$ is on-shell equivalent to zero.\footnote{Objects that are on-shell equivalent to zero are referred to ``evanescent'' in Ref.~\cite{Cohen:2023ekv}. We choose not to use ``evanescent'' here to avoid confusion with the more common use of the term as referring to operators that are present in dimensional regularization but vanish in four-dimensional spacetime.} Two other examples of objects that are on-shell equivalent to zero are: 
\begin{equation}
U_{,\s j} \,\oseq\, 0 \qquad \text{and} \qquad
\Delta^{-1}_{ji_a}(p_a) \,\oseq\, 0 \,,
\label{eq:oseq_0}
\end{equation}
where $p_a$ is an external momentum. The first derivative of the potential vanishes as soon as the vacuum limit is taken: $\overline U_{,\s j} =0$. The inverse propagator vanishes after all three operations listed above are performed:
\begin{equation}
\lim\limits_{p_a^2\to m_{\alpha_a}^2}\overline e_{\alpha_a}{}^{i_a}\,\overline\Delta^{\,-1}_{ji_a}(p_a) = \lim\limits_{p_a^2\to m_{\alpha_a}^2}\overline\Delta^{\,-1}_{\alpha_a\beta_a}(p_a) \,\overline e^{\beta_a}{}_j = 0 \,.
\end{equation}

We say an object on the target manifold is {\it on-shell covariant} if it transforms as a tensor up to inhomogeneous terms that are on-shell equivalent to zero. On-shell covariance of amputated correlation functions was discussed previously in the context of functional geometry~\cite{Cohen:2022uuw,Cohen:2023ekv,Cohen:2024bml}. Here we see its counterpart in field space geometry: $\M_{i_1\dots i_n}$ are generally not tensors on the target manifold, but must be on-shell covariant. Two properties of on-shell covariant objects immediately follow from this definition: sums and products of on-shell covariant objects are on-shell covariant; if an object is on-shell equivalent to a tensor, it must be on-shell covariant.

\subsection{On-shell covariant building blocks}
\label{sec:field_bb}

We can make the on-shell covariance of $\M_{i_1\dots i_n}(\phi\,;p_1, \dots, p_n)$ manifest by expressing them in terms of on-shell covariant building blocks. Let us first illustrate this explicitly for $n=3$ and 4, before generalizing to arbitrary $n$. From the standard Feynman rules, we obtain:
\begin{subequations}
\label{eq:M3_M4}
\begin{align}
    \M_{i_1i_2i_3}(p_1, p_2, p_3) &= V_{i_1i_2i_3}(p_1, p_2, p_3) \,,\\[4pt]
    \M_{i_1i_2i_3i_4}(p_1, p_2, p_3, p_4) &= V_{i_1i_2i_3i_4}(p_1, p_2, p_3, p_4) \notag\\[5pt]
    & \quad\; -\Bigl[\Delta^{\,jk}(p_{12}) \, V_{ji_1 i_2} (-p_{12}, p_1, p_2) \, V_{k i_3 i_4} (p_{12}, p_3, p_4)\Bigr]_\text{3 terms} \,,\notag\\[2pt]
\end{align}
\end{subequations}
where $p_{ab} \equiv p_a+p_b$, and $V_{ijk\dots}$ are the vertex functions with all momenta incoming:
\begin{subequations}
\label{eq:V3_V4}
\begin{align}
    V_{i_1 i_2 i_3} (p_1, p_2, p_3) &= - \Bigl[(p_1\cdot p_2) \,g_{i_1 i_2, i_3} \Bigr]_\text{3 terms} -U_{,i_1 i_2 i_3} \notag\\[5pt]
    &= \frac{1}{2}\,\Bigl[p_1^2\, (g_{i_1 i_2, i_3} + g_{i_1 i_3, i_2} - g_{i_2 i_3, i_1})\Bigr]_\text{3 terms} - U_{,i_1 i_2 i_3} \,,\\[10pt]
    V_{i_1 i_2 i_3 i_4} (p_1, p_2, p_3, p_4) &= -\Bigl[ (p_1\cdot p_2)\, g_{i_1 i_2, i_3 i_4} \Bigr]_\text{6 terms} -U_{,i_1 i_2 i_3 i_4} \notag\\[5pt]
	&= -\frac{1}{2} \,\Bigl[ p_{12}^2\, (g_{i_1 i_2, i_3 i_4} + g_{i_3 i_4, i_1 i_2})\Bigr]_\text{3 terms} \notag\\[5pt]
	&\quad +\frac{1}{2}\, \Bigl[ p_1^2\, (g_{i_1 i_2, i_3 i_4} + g_{i_1 i_3, i_2 i_4} + g_{i_1 i_4, i_2 i_3}) \Bigr]_\text{4 terms} -U_{,i_1 i_2 i_3 i_4} \,.
\end{align}
\end{subequations}
Here and in what follows, we leave the $\phi$ arguments implicit to reduce notational clutter. It is understood that all quantities are evaluated at a general point $\phi$ on the target manifold, unless specified otherwise.

We now consider how the objects entering \cref{eq:V3_V4} transform under nonderivative field redefinitions. The internal propagator is the inverse of $\Delta^{-1}_{ij}(p)$ defined in \cref{eq:inv_prop}, which transforms as
\begin{equation}
\widetilde\Delta^{-1}_{ij}(p) = \frac{\partial\phi^k}{\partial\widetilde\phi^i}\,\frac{\partial\phi^l}{\partial\widetilde\phi^j}\,\Delta^{-1}_{kl}(p) -\frac{\partial^2\phi^k}{\partial\widetilde\phi^i\partial\widetilde\phi^j}\, U_{,k} \,.
\label{eq:Delta_transform}
\end{equation}
This is not a tensorial transformation. However, the inhomogeneous term on the right-hand side of \cref{eq:Delta_transform} is on-shell equivalent to zero because $U_{,k} \oseq 0$. Therefore, $\Delta^{-1}_{ij}(p)$ is on-shell covariant, and so is $\Delta^{ij}(p)$. A quicker way to reach the same conclusion is to note that $\Delta^{ij}(p)$ is on-shell equivalent to a tensor: 
\begin{equation}
\Delta^{\,ij}(p) \oseq \DDelta^{\,ij}(p) \,, 
\end{equation}
where $\DDelta^{\,ij}(p)$ (note boldface) is the inverse of
\begin{equation}
\DDelta^{-1}_{ij}(p) \equiv g_{ij}\, p^2 - U_{;ij} \,.
\end{equation}
Note that $U_{;ij} = U_{,ij} -\Gamma^k_{ij}\,U_{,k}$ is symmetric since the Levi-Civita connection satisfies $\Gamma^k_{ij} = \Gamma^k_{ji}$.

The vertex functions in \cref{eq:V3_V4}, on the other hand, do not have simple transformation properties under field redefinitions. Only when they are assembled into $\M_{i_1i_2i_3}$ and $\M_{i_1i_2i_3i_4}$ according to \cref{eq:M3_M4} do we know that the results must be on-shell covariant. However, we can regroup the terms in \cref{eq:M3_M4} such that the individual building blocks are on-shell covariant. Using \cref{eq:Gamma_R,eq:U_cd}, we find, after some algebra: 
\begin{subequations}
\label{eq:M3_M4_fs}
\begin{align}
    \M_{i_1i_2i_3}(p_1, p_2, p_3) &= \V_{i_1i_2i_3}(p_1, p_2, p_3) \,,\\[10pt]
    \M_{i_1i_2i_3i_4}(p_1, p_2, p_3, p_4) &= \V_{i_1i_2i_3i_4}(p_1, p_2, p_3, p_4) \notag\\[5pt]
    & \quad\; -\Bigl[\Delta^{\,jk}(p_{12}) \, \V_{ji_1 i_2} (-p_{12}, p_1, p_2) \, \V_{k i_3 i_4} (p_{12}, p_3, p_4)\Bigr]_\text{3 terms} \,,\notag\\[1pt]
\end{align}
\end{subequations}
where
\begin{subequations}
\label{eq:Vcal_fs}
\begin{align}
    \V_{i_1i_2i_3}(p_1, p_2, p_3) &= - U_{;(i_1 i_2 i_3)} +\,\Gamma^j_{(i_1i_2i_3)} U_{,\s j} \notag\\[5pt]
    &\qquad + 
    \Gamma^j_{i_2 i_3} \, \Delta^{-1}_{ji_1}(p_1) + \Gamma^j_{i_1 i_3} \, \Delta^{-1}_{ji_2}(p_2) + \Gamma^j_{i_1 i_2} \, \Delta^{-1}_{ji_3}(p_3) \,,\\[10pt]
    \V_{ji_1 i_2} (p_j, p_1, p_2) &= -U_{;(ji_1 i_2)} +\Gamma^l_{(ji_1i_2)} U_{,l} \notag\\[5pt]
    &\qquad +\Gamma^l_{j i_2}\,\Delta^{-1}_{li_1}(p_1) +\Gamma^l_{j i_1}\,\Delta^{-1}_{li_2}(p_2) \,,\\[10pt]
    \V_{i_1i_2i_3i_4}(p_1, p_2, p_3, p_4) &= - \frac{2}{3}\,\Bigl[\, p_{12}^2\, R_{i_1 (i_3 i_4) i_2} \Bigr]_\text{3 terms} - U_{;(i_1 i_2 i_3 i_4)} +\Gamma^j_{(i_1i_2i_3i_4)} U_{,\s j} \notag\\[5pt]
    &\qquad + \Bigl[ \Gamma^j_{(i_2 i_3 i_4)}\, \Delta^{-1}_{ji_1}(p_1)\Bigr]_\text{4 terms}  \,.
\end{align}
\end{subequations}
Here and in what follows, parentheses denote full symmetrization of indices, e.g., $R_{i_1 (i_3 i_4) i_2} = \frac{1}{2}\,(R_{i_1i_3i_4i_2}+R_{i_1i_4i_3i_2})$, and $U_{;(i_1i_2i_3)} = \frac{1}{3!}\bigl(U_{;i_1i_2i_3}\bigr)_\text{6 terms}$. The first two equations in Eqs.~(\ref{eq:Vcal_fs}) can be expressed in the general form:
\begin{equation}
    \V_{i_1i_2i_3} (p_1, p_2, p_3) = - U_{;(i_1 i_2 i_3)} +\,\Gamma^j_{(i_1i_2i_3)} U_{,\s j} + \sum_{a\in\text{ext}} \Gamma^j_{i_1\dots\slashed{i_a}\dots i_3}\, \Delta^{-1}_{ji_a}(p_a) \,,
    \label{eq:Vcal_fs_3}
\end{equation}
where ``$a\in\text{ext}$'' means $a$ runs over the subset of $\{1,2,3\}$ for which $p_a$ is an external momentum.

Note that \cref{eq:M3_M4_fs} takes exactly the same form as the standard expressions \cref{eq:M3_M4} for the 3- and 4-point amputated correlation functions, but with the Feynman vertices $V_{i_1\dots i_m}$ replaced by a new type of vertices $\V_{i_1\dots i_m}$. In other words, we have just seen that:
\vspace{5pt}
\begin{equation}
\boxed{\rule[-1.5ex]{0pt}{4.5ex}\quad
\M_{i_1\dots i_n} (p_1, \dots , p_n) = \Mfun_{i_1\dots i_n} \bigl(\Delta^{ij}, \{V_{j_1\dots j_m}\} \bigr)
= \Mfun_{i_1\dots i_n} \bigl(\Delta^{ij}, \{\V_{j_1\dots j_m}\} \bigr)\,,\quad}
\label{eq:Mfun_fs}
\vspace{7pt}
\end{equation}
for $n=3, 4$. We emphasize that the $\phi$ arguments are implicit, i.e., this equation is satisfied at any point on the target manifold. In \cref{sec:field_nt}, we will show that \cref{eq:Mfun_fs} holds for all $n\ge 3$, with properly defined $\V_{i_1\dots i_m}$ (see \cref{eq:Vcal_fs_gen}). The new vertices $\V_{i_1\dots i_m}$ are not tensors on the entire target manifold because they contain nontensorial terms of the form:
\begin{equation}
\Gamma^j_{(i_1\dots i_m)}\, U_{,\s j} \qquad \text{and}\qquad \Gamma^j_{(i_1\dots \slashed{i_a}\dots i_m)}\, \Delta^{-1}_{ji_a}(p_a) \quad (a\in\text{ext})\,.
\label{eq:nt}
\end{equation}
Importantly, these nontensorial terms are on-shell equivalent to zero because of \cref{eq:oseq_0}. As a result, the $\V_{i_1\dots i_m}$ vertices are on-shell equivalent to a set of tensors (which have appeared previously in on-shell amplitudes expressions~\cite{Cohen:2021ucp,Cheung:2021yog}):
\begin{subequations}
	\label{eq:Vbar_OS}
	\begin{align}
	\V_{i_1i_2i_3}(p_1, p_2, p_3) &\oseq - U_{;(i_1i_2i_3)} \,,\\[5pt]
	\V_{i_1i_2i_3i_4}(p_1, p_2, p_3, p_4) &\oseq -\frac{2}{3}\,\Bigl[\, p_{12}^2\, R_{i_1 (i_3 i_4) i_2} \Bigr]_\text{3 terms} - U_{;(i_1i_2i_3i_4)} \,,
	\end{align}
\end{subequations}
and similarly for higher-point $\V_{i_1\dots i_m}$ vertices as we will see below. In other words, the $\V_{i_1\dots i_m}$ vertices are on-shell covariant. \cref{eq:Mfun_fs} therefore gives a manifestly on-shell covariant expression for the $n$-point amputated correlation function for an EFT of the form \cref{eq:L_2d} (the special cases of $n=3, 4$ are given in \cref{eq:M3_M4_fs}).

\subsection{Origin of nontensorial terms}
\label{sec:field_nt}

There is one special case where the nontensorial terms \cref{eq:nt} vanish even away from the vacuum and on-shell limits: if we are in a field basis that corresponds to a set of Riemann normal coordinates at $\phi$, (generalized) Christoffel symbols evaluated at $\phi$ would vanish upon symmetrization of its lower indices. In this subsection, we will show explicitly how the presence of nontensorial terms in the expressions of $\M_{i_1\dots i_n}$ can be understood as coming from a mismatch between a general field basis and Riemann normal coordinates at $\phi$. We will also provide a proof of \cref{eq:Mfun_fs} for general $n$-point amputated correlation functions. 

A set of Riemann normal coordinates at $\phi$ can be constructed as follows. Denote the geodesic from $\phi$ to any point $\phi+\eta$ in the vicinity of $\phi$ by $\gamma(\lambda)$, which satisfies 
\begin{equation}
    \frac{\dd^2\gamma^i}{\dd\lambda^2} + \Gamma^i_{jk}\,\frac{\dd\gamma^j}{\dd\lambda}\,\frac{\dd\gamma^k}{\dd\lambda} = 0 \,,
    \label{eq:geodesic}
\end{equation}
with $\gamma(0) = \phi$, $\gamma(1) = \phi+\eta$. The Riemann normal coordinates for the point $\phi + \eta$ are components of the tangent vector to the geodesic at the origin $\xi^i = \frac{\dd\gamma^i}{\dd\lambda}(0)$. The relation between $\eta^i$ and $\xi^i$ can be derived by solving \cref{eq:geodesic} order by order:
\begin{equation}
    \eta^i = \xi^i -\sum_{m=2}^\infty \frac{1}{m!}\,\Gamma^i_{(i_1\dots i_m)}\, \xi^{i_1}\dots \xi^{i_m}\,,
    \label{eq:eta_to_xi}
\end{equation}
where $\Gamma^i_{(i_1\dots i_m)}$ are all evaluated at $\phi$. It follows that $\Gamma^i_{(i_1\dots i_m)}$ must vanish if we are already in a basis where $\eta^i$ are a set of Riemann normal coordinates at $\phi$.

On the other hand, if we start from a general basis, we can use \cref{eq:eta_to_xi} to rewrite the action in \cref{eq:L_2d_expand_phi} in terms of the Riemann normal coordinates $\xi^i$:
\begin{align}
    S[\phi+\eta] = \;& \int_x \,\biggl[ \,\frac{1}{2} \biggl( g_{ij} +\frac{1}{3}\,R_{i(kl)j} \,\xi^k \xi^l + \cdots \biggr) \, (\partial_\mu\xi^i) (\partial^\mu\xi^j) \notag\\[5pt]
	& -U -U_{;i}\, \xi^i
	-\frac{1}{2}\,U_{;ij} \,\xi^i\xi^j -\frac{1}{3!}\, U_{;(ijk)}\,\xi^i\xi^j\xi^k -\cdots \biggr]\,.
    \label{eq:L_xi}
\end{align}
The coefficient functions in \cref{eq:L_xi} are all tensors, which is expected since $\xi^i$ are components of the tangent vector to the geodesic. As a result, Feynman rules for the $\xi$ fields as derived from $S[\phi+\eta]$ are written entirely in terms of tensors. The propagator is given by $\DDelta^{ij}(p)$, and the vertices are given by:
\begin{subequations}
\label{eq:V_RNC}
\begin{align}
    \VV_{i_1 i_2 i_3}(p_1, p_2, p_3) &= -U_{;(i_1 i_2 i_3)} \,,\\[2pt]
    \VV_{i_1 i_2 i_3 i_4} (p_1, p_2, p_3, p_4) &= - \frac{2}{3}\,\Bigl[(p_1\cdot p_2) \, R_{i_1(i_3 i_4)i_2}\Bigr]_\text{6 terms} -U_{;(i_1 i_2 i_3 i_4)} \notag\\[5pt]
    &= -\frac{2}{3}\,\Bigl[p_{12}^2 \, R_{i_1(i_3 i_4)i_2}\Bigr]_\text{3 terms} -U_{;(i_1 i_2 i_3 i_4)} \,,
\end{align}
\end{subequations}
etc., where we have used $R_{i_1 i_2 i_3 i_4} + R_{i_1 i_3 i_4 i_2} + R_{i_1 i_4 i_2 i_3} = 0$ to arrive at the last line. Expressions for higher-point vertices can be derived using formulae for the derivatives of the metric and potential in Riemann normal coordinates in App.~A of Ref.~\cite{Cohen:2021ucp} (see also Refs.~\cite{Muller:1997zk,Hatzinikitas:2000xe}). We will present another algorithm that allows us to easily obtain higher-point vertices in \cref{sec:geo_fs}. The result takes the following form:
\begin{equation}
    \VV_{i_1 \dots i_m}(p_1, \dots, p_m) = \R_{i_1\dots i_m}(p_1, \dots, p_m) -U_{;(i_1 \dots i_m)} \,,
\end{equation}
where $\R_{i_1\dots i_m}$ is a symmetric tensor built from the Riemann curvature tensor and its covariant derivatives (see \cref{eq:R_def}).

To derive full (as opposed to connected) correlation functions of $\eta$, we can relate them to correlation functions of $\xi$ using \cref{eq:eta_to_xi}:
\begin{align}
\langle \eta^{i_1}\dots \eta^{i_n}\rangle_\phi &= \langle\xi^{i_1} \dots \xi^{i_n} \rangle_\phi \notag\\[5pt]
&\quad\;\; -\sum_{a=1}^n \sum_{m=2}^\infty \frac{1}{m!} \, \Gamma^{i_a}_{(j_1\dots j_m)}\, \bigl\langle \xi^{i_1} \dots \slashed{\xi}^{i_a} (\xi^{j_1}\dots \xi^{j_m}) \dots \xi^{i_n} \bigr\rangle_\phi + \cdots\,,
\label{eq:eta_cf}
\end{align}
where $\slashed{\xi}^{i_a} (\xi^{j_1}\dots \xi^{j_m})$ means $\xi^{i_a}$ is replaced by $\xi^{j_1}\dots \xi^{j_m}$, and the ellipses at the end of the equation represent terms where more than one $\xi$'s in the correlation function are replaced by products of $\xi$'s. The correlation functions in \cref{eq:eta_cf} are all evaluated with respect to the following action plus source term:\footnote{The path integral measures $\D\eta$ and $\D\xi$ differ by a Jacobian, which is a functional determinant and can be neglected for tree-level calculations. Even at loop level, the Jacobian factor for a nonderivative field redefinition like \cref{eq:eta_to_xi} only gives rise to scaleless integrals which vanish in dimensional regularization (see e.g., Refs.~\cite{tHooft:1973wag,Arzt:1993gz,Criado:2018sdb,Cohen:2024fak}).}
\begin{align}
&S[\phi+\eta] + \int_x U_{,i}\,\eta^i \notag\\[5pt]
&\hspace{20pt}= S[\phi+\eta] + \int_x U_{,i}\,\xi^i - \int_x\,\sum_{m=2}^\infty \frac{1}{m!} \, \bigl(\Gamma^i_{(i_1\dots i_m)}\, U_{,i} \bigr)\, \xi^{i_1} \dots \xi^{i_m} \,.
\label{eq:S_plus_Jeta}
\end{align}
We see that when computing the $\xi$ correlation functions appearing on the right-hand side of \cref{eq:eta_cf}, we must include the extra terms $-\int_x \sum\limits_{m=2}^\infty \frac{1}{m!} \, \bigl(\Gamma^i_{(i_1\dots i_m)}\, U_{,i} \bigr)\, \xi^{i_1} \dots \xi^{i_m}$ in addition to the original action. In particular, the quadratic term restores the propagator to its standard nontensorial form:
\begin{align}
S[\phi+\eta]\Bigr|_{\O(\xi^2)} -\frac{1}{2}\,\int_x \Gamma^i_{jk}\, U_{,i} \, \xi^j\xi^k  &= \frac{1}{2}\int_p \,\xi^j(-p) \,\bigl[\DDelta^{-1}_{jk}(p) -\Gamma^i_{jk}\,U_{,i} \bigr] \,\xi^k(p) \notag\\[5pt]
&= \frac{1}{2}\int_p \,\xi^j(-p) \,\Delta^{-1}_{jk}(p)\,\xi^k(p) \,.
\label{eq:S_xi2}
\end{align}
Meanwhile, the cubic and higher terms in \cref{eq:S_plus_Jeta} amount to an extra contribution to the vertices:
\begin{equation}
\delta\VV_{i_1\dots i_m} = -\Gamma^i_{(i_1\dots i_m)}\, U_{,i}\,.
\label{eq:deltaV}
\end{equation}

Now let us take a closer look at the additional terms in the second line of \cref{eq:eta_cf}, which give the difference between $\langle \eta^{i_1}\dots \eta^{i_n}\rangle$ and $\langle \xi^{i_1}\dots \xi^{i_n}\rangle$. These terms involve correlation functions between fields and local operators. For example, consider the case of $n=4$. The $a=1, m=2$ term on the right-hand side of \cref{eq:eta_cf} can be represented as:
\begin{equation}
\begin{tikzpicture}[baseline = (v)]
\begin{feynman}
    \vertex[blob, minimum size = 25pt] (v) {};
    \vertex[above = 18pt of v] (vt) {};
    \vertex[left = 18pt of vt, crossed dot, minimum size = 6pt] (vtl) {};
    \vertex[above = 1pt of vtl] (vtlt) {};
    \vertex[left = 1pt of vtlt] (vtltl) {};
    \vertex[above = 32pt of v] (t) {};
    \vertex[left = 32pt of t] (phi1) {\footnotesize $i_1$};
    \vertex[right = 32pt of t] (phi4) {\footnotesize $i_4$};
    \vertex[below = 32pt of v] (b) {};
    \vertex[left = 32pt of b] (phi2) {\footnotesize $i_2$};
    \vertex[right = 32pt of b] (phi3) {\footnotesize $i_3$};
    \diagram*{ (phi1) -- (vtl), 
    (vtltl) -- [quarter left] (v) -- [quarter left] (vtltl), 
    (v)-- (phi2), 
    (phi4) -- (v) -- (phi3) };
\end{feynman}
\end{tikzpicture}
\end{equation}
where $\otimes$ denotes an insertion of local operator $-\frac{1}{2}\,\Gamma^{i_1}_{j_1j_2}\,\xi^{j_1}\xi^{j_2}$. The blob represents a 5-point full (as opposed to connected) correlation function of $\xi$, which can be computed by inserting vertices and performing Wick contractions as usual. If we require the full diagram to be a tree diagram, then the only possibility is to insert a 3-point vertex and obtain:
\begin{equation}
\begin{tikzpicture}[baseline = (v12)]
\begin{feynman}
\vertex[crossed dot, minimum size = 6pt] (v12) {};
\vertex[left = 15pt of v12] (l) {};
\vertex[above = 25pt of l] (phi1) {\footnotesize $i_1$};
\vertex[below = 25pt of l] (phi2) {\footnotesize $i_2$};
\vertex[right = 25pt of v12, dot, minimum size = 0pt] (v34) {};
\vertex[right = 15pt of v34] (r) {};
\vertex[above = 25pt of r] (phi4) {\footnotesize $i_4$};
\vertex[below = 25pt of r] (phi3) {\footnotesize $i_3$};
\diagram*{ (phi1) -- (v12) -- (phi2), (v12) -- (v34), (phi3) -- (v34) -- (phi4) };
\end{feynman}
\end{tikzpicture}
\;\; + \;\;
\begin{tikzpicture}[baseline = (v12)]
\begin{feynman}
\vertex[crossed dot, minimum size = 6pt] (v12) {};
\vertex[left = 15pt of v12] (l) {};
\vertex[above = 25pt of l] (phi1) {\footnotesize $i_1$};
\vertex[below = 25pt of l] (phi2) {\footnotesize $i_3$};
\vertex[right = 25pt of v12, dot, minimum size = 0pt] (v34) {};
\vertex[right = 15pt of v34] (r) {};
\vertex[above = 25pt of r] (phi4) {\footnotesize $i_4$};
\vertex[below = 25pt of r] (phi3) {\footnotesize $i_2$};
\diagram*{ (phi1) -- (v12) -- (phi2), (v12) -- (v34), (phi3) -- (v34) -- (phi4) };
\end{feynman}
\end{tikzpicture}
\;\; + \;\;
\begin{tikzpicture}[baseline = (v12)]
\begin{feynman}
\vertex[crossed dot, minimum size = 6pt] (v12) {};
\vertex[left = 15pt of v12] (l) {};
\vertex[above = 25pt of l] (phi1) {\footnotesize $i_1$};
\vertex[below = 25pt of l] (phi2) {\footnotesize $i_4$};
\vertex[right = 25pt of v12, dot, minimum size = 0pt] (v34) {};
\vertex[right = 15pt of v34] (r) {};
\vertex[above = 25pt of r] (phi4) {\footnotesize $i_2$};
\vertex[below = 25pt of r] (phi3) {\footnotesize $i_3$};
\diagram*{ (phi1) -- (v12) -- (phi2), (v12) -- (v34), (phi3) -- (v34) -- (phi4) };
\end{feynman}
\end{tikzpicture}
\;.
\label{eq:tree_diag}
\end{equation}
The diagrams in \cref{eq:tree_diag} have the same topology as the standard set of $s, t, u$-channel Feynman diagrams contributing to the 4-point correlation function, but with one of the 3-point vertices replaced by the local operator insertion discussed above. Concretely, let us work through the first diagram for example: 
\begin{align}
\begin{tikzpicture}[baseline = (v12)]
\begin{feynman}
\vertex[crossed dot, minimum size = 6pt] (v12) {};
\vertex[left = 15pt of v12] (l) {};
\vertex[above = 25pt of l] (phi1) {\footnotesize $i_1$};
\vertex[below = 25pt of l] (phi2) {\footnotesize $i_2$};
\vertex[right = 25pt of v12, dot, minimum size = 0pt] (v34) {};
\vertex[right = 15pt of v34] (r) {};
\vertex[above = 25pt of r] (phi4) {\footnotesize $i_4$};
\vertex[below = 25pt of r] (phi3) {\footnotesize $i_3$};
\diagram*{ (phi1) -- (v12) -- (phi2), (v12) -- (v34), (phi3) -- (v34) -- (phi4) };
\end{feynman}
\end{tikzpicture}
\;\; = \;\;& -\Gamma^{i_1}_{j_1j_2} \,i\Bigl[\VV_{kj_3j_4}(p_{12}, p_3, p_4) + \delta\VV_{kj_3j_4}\Bigr] \notag\\[-20pt]
&\quad i\Delta^{j_1 k}(p_{12})\, i\Delta^{i_2j_2}(p_2) \, i\Delta^{i_3j_3}(p_3)\, i\Delta^{i_4j_4}(p_4) \notag\\[8pt]
=\;\;& i \Bigl[\Gamma^{m}_{lj_2} \,\Delta^{-1}_{mj_1}(p_1)\Bigr]\,i\Bigl[\VV_{kj_3j_4}(p_{12}, p_3, p_4) + \delta\VV_{kj_3j_4}\Bigr] \notag\\[5pt]
&\quad i\Delta^{kl}(p_{12})\, i\Delta^{i_1j_1}(p_1)\, i\Delta^{i_2j_2}(p_2) \, i\Delta^{i_3j_3}(p_3)\, i\Delta^{i_4j_4}(p_4)\,,
\label{eq:new_V3}
\end{align}
where symmetry factors from the vertices are canceled by combinatoric factors from Wick contractions. From \cref{eq:new_V3} it is clear that the contribution from this diagram to the 4-point $\eta$ correlation function can be reproduced by including an extra term $\Gamma^{m}_{lj_2} \,\Delta^{-1}_{mj_1}(p_1)$ in the 3-point vertex. 

Following the same strategy, we see that each additional term on the right-hand side of \cref{eq:eta_cf} yields a similar set of diagrams, which have the same topology as the standard set of tree-level Feynman diagrams contributing to the $n$-point correlation function, but with one or more of the vertices replaced by $\Gamma^j_{(i_1\dots \slashed{i_a}\dots i_m)}\, \Delta^{-1}_{ji_a}(p_a)$, where $p_a$ represents an external momentum. The same contributions can be reproduced if we stipulate that for each $m$-point vertex in a Feynman diagram, we must include an extra term $\Gamma^j_{(i_1\dots \slashed{i_a}\dots i_m)}\, \Delta^{-1}_{ji_a}(p_a)$ for each external particle (if any) that is directly connected to the vertex. 

Combining the analysis above and the discussion around \cref{eq:deltaV,eq:S_xi2}, we see that correlation functions of $\eta$ can be obtained from the standard set of Feynman diagrams, with propagators given by $\Delta^{ij}(p)$ and vertices given by:
\begin{align}
    \V_{i_1\dots i_m} (p_1, \dots, p_m) &= \VV_{i_1\dots i_m} (p_1,\dots, p_m) + \delta\VV_{i_1\dots i_m} + \sum_{a\in\text{ext}} \Gamma^j_{(i_1\dots \slashed{i_a}\dots i_m)}\, \Delta^{-1}_{ji_a}(p_a) \notag\\[5pt]
    &= \R_{i_1\dots i_m}(p_1, \dots, p_m) -U_{;(i_1 \dots i_m)} \notag\\[5pt]
    &\quad\; - \Gamma^j_{(i_1\dots i_m)}\, U_{,\s j} + \sum_{a\in\text{ext}} \Gamma^j_{(i_1\dots \slashed{i_a}\dots i_m)}\, \Delta^{-1}_{ji_a}(p_a)\,,
    \label{eq:Vcal_fs_gen}
\end{align}
where $\R_{i_1i_2i_3} = 0$, $\R_{i_1i_2i_3i_4} = -\frac{2}{3}\bigl[ p_{12}^2\, R_{i_1(i_3i_4)i_2}\bigr]_\text{3 terms}$, and $\R_{i_1\dots i_m}$ with higher $m$ can be constructed recursively in terms of the Riemann curvature tensor and its covariant derivatives, as we will discuss in detail in \cref{sec:geo_fs}. We have therefore proved that \cref{eq:Mfun_fs}, which expresses $\M_{i_1\dots i_n}$ in terms of on-shell covariant building blocks, holds for general $n$-point amputated correlation functions. For $m=3$ and $4$, \cref{eq:Vcal_fs_gen} reproduces our previous results in \cref{eq:Vcal_fs,eq:Vcal_fs_3} obtained by manually regrouping terms in $\M_{i_1i_2i_3}$ and $\M_{i_1i_2i_3i_4}$. As discussed below \cref{eq:nt}, the nontensorial terms in the last line of \cref{eq:Vcal_fs_gen} are on-shell equivalent to zero, so they do not contribute to on-shell amplitudes. This is consistent with the fact that $\eta$ and $\xi$ interpolate the same one-particle states.

We end this section by offering a slightly more general perspective on the appearance of nontensorial terms. At a given point $\phi$ on the target manifold, we can define a set of normal coordinates with respect to any connection, not just the Levi-Civita connection derived from the metric (in which case the normal coordinates are Riemann normal coordinates). The relation between a general basis $\eta$ and normal coordinates $\xi$ is given by the same \cref{eq:eta_to_xi}, with $\Gamma^i_{(i_1\dots i_m)}$ defined from the chosen connection by recursively applying \cref{eq:Gamma_def}. Expanding the action in the $\xi$ basis, one again obtains tensorial coefficients, and hence tensorial propagator and vertices. However, for a general connection, the expansion coefficients contain additional terms that involve the torsion and covariant derivatives of the metric.\footnote{For example, if one chooses a metric-compatible connection that has torsion, \cref{eq:L_xi} would be modified to~\cite{katanaev2018normal}:
\begin{align}
S[\phi+\eta] =\;& \int_x \,\biggl\{\,\frac{1}{2}\, \biggl[ \, g_{ij} - T_{(ij)k}\,\xi^k \notag\\[5pt]
&+ \biggl(\frac{1}{3}\,R_{i(kl)j} - \frac{2}{3}\,T_{(ij)(k;l)} + \frac{1}{4}\,T^m{}_{ik}T_{mjl} - \frac{1}{3}\,T^m{}_{ik}T_{jlm}\biggr) \,\xi^k\xi^l + \cdots\biggr](\partial_\mu\xi^i)(\partial^\mu\xi^j) \notag\\[5pt]
	& -U -U_{;i}\, \xi^i
	-\frac{1}{2}\,U_{;ij} \,\xi^i\xi^j -\frac{1}{3!}\, U_{;(ijk)}\,\xi^i\xi^j\xi^k -\cdots \biggr\}\,,
\end{align}
where $T^i{}_{jk} \equiv \Gamma^i_{jk}-\Gamma^i_{kj}$ and $T_{ijk} = g_{il}\, T^l{}_{jk}$.} On the other hand, the fact that \cref{eq:eta_to_xi} holds regardless of the choice of connection implies that the difference between the $\eta$ and $\xi$ bases can be accommodated by exactly the same nontensorial terms in the vertices as shown in the last line of \cref{eq:Vcal_fs_gen}, with $\Gamma^i_{(i_1\dots i_m)}$ computed from the chosen connection. Therefore, the presence of nontensorial terms in $\M_{i_1\dots i_n}$ can be attributed to a mismatch between a general field basis and normal coordinates at $\phi$ with respect to any connection.

It is useful to keep in mind that the notion of (on-shell) covariance relies on a connection, but not necessarily a metric. In field space geometry, we start out by defining a metric, and then simply stick to the unique Levi-Civita connection associated with the metric. There is no obstacle to working with a different connection, which would define a different set of on-shell covariant vertices. But there does not seem to be any advantage either. However, the conceptual separation between a metric and a connection is useful when we generalize the discussion to functional geometry, where the definition of a metric is less obvious, as we will see in the next two sections.

\section{EFT amplitudes on the field configuration manifold}
\label{sec:functional}

The field space geometry framework discussed in the previous section has two limitations: it concerns EFT actions truncated at $\O(\partial^2)$, and manifests on-shell covariance under nonderivative field redefinitions only. As mentioned in the Introduction, general field redefinitions (with or without derivatives) can be viewed as coordinate transformations on the field configuration manifold. In this section, we generalize the discussion of \cref{sec:field} to find on-shell covariant building blocks of amplitudes on the field configuration manifold for general scalar EFT actions.

\subsection{Field configuration manifold}

The field configuration manifold is charted by $\phi^i(x)$ or, more conveniently $\phi^i(p)$. Unlike in field space geometry, positions or momenta are not external quantities, but must be considered as part of the coordinates of the field configuration manifold; in other words, we should think of $x$ in $\phi^i(x)$ as an additional index, which is on the same footing as the flavor index $i$, and similarly for $p$ in $\phi^i(p)$. To emphasize this point, we collect both flavor and position/momentum indices and write: 
\begin{equation}
\phi^{(ix)} \equiv \phi^i(x)\qquad
\text{and}\qquad
\phi^{(ip)} \equiv \phi^i(p)\,,
\end{equation}
and similarly for other objects on the field configuration manifold introduced below. The target manifold in field space geometry can be viewed as a submanifold of the field configuration manifold, which corresponds to $\phi^{(ix)} = \phi^i =$ constant, or equivalently, $\phi^{(ip)} = \phi^i\,(2\pi)^d\delta^d(p)$. We assume that the vacuum preserves spacetime translation symmetry, so it corresponds to a point $\overline\phi$ on the constant field submanifold.

In what follows, we will encounter expressions with many flavor and momentum indices. To reduce notational clutter, we will often use $(i_1p_1), (i_2p_2), \dots$ for these indices, and abbreviate $(i_ap_a) \to a$ as in DeWitt notation, e.g.,
\begin{equation}
\phi^a \equiv \phi^{(i_ap_a)}\,.
\end{equation}
In this notation, derivatives on the field configuration manifold are normalized as
\begin{equation}
{}^{}_{,a} \equiv \,{}^{}_{,(i_a p_a)}\, \equiv \frac{\partial}{\partial\phi^a} \equiv \frac{\partial}{\partial\phi^{(i_ap_a)}} \equiv  (2\pi)^d\frac{\delta}{\delta\phi^{i_a}(p_a)} \,,
\label{eq:diff}
\end{equation}
where $\delta$ denotes the standard functional derivative, and we use $\partial$ for its normalized version in momentum space. The identity on the field configuration manifold is normalized as:
\begin{equation}
\delta^a_b \equiv  \delta^{i_a}_{i_b}\, \delta^{p_a}_{p_b} \equiv \frac{\partial\phi^a}{\partial\phi^b} = \delta^{i_a}_{i_b}\, (2\pi)^d\delta^d(p_a-p_b) \,.
\end{equation}
We will also encounter more general momentum-conserving $\delta$-functions, which we abbreviate as:
\begin{equation}
\delta^{q_1\dots q_m}_{p_1\dots p_n} \equiv (2\pi)^d \,\delta^d(q_1 + \cdots + q_m - p_1-\cdots - p_n) \,.
\end{equation}
When we contract indices on the field configuration manifold, an integral over momentum is implied, in addition to a sum over flavor indices:
\begin{equation}
{}^a{}_a \equiv \,{}^{(i_a p_a)}{}_{(i_a p_a)}\, \equiv \sum_{i_a} \int_{p_a} = \sum_{i_a} \int \frac{\dd^d p^{}_a}{(2\pi)^d} \,.
\label{eq:contract}
\end{equation}

Let us demonstrate all this notation by considering the expansion of a general EFT action, which is a scalar on the field configuration manifold, around a point $\phi$. The result can be compactly written as:
\begin{align}
S[\phi+\eta] &= \sum_{n=0}^\infty \frac{1}{n!}\,S_{,1\dots n}[\phi]\, \eta^1\dots \eta^n \notag\\[5pt]
&\equiv S[\phi] + S_{,a}[\phi]\,\eta^a +\frac{1}{2}\, \Delta^{-1}_{ab}[\phi]\,\eta^a\eta^b +\sum_{n=3}^\infty \frac{1}{n!}\,V_{1\dots n}[\phi]\, \eta^1\dots \eta^n \,,
\label{eq:S_expand}
\end{align}
where we have defined:
\begin{equation}
\Delta^{-1}_{ab}[\phi] \equiv S_{,ab}[\phi] 
\qquad \text{and}\qquad
V_{1\dots n}[\phi] \equiv S_{,1\dots n}[\phi]\,.
\label{eq:Delta_V}
\end{equation}
These are functionals of $\phi$. For EFT actions of the form \cref{eq:L_2d} and on the constant field submanifold, $\Delta^{-1}_{ab}$ and $V_{1\dots n}$ reduce to quantities defined in \cref{sec:field}:
\begin{subequations}
\begin{align}
\Delta^{-1}_{ab} [\phi] \bigr|_{\phi(x) =\phi} &= \Delta^{-1}_{i_ai_b}(\phi\,; p_a)\,\delta_{p_a p_b}\,,\label{eq:invprop_fsl}\\[5pt]
V_{1\dots n} [\phi] \bigr|_{\phi(x) =\phi} &= V_{i_1\dots i_n}(\phi\,; p_1,\dots, p_n) \, \delta_{p_1\dots p_n}\,.
\end{align}
\end{subequations}
We use the same symbols for the propagator and vertices on the field configuration manifold and the constant field submanifold, keeping in mind that they differ by momentum-conserving $\delta$-functions on the constant field submanifold. It should be clear from the index notation whether we are referring to a quantity on the target manifold or the full field configuration manifold. For nonconstant field configurations $\phi(x)$, $\Delta^{-1}_{ab} [\phi]$ and $V_{1\dots n} [\phi]$ are generally not proportional to momentum-conserving $\delta$-functions because a nonconstant background field breaks spacetime translation symmetry.

Note that at this point, we have not introduced any Riemannian structure. All we have assumed so far is that the field configuration manifold is a differentiable manifold, where differentiation is implemented by the familiar functional derivative. We have not defined a metric on this manifold. As discussed at the end of \cref{sec:field}, what we really need to construct on-shell covariant vertices is a connection. In field space geometry, it is natural to use the Levi-Civita connection derived from the metric, which is naturally defined from the two-derivative terms in the action. In functional geometry, the identification of a metric is more subtle, as we will see in \cref{sec:geometry}. In this section, we will focus on obtaining on-shell covariant building blocks of EFT amplitudes on the field configuration manifold, and will stay agnostic about the existence of a metric on this manifold.

\subsection{Amplitudes and correlation functions}

In \cref{sec:field}, we studied correlation functions as quantities on the target manifold: at any point $\phi$ on the target manifold, we expanded the action as in \cref{eq:L_2d_expand_phi}, and defined correlation functions in the presence of a constant source $\J_i(x) = -U_{,i}(\phi)$ to cancel the linear term in the expansion; see \cref{eq:ccf_phi}. The result was equivalent to defining correlation functions via the standard Feynman rules. A natural generalization of this prescription would be to expand the action around a point $\phi$ on the field configuration manifold as in \cref{eq:S_expand}, and define correlation functions in the presence of a nonconstant source that cancels the linear term, $\J_i(x) = -\frac{\delta S}{\delta\phi^i(x)}$, or equivalently, $\J_a = -S_{,a}[\phi]$:
\begin{equation}
\bigl\langle \eta^1 \dots \, \eta^n\bigr\rangle_{\text{c}, \,\phi} 
\equiv \frac{1}{i^n}\,\frac{\partial}{\partial\J_1} \cdots \frac{\partial}{\partial\J_n} \log \int\D\eta\, e^{i \bigl(S[\phi+\eta] +\J_a\, \eta^a\bigr)} \biggr|_{\J_a = -S_{,a}[\phi]} \,.
\end{equation}
The result is again equivalent to defining correlation functions via the standard Feynman rules, with propagator given by $\Delta^{ab}[\phi]$, i.e., the inverse of $\Delta^{-1}_{ab}[\phi]$ (which should exist in a finite neighborhood of $\overline\phi$ for any healthy theory):
\begin{equation}
\Delta^{ab}[\phi]\,\Delta^{-1}_{bc}[\phi] = \delta^a_c \,,
\label{eq:Delta_Delta_inv}
\end{equation}
and vertices given by $V_{1\dots n}$. The only difference compared to the target manifold case in \cref{sec:field} is that momentum conservation at each vertex is replaced by index contraction between the vertex and the propagators connected to it---see e.g., \cref{eq:M3_M4_fun} below. As in \cref{sec:field}, we will often leave the $\phi$ arguments implicit in what follows to reduce notational clutter. It is understood that all quantities on the field configuration manifold are evaluated at a general point $\phi$ unless specified otherwise.

Generalizing \cref{eq:M_def}, we can define amputated correlation functions $\M_{1\dots n}$ on the field configuration manifold by factoring out the functional version of external propagators:
\begin{equation}
\bigl\langle \eta^1 \cdots\, \eta^n\bigr\rangle_{\text{c}, \,\phi} \equiv i^n \Delta^{11'}[\phi] \cdots \Delta^{nn'}[\phi]\, i\M_{1'\cdots n'}[\phi] \,.
\label{eq:M_def_fun}
\end{equation}
For example, the 3- and 4-point amputated correlation functions are given by:
\begin{subequations}
	\label{eq:M3_M4_fun}
	\begin{align}
	\M_{123} &= V_{123} \,,\\[5pt]
	\M_{1234} &= V_{1234} - \bigl(\Delta^{ab}\, V_{a12} \, V_{b34}\bigr)_\text{3 terms} \,.
	\end{align}
\end{subequations}
In the constant field limit and for actions of the form \cref{eq:L_2d}, 
\begin{subequations}
\begin{align}
\Delta^{aa'}[\phi] \bigr|_{\phi(x) =\phi} &= \Delta^{i_{a^{\vphantom{\prime}}} i_{a'}}(\phi\,; p_a)\,\delta^{p_{a^{\vphantom{\prime}}}p_{a'}} \,,\\[5pt]
\M_{1'\cdots n'}[\phi] \bigr|_{\phi(x) =\phi} &= \M_{i_{1'}\dots i_{n'}}(\phi\,; p_{1'}, \dots, p_{n'})\, \delta_{p_{1'}\dots p_{n'}} \,,
\label{eq:M_fsl}
\end{align}
\end{subequations}
so we reproduce \cref{eq:M_def} from \cref{eq:M_def_fun}.

To derive scattering amplitudes from correlation functions, we again need to find the particle states. Let us label the one-particle states by $|\alpha, P\rangle$, where $P$ is an on-shell 4-momentum, i.e., $P^2 = m_\alpha^2$. Here and in what follows, we use capital letters for on-shell momenta. For EFT actions of the form \cref{eq:L_2d}, we were able to use a vielbein obtained from the target manifold metric to find the mass basis. At tree level, the vielbein can also be extracted from the overlap between states created by the fluctuation fields $\eta$ around the vacuum point $\overline\phi$ and one-particle states \cite{Cheung:2021yog}:
\begin{equation}
\langle 0 | \eta^i(x) | \alpha, P \rangle = \overline e_\alpha{}^i\, e^{-iP\cdot x}\,.
\label{eq:e_def}
\end{equation}
For more general EFT actions, we can use the same \cref{eq:e_def} to define a set of objects $\overline e_\alpha{}^i$, although they may not represent a vielbein. Then just as in field space geometry, $\overline e^\alpha{}_i\,\eta^i$ (where $\overline e^\alpha{}_i$ is the inverse of $\overline e_\alpha{}^i$) are the fields that create mass eigenstates from the vacuum, whose wave functions are plane waves with unit normalization:
\begin{equation}
\langle 0 | \,\overline e^\alpha{}_i\,\eta^i(x) | \alpha, P \rangle = e^{-iP\cdot x}\,.
\end{equation}
The propagator in the mass basis is given by: 
\begin{align}
\overline\Delta^{(\alpha p)(\beta q)} &\equiv \overline e^\alpha{}_i\,\overline e^\beta{}_j\, \overline\Delta^{(ip)(jq)} =  \overline e^\alpha{}_i\,\overline e^\beta{}_j\, \langle \eta^{(ip)}\, \eta^{(jq)} \rangle_{\overline\phi}^{} \notag\\[5pt]
&= \delta^{\alpha\beta}\delta^{pq}\, \frac{i}{p^2-m_\alpha^2} + \text{(non-singular)} \,.
\label{eq:prop}
\end{align}
As in \cref{sec:field}, we use bar to denote quantities evaluated at $\phi = \overline\phi$. 

To write on-shell amplitudes in terms of quantities defined on the field configuration manifold, let us define:
\begin{equation}
\A_{(\alpha_1 P_1) \dots (\alpha_n P_n)} \equiv \delta_{P_1\dots  P_n}\, \A_{\alpha_1\dots \alpha_n}(P_1,\dots, P_n)\,.
\end{equation}
The intuition is that $\A_{(\alpha_1 P_1) \dots (\alpha_n P_n)}$ should be related to $\overline\M_{1\dots n} \equiv \overline\M_{(i_1 p_1)\dots (i_n p_n)}$ via a set of objects connecting the fields labeled by $(i_a p_a)$ to the particle states labeled by $(\alpha_a P_a)$, with $P_a$ a set of on-shell momenta. To see that this is indeed the case, we apply the LSZ formula to extract residues of correlation functions among mass eigenstate fields as follows: 
\begin{align}
\A_{(\alpha_1 P_1) \dots (\alpha_n P_n)} &= \frac{1}{i^{n+1}} \lim_{p_a\to P_a} \overline\Delta^{\,-1}_{(\alpha_1 p_1)(\beta_1 q_1)} \dots\, \overline\Delta^{\,-1}_{(\alpha_n p_n)(\beta_n q_n)} \bigl\langle \overline e^{\beta_1}{}_{j_1} \,\eta^{(j_1 q_1)}\dots \, \overline e^{\beta_n}{}_{j_n} \,\eta^{(j_n q_n)} \bigr\rangle_{\text{c},\,\overline\phi} \notag\\[5pt]
&=  \frac{1}{i^{n+1}} \lim_{p_a\to P_a} \overline e_{\alpha_1}{}^{i_1}\, \overline\Delta^{\,-1}_{(i_1 p_1)(j_1 q_1)} \dots \, \overline e_{\alpha_n}{}^{i_n}\, \overline\Delta^{\,-1}_{(i_n p_n)(j_n q_n)} \bigl\langle \eta^{(j_1 q_1)}\dots\,\eta^{(j_n q_n)} \bigr\rangle_{\text{c},\,\overline\phi} \notag\\[5pt]
&= \lim_{p_a\to P_a} \overline e_{\alpha_1}{}^{i_1} \dots \, \overline e_{\alpha_n}{}^{i_n}\, \overline\M_{(i_1 p_1) \dots (i_n p_n)} \notag\\[5pt]
&= \overline e_{\alpha_1}{}^{i_1} \, \delta_{P_1}^{p_1} \dots \, \overline e_{\alpha_n}{}^{i_n} \, \delta_{P_n}^{p_n} \,\overline\M_{(i_1 p_1) \dots (i_n p_n)} \,.
\label{eq:A_fun_0}
\end{align}
We have used the relation between flavor- and mass-basis propagators in \cref{eq:prop} to arrive at the second line, and used \cref{eq:M_def_fun} to arrive at the third line. Finally, we have rewritten $\lim\limits_{p_a\to P_a}$ as integrals over momentum space $\delta$-functions (recall from \cref{eq:contract} that contracted momentum indices imply integration). The objects connecting $\A_{(\alpha_1 P_1) \dots (\alpha_n P_n)}$ and $\overline\M_{(i_1 p_1) \dots (i_n p_n)}$ in \cref{eq:A_fun_0} are nothing but the wave function factors in momentum space:
\begin{equation}
\overline e_\alpha{}^i\, \delta_P^p = \langle 0 | \eta^{(ip)} | \alpha, P \rangle \equiv \Psi^{(ip)}_{(\alpha P)} \,,
\label{eq:wf}
\end{equation}
We can now write \cref{eq:A_fun_0} in a compact form, applying our shorthand notation $(i_a p_a) \to a$:
\begin{equation}
\A_{(\alpha_1 P_1) \dots (\alpha_n P_n)} = \Psi^1_{(\alpha_1 P_1)} \,\dots \, \Psi^n_{(\alpha_n P_n)}\, \overline\M_{1\dots n} \,.
\label{eq:A_fun}
\end{equation}
This is the functional version of \cref{eq:A_fs}.

\subsection{On-shell covariance}
\label{sec:functional_osc}

From \cref{eq:A_fun} we see that on-shell amplitudes $\A_{(\alpha_1 P_1)\dots(\alpha_n P_n)}$ can be obtained from amputated correlation functions $\M_{1\dots n}[\phi]$ on the field configuration manifold by
\begin{itemize}
	\item[1)] taking the vacuum limit $\phi\to\overline\phi$, and
	\item[2)] contracting with wave function factors $\Psi^{(ip)}_{(\alpha P)}$ defined in \cref{eq:wf}, which implements both the on-shell limit and projection onto mass eigenstates.
\end{itemize}
Under a general field redefinition, $\phi^{(ip)} = F^{(ip)}[\widetilde\phi]$ with $F^{(ip)}$ a set of analytic functionals, which maps $\widetilde\phi$ to $\phi$ and $\widetilde\phi+\widetilde\eta$ to $\phi + \eta$, we have:
\begin{equation}
\eta^{(ip)} = F^{(ip)}[\widetilde{\phi}+\widetilde{\eta}]-F^{(ip)}[\widetilde{\phi}] = \frac{\partial\phi^{(ip)}}{\partial\widetilde{\phi}^{(jq)}} \, \widetilde{\eta}^{(jq)} + \O(\widetilde\eta^2)\,.
\end{equation}
The wave function factor therefore transforms as:
\begin{equation}
\widetilde\Psi^{(ip)}_{(\alpha P)} = \frac{\partial\widetilde\phi^{(ip)}}{\partial\phi^{(jq)}} \biggr|_{\phi=\overline\phi} 
\,\Psi^{(jq)}_{(\alpha P)} +\langle 0 | \O(\eta^2) | \alpha, P\, \rangle \,.
\label{eq:wf_transform}
\end{equation}
The $\O(\eta^2)$ terms can have a nontrivial overlap with one-particle states, but only at loop level. Therefore, as far as tree-level calculations are concerned, the wave function factors on the right-hand side of \cref{eq:A_fun} transform as vectors at the vacuum point $\overline\phi$, for a fixed set of particle states labeled by $(\alpha_1 P_1)\dots (\alpha_n P_n)$.

Following the same logic as in \cref{sec:field_osc}, we conclude that amputated correlation functions $\M_{1\dots n}[\phi]$ defined on the field configuration manifold must transform as tensors in the vacuum and on-shell limits. More precisely, under a general field redefinition $\phi^{(ip)} = F^{(ip)}[\widetilde\phi]$ which can involve derivatives, we have
\begin{equation}
\widetilde\M_{1\dots n}[\widetilde\phi] = \frac{\partial\phi^{1'}}{\partial\widetilde\phi^1} \,\cdots\, \frac{\partial\phi^{n'}}{\partial\widetilde\phi^n} \, \M_{1'\dots n'}\bigl[F [\widetilde\phi]\bigr] + X_{1\dots n}[\widetilde\phi] \,,
\end{equation}
where $X_{1\dots n}$ satisfies
\begin{equation}
\widetilde\Psi^1_{(\alpha_1 P_1)}\dots \widetilde\Psi^n_{(\alpha_n P_n)}\, \overline X_{1\dots n} = 0 \,.
\label{eq:X_fun_long}
\end{equation}
This implies that the {\it on-shell equivalence} relation ``\,$\oseq$\,'' that we introduced in \cref{sec:field_osc} can be naturally generalized to the field configuration manifold: we say that two objects on the field configuration manifold are on-shell equivalent if they are equal upon performing the two operations listed at the beginning of this subsection. With this definition, we can abbreviate \cref{eq:X_fun_long} as:
\begin{equation}
X_{1\dots n} \oseq\, 0\,.
\end{equation}
In addition to $X_{1\dots n}$, two other objects that are on-shell equivalent to zero are:
\begin{equation}
S_{,c} \,\oseq\, 0 \qquad \text{and} \qquad 
\Delta^{-1}_{ca} \,\oseq\, 0\,,
\label{eq:oseq_0_fun}
\end{equation}
where $p_a$ is an external momentum.

We say an object on the field configuration manifold is {\it on-shell covariant} if it transforms as a tensor up to inhomogeneous terms that are on-shell equivalent to zero. $\M_{1\dots n}$ are generally not tensors, but must be on-shell covariant~\cite{Cohen:2022uuw,Cohen:2023ekv}. Just like on the target manifold, sums and products of on-shell covariant objects on the field configuration manifold are on-shell covariant. If an object is on-shell equivalent to a tensor, it must be on-shell covariant.

\subsection{On-shell covariant building blocks}
\label{sec:functional_bb}

In \cref{sec:field}, we saw that for an EFT action truncated at $\O(\partial^2)$, amputated correlation functions on the target manifold $\M_{i_1\dots i_n}$ can be expressed in terms of elementary building blocks $\Delta^{ij}$ and $\{\V_{j_1\dots j_m}\}$, each of which is on-shell covariant under nonderivative field redefinitions $\phi^i= f^i(\widetilde\phi)$. Here we have seen that for a general EFT action, amputated correlation functions on the field configuration manifold $\M_{1\dots n} \equiv \M_{(i_1p_1)\dots (i_np_n)}$ must be on-shell covariant under general field redefinitions $\phi^{(ip)}= F^{(ip)}[\widetilde\phi]$. The next step is to show that we can express $\M_{1\dots n}$ in terms of on-shell covariant building blocks on the field configuration manifold. 

Up to this point, our discussion in this section has only relied on the existence of the field configuration manifold, upon which we can define functions (which are functionals of $\phi$) and take their derivatives (which are the ordinary functional derivatives, up to normalization). In particular, the EFT action $S[\phi]$ is a scalar on the field configuration manifold, and its derivatives give the standard Feynman propagator and vertices; see \cref{eq:Delta_V}. We would like to construct on-shell covariant versions of the vertices just as in \cref{sec:field}. For this we need to introduce some additional structures on the field configuration manifold, namely a connection with which we can covariantize derivatives.

Suppose we have a connection $\Gamma^a_{bc}$ on the field configuration manifold. For a tensor $T^{a\dots}{}_{b\dots}$, we can take its covariant derivative with respect to this connection following the standard formula:
\begin{equation}
T^{a\dots}{}_{b\dots;c} \equiv T^{a\dots}{}_{b\dots,c} + \bigl(\Gamma^a_{cd} \,T^{d\dots}{}_{b\dots} + \,\cdots \bigr) - \bigl(\Gamma^d_{cb}\, T^{a\dots}{}_{d\dots} +\,\cdots\bigr) \,.
\end{equation}
Note that we do not assume the connection $\Gamma^a_{bc}$ is torsion-free and we must therefore be careful about the ordering of its lower indices. Under a coordinate transformation on the field configuration manifold, the connection transforms as:
\begin{equation}
\widetilde\Gamma^a_{bc} = \frac{\partial\widetilde\phi^a}{\partial\phi^d}\,\frac{\partial\phi^e}{\partial\widetilde\phi^b}\,\frac{\partial\phi^f}{\partial\widetilde\phi^c} \,\Gamma^d_{ef} +\frac{\partial\widetilde\phi^a}{\partial\phi^d} \, \frac{\partial^2\phi^d}{\partial\widetilde\phi^b\partial\widetilde\phi^c} \,.
\label{eq:connection_transform}
\end{equation}
We leave the explicit expression of the connection unspecified at this point. In principle, we could define many different connections on any manifold, each of which leads to a distinct notion of covariant differentiation. Our discussion in this section is valid for any connection on the field configuration manifold. As we will see shortly, just like in field space geometry, any connection defines a set of on-shell covariant building blocks for $\M_{1\dots n}$.

Given a connection $\Gamma^a_{bc}$, we can now repeat the analysis in \cref{sec:field_nt} on the field configuration manifold. Expanding the fields around a general point $\phi$ on the field configuration manifold, we can solve for the geodesic from $\phi$ to $\phi+\eta$ and obtain:
\begin{equation}
\eta^a = \xi^a -\sum_{m=2}^\infty \frac{1}{m!}\,\Gamma^a_{(1\dots m)}\, \xi^{1}\dots \xi^{m}\,,
\label{eq:eta_to_xi_fun}
\end{equation}
where $\xi$ are the normal coordinates at $\phi$ with respect to the connection $\Gamma^a_{bc}$, and $\Gamma^a_{1\dots m}$ are defined recursively via
\begin{equation}
\Gamma^a_{1\dots m (m+1)} \equiv \Gamma^a_{1\dots m, m+1} - \sum_{b=1}^m \Gamma^c_{(m+1) b} \, \Gamma^a_{1\dots \slashed{b} c\dots m} \,.
\label{eq:Gamma_def_fun}
\end{equation}
Similarly to \cref{eq:Gamma_def}, the notation $\slashed{b}c$ means the index $b$ is replaced by $c$. In terms of $\xi$, the expansion of the action has a manifestly covariant expression: 
\begin{equation}
S[\phi+\eta] = \sum_{m=0}^\infty \frac{1}{m!}\, S_{;(1\dots m)} \, \xi^1 \dots \xi^m \,.
\end{equation}
That the expansion coefficients of a scalar function in normal coordinates are the symmetrized covariant derivatives can be proved following the same steps as in App.~A of Ref.~\cite{Cohen:2021ucp}; they give rise to covariant Feynman rules for the correlation functions in the $\xi$ basis~\cite{Volkov:1973vd,Finn:2019aip,Finn:2020nvn}. In the original $\eta$ basis, including the source term with $\J_a = -S_{,a}$, we have
\begin{equation}
S[\phi+\eta] + \J_a\eta^a = S +\frac{1}{2}\,\Delta^{-1}_{ab}\,\xi^a\xi^b + \sum_{m=3}^\infty \frac{1}{m!}\, \Bigl[ S_{;(1\dots m)}  +\Gamma^a_{(1\dots m)} \, S_{,a}\Bigr]\, \xi^1 \dots \xi^m \,,
\label{eq:S_plus_Jeta_fun}
\end{equation}
where we have used $S_{;(12)} = S_{,12} -\Gamma^a_{(12)}\,S_{,a} = \Delta^{-1}_{12} -\Gamma^a_{(12)}\,S_{,a}$ to simplify the quadratic term. In complete analogy to \cref{eq:eta_cf}, we can compute correlation functions of $\eta$ in a general background by relating them to correlation functions of $\xi$:
\begin{align}
\langle \eta^1\dots \eta^n\rangle_\phi &= \langle\xi^1 \dots \xi^n \rangle_\phi \notag\\[5pt]
&\quad\;\; -\sum_{a=1}^n \sum_{m=2}^\infty \frac{1}{m!} \, \Gamma^a_{(1'\dots m')}\, \bigl\langle \xi^1 \dots \slashed{\xi}^a (\xi^{1'}\dots \xi^{m'}) \dots \xi^n \bigr\rangle_\phi + \cdots \,,
\label{eq:eta_cf_fun}
\end{align}
where $\slashed{\xi}^{a} (\xi^{1'}\dots \xi^{m'})$ means $\xi^{a}$ is replaced by $\xi^{1'}\dots \xi^{m'}$, and the ellipses at the end of the equation represent terms where more than one $\xi$'s in the correlation function are replaced by products of $\xi$'s. All correlation functions in \cref{eq:eta_cf_fun} are evaluated with respect to the action plus source term shown in \cref{eq:S_plus_Jeta_fun}.\footnote{As in \cref{sec:field_nt}, we can neglect the difference between the path integral measures $\D\eta$ and $\D\xi$ for tree-level calculations. Care must be taken if one wishes to go beyond tree level because, depending on the connection chosen, the Jacobian factor from \cref{eq:eta_to_xi_fun} may have nontrivial momentum dependence and give rise to loop integrals that do not vanish even in dimensional regularization; see e.g., Ref.~\cite{Cohen:2024fak}.} Following the same logic as in \cref{sec:field_nt}, we conclude that correlation functions of $\eta$ can be obtained from the standard set of Feynman diagrams, with propagators given by $\Delta^{ab}$, which is the inverse of
\begin{equation}
\Delta^{-1}_{ab} = S_{,ab} = S_{;(ab)} +\Gamma^c_{(ab)}\, S_{,c}\,,
\label{eq:Delta_inv_fun}
\end{equation}
and vertices given by:
\begin{equation}
\V_{1\dots m} = S_{;(1\dots m)} + \Gamma^c_{(1\dots m)}\, S_{,c} +\sum_{a\in\text{ext}} \Gamma^c_{(1\dots \slashed{a}\dots m)}\, \Delta^{-1}_{ca} \,.
\label{eq:Vcal_fun}
\end{equation}
In other words, we now have the functional version of \cref{eq:Mfun_fs}:
\begin{equation}
\boxed{\rule[-1.5ex]{0pt}{4.5ex}\quad
\M_{1\dots n} = \Mfun_{1\dots n}\bigl(\Delta^{ab}, \{ V_{a_1\dots a_m}\}\bigr) = \Mfun_{1\dots n}\bigl(\Delta^{ab}, \{ \V_{a_1\dots a_m}\}\bigr) \,.
\quad}
\label{eq:Mfun_fg}
\end{equation}

For example, the 3- and 4-point amputated correlation functions are given by (compare \cref{eq:M3_M4_fun}):
\begin{subequations}
\label{eq:M3_M4_fg}
\begin{align}
\M_{123} & = \V_{123} \,,\\[5pt]
\M_{1234} &= \V_{1234} - \bigl(\Delta^{ab}\, \V_{a12} \, \V_{b34}\bigr)_\text{3 terms}\,,
\end{align}
\end{subequations}
where
\begin{subequations}
\begin{align}
\V_{123} &= S_{;(123)} + \Gamma^c_{(123)}\, S_{,c} + \Gamma^c_{(23)}\, \Delta^{-1}_{c1} + \Gamma^c_{(13)}\, \Delta^{-1}_{c2} +\Gamma^c_{(12)}\, \Delta^{-1}_{c3} \,,\\[5pt]
\V_{a12} &= S_{;(a12)} + \Gamma^c_{(a12)}\, S_{,c} + \Gamma^c_{(a2)}\, \Delta^{-1}_{c1} + \Gamma^c_{(a1)}\, \Delta^{-1}_{c2} \,,\\[5pt]
\V_{b34} &= S_{;(b34)} + \Gamma^c_{(b34)}\, S_{,c} + \Gamma^c_{(b4)}\, \Delta^{-1}_{c3} + \Gamma^c_{(b3)}\, \Delta^{-1}_{c4} \,,\\[5pt]
\V_{1234} &= S_{;(1234)} +\Gamma^c_{(1234)}\, S_{,c} + \Bigl[\Gamma^c_{(234)}\, \Delta^{-1}_{c1} \Bigr]_\text{4 terms}\,.
\end{align}
\end{subequations}

The propagator $\Delta^{ab}$ and vertices $\V_{1\dots m}$ defined in \cref{eq:Delta_inv_fun,eq:Vcal_fun} are not tensors, but they are all on-shell covariant. The nontensorial terms, which again come from the mismatch between a general field basis $\eta$ and normal coordinates $\xi$, are of the form:
\begin{equation}
\Gamma^c_{(1\dots m)} \,S_{,c} \qquad \text{and}\qquad \Gamma^c_{(1\dots \slashed{a}\dots m)}\, \Delta^{-1}_{ca} \quad (a\in\text{ext})\,.
\end{equation}
So, as long as $\Gamma^c_{(1\dots m)}$ and $\Gamma^c_{(1\dots \slashed{a}\dots m)}$ are not singular, the nontensorial terms are on-shell equivalent to zero because of \cref{eq:oseq_0_fun}. As a result, $\Delta^{ab}$ and $\V_{1\dots m}$ are on-shell equivalent to a set of tensors:
\begin{subequations}
\label{eq:Delta_V_oseq}
\begin{align}
\Delta^{ab} &\oseq \DDelta^{ab}\,,\\[5pt]
\V_{1\dots m} &\oseq S_{;(1\dots m)}\,,
\end{align}
where $\DDelta^{ab}$ (note boldface) is the inverse of
\begin{equation}
\DDelta^{-1}_{ab} \equiv S_{;(ab)}\,.
\end{equation}
\end{subequations}
A subtlety is that some choices of the functional connection $\Gamma^c_{(1\dots \slashed{a}\dots m)}$ may have a pole in the momentum $p_c$, in which case $\Gamma^c_{(1\dots \slashed{a}\dots m)}\, \Delta^{-1}_{ca}$ can be nonzero in the on-shell limit, meaning $\V_{1\dots m}$ is not on-shell equivalent to $S_{;(1\dots m)}$. This happens, for example, in any massive theory for the set of connection choices discussed in Ref.~\cite{Cohen:2024bml}; more on this in \cref{sec:geo_metric_1}. However, we show in \cref{app:osc} that even in this case, the $\V_{1\dots m}$ are still on-shell covariant under field redefinitions that do not artificially introduce single-particle poles. Therefore, \cref{eq:Mfun_fg} gives a manifestly on-shell covariant expression for the $n$-point amputated correlation function in a general scalar EFT; the special cases of $n=3, 4$ are given in \cref{eq:M3_M4_fg}. 

Having understood how EFT amplitudes can be constructed from on-shell covariant building blocks on the field configuration manifold, we now turn to the question of defining a Riemannian geometry on this manifold.

\section{Geometry of the field configuration manifold}
\label{sec:geometry}

In the previous section, we showed that EFT amplitudes can be expressed in terms of on-shell covariant propagators and vertices on the field configuration manifold; see \cref{eq:Mfun_fg}. Throughout that discussion, we stayed agnostic about the existence of a metric on the field configuration manifold. The point we wish to emphasize is that \cref{eq:Mfun_fg} is valid---and the propagator $\Delta^{ab}$ and vertices $\V_{1\dots m}$ are on-shell covariant under nonsingular field redefinitions---for {\it any\,} $\Gamma^a_{bc}$ that transforms as a connection, which may or may not be associated with a metric. Nevertheless, given the similarity between on-shell covariant amplitude expressions on both the target manifold and the field configuration manifold---see \cref{eq:Mfun_fs,eq:Mfun_fg}---one would naturally wonder whether we can identify a metric on the field configuration manifold that is a natural generalization of the field space geometry metric. This is the question we aim to address in this section. We will first review the idea introduced in Ref.~\cite{Cheung:2022vnd} and further discussed in Ref.~\cite{Cohen:2024bml}, and discuss its features and limitations. We then advocate a variation on this proposal that gives an algorithm to unambiguously define a metric on the field configuration manifold, which renders the reduction from functional geometry to field space geometry transparent.

\subsection{Preliminaries}
\label{sec:geo_pre}

A key observation that allowed us to identify a metric in field space geometry is that $\partial_\mu\phi^i$ transforms like a vector under a nonderivative field redefinition $\phi^i = f^i(\widetilde\phi)$:
\begin{equation}
\partial_\mu\phi^i = \frac{\partial\phi^i}{\partial\widetilde\phi^j}\, \partial_\mu\widetilde\phi^j\,.
\label{eq:dphi_transform_fs}
\end{equation}
Now consider a local field redefinition that involves derivatives:
\begin{equation}
\phi^i = f^i\bigl(\widetilde\phi\,,\, \partial_\mu\widetilde\phi\,,\, \partial_\mu\partial_\nu\widetilde\phi\,,\, \dots \bigr) \,.
\label{eq:redef_local}
\end{equation}
Here ``local'' means $\phi^i(x)$ is a polynomial function of $\widetilde\phi^j(x)$, $\partial_\mu\widetilde\phi^j(x)$, $\partial_\mu\partial_\nu\widetilde\phi^j(x)$, etc., all evaluated at the same spacetime point $x$. Taking a spacetime derivative, we find:
\begin{equation}
\partial_\mu\phi^i = \frac{\partial f^i}{\partial\widetilde\phi^j} \,\partial_\mu\widetilde\phi^j + \frac{\partial f^i}{\partial(\partial_\nu\widetilde\phi^j)} \,\partial_\mu\partial_\nu\widetilde\phi^j + \frac{\partial f^i}{\partial(\partial_\nu\partial_\rho\widetilde\phi^j)} \,\partial_\mu\partial_\nu\partial_\rho\widetilde\phi^j + \cdots \,,
\label{eq:dphi_transform_expand}
\end{equation}
where all terms are evaluated at $x$. Comparing with 
\begin{equation}
\frac{\delta\phi^i(x)}{\delta\widetilde\phi^j(y)} = \frac{\partial f^i}{\partial\widetilde\phi^j}\biggr|_x \delta^d(x-y) + \frac{\partial f^i}{\partial(\partial_\nu\widetilde\phi^j)}\biggr|_x \partial_\nu^x\,\delta^d(x-y) + \frac{\partial f^i}{\partial(\partial_\nu\partial_\rho\widetilde\phi^j)}\biggr|_x \partial^x_\nu\,\partial^x_\rho\,\delta^d(x-y) +\cdots\,,
\end{equation}
where the superscripts $x$ on $\partial$ denote that partial derivatives are taken with respect to $x$, we see that \cref{eq:dphi_transform_expand} is equivalent to a vector transformation law on the field configuration manifold: 
\begin{equation}
\partial_\mu\phi^i(x) = \int_y\, \frac{\delta\phi^i(x)}{\delta\widetilde\phi^j(y)}\,\partial_\mu\widetilde\phi^j(y)\,,
\label{eq:dphi_transform_x}
\end{equation}
thereby generalizing \cref{eq:dphi_transform_fs}.

We can in fact relax the requirement that the field redefinition be local. The essential property of the field redefinitions we consider is that they are translation-invariant. Denoting the infinitesimal spacetime translation operator by $\T_\epsilon$, we have
\begin{equation}
\T_\epsilon\, \phi^i(x) = \phi^i(x+\epsilon) = \phi^i(x) + \epsilon^\mu\,\partial_\mu\phi^i(x) + \O(\epsilon^2)\,.
\label{eq:T_phi}
\end{equation}
The left-hand side of this equation should be understood as the function $\T_\epsilon\,\phi^i$, which is obtained by acting $\T_\epsilon$ on the function $\phi^i$, evaluated at $x$. We say a field redefinition $\phi^i = F^i[\widetilde\phi]$ is {\it translation-invariant} if it commutes with spacetime translation:
\begin{equation}
\T_\epsilon\, \phi^i = \T_\epsilon\, F^i[\widetilde\phi] = F^i[\T_\epsilon\,\widetilde\phi] \,.
\label{eq:TF_commute}
\end{equation}
Any local field redefinitions of the form \cref{eq:redef_local} trivially commute with spacetime translation, but the set of translation-invariant field redefinitions is larger than the set of local field redefinitions. 

To prove the vector transformation of $\partial_\mu\phi^i$ on the field configuration manifold under translation-invariant field redefinitions, we can expand both sides of \cref{eq:TF_commute} in $\epsilon$. The left-hand side, evaluated at $x$, is already given by \cref{eq:T_phi}. The right-hand side is:
\begin{equation}
F^i[\T_\epsilon\,\widetilde\phi] = F^i[\widetilde\phi] + \int_y\,\frac{\delta F^i}{\delta\widetilde\phi^j(y)} \,\epsilon^\mu\,\partial_\mu\widetilde\phi^j(y) +\O(\epsilon^2) \,.
\end{equation}
Evaluating this expression at $x$, we see that \cref{eq:TF_commute} implies that
\begin{equation}
\T_\epsilon\, \phi^i(x) = \phi^i(x) + \epsilon^\mu \int_y\, \frac{\delta\phi^i(x)}{\delta\widetilde\phi^j(y)}\, \partial_\mu\widetilde\phi^j(y) +\O(\epsilon^2) \,.
\label{eq:T_phi_2}
\end{equation}
Comparing \cref{eq:T_phi,eq:T_phi_2}, we immediately reproduce \cref{eq:dphi_transform_x}.

Let us introduce some notation for later convenience. We denote:
\begin{equation}
\dphi{\mu}^{(ix)} \equiv \partial_\mu\phi^i(x)\,,
\end{equation}
to be read as the $(ix)$ component of the vector function $\dphi{\mu}$ on the field configuration manifold (which is a functional of $\phi$). Upon Fourier transformation:
\begin{equation}
\dphi{\mu}^a = \dphi{\mu}^{(i_a p_a)} = -ip_{a\mu} \phi^{(i_a p_a)} = -ip_{a\mu} \phi^a\,.
\end{equation}
Note that there is no sum over $a$ when $p_{a\mu}$ appears as a multiplicative factor in an expression. The vector transformation law \cref{eq:dphi_transform_x} can then be written concisely as
\begin{equation}
\dphi{\mu}^a = \frac{\partial\phi^a}{\partial\widetilde\phi^b}\; \dphit{\mu}^b\,.
\label{eq:dphi_transform_p}
\end{equation}

It is a useful exercise to check \cref{eq:dphi_transform_p} by considering field redefinitions in momentum space~\cite{Cohen:2024bml}. Assuming $\phi^i = F^i[\widetilde\phi]$ is nonsingular at $\widetilde\phi=0$, we can expand:
\begin{equation}
\phi^a = (c_0)^a + (c_1)^a_1 \,\widetilde\phi^1 + (c_2)^a_{12}\, \widetilde\phi^1 \widetilde\phi^2 + \cdots = \sum_{n=0}^\infty (c_n)^a_{1\dots n} \,\widetilde\phi^1 \cdots \widetilde\phi^n\,,
\end{equation}
where, consistent with our notation, $(c_n)^a_{1\dots n}$ carries flavor indices $i_a, i_1, \dots, i_n$ and is also a function of momenta $p_a, p_1, \dots, p_n$. For translation-invariant field redefinitions, these coefficients must satisfy momentum conservation:
\begin{equation}
(c_n)^a_{1\dots n} \propto \delta^{p_a}_{p_1\dots p_n} \,.
\end{equation}
As a result, 
\begin{equation}
\dphi{\mu}^a =-i\,\sum_{n=0}^\infty (c_n)^a_{1\dots n} \,p_{a\mu}\,\widetilde\phi^1 \cdots \widetilde\phi^n
=-i\,\sum_{n=0}^\infty (c_n)^a_{1\dots n} \,(p_1+\cdots +p_n)_\mu\,\widetilde\phi^1 \cdots \widetilde\phi^n \,,
\end{equation}
which is equal to:
\begin{equation}
\frac{\partial\phi^a}{\partial\widetilde\phi^b} \;\dphit{\mu}^b = -i\sum_{n=0}^\infty \sum_{b=1}^n (c_n)^a_{1\dots n} \,\widetilde\phi^1 \cdots p_{b\mu}\,\widetilde\phi^b \cdots \widetilde\phi^n \,,
\end{equation}
where the expression on the right-hand side means we are replacing 
$\widetilde\phi^b$ in $\widetilde\phi^1 \cdots \widetilde\phi^n$ by $p_{b\mu}\,\widetilde\phi^b$. This verifies \cref{eq:dphi_transform_p}. If we further restrict ourselves to local field redefinitions, $(c_n)^a_{1\dots n}$ would be polynomial functions of momenta. From the discussion above it is clear that the vector transformation of $\dphi{\mu}^a$ relies only on translation invariance but does not require the field redefinition to be local. On the other hand, it is worth noting that there exist many field redefinitions that are not translation-invariant, under which $\dphi{\mu}^a$ does not transform like a vector. A simple example is $\phi^{i_a}(x) = \varphi^{i_a}(x) + \widetilde\phi^{i_a}(x)$, with $\varphi^{i_a}(x)$ a spacetime-dependent background field. In this case, $(c_0)^a = \int_x e^{ip_a\cdot x}\,\varphi^{i_a}(x)$ is not proportional to $\delta^{p_a}$ for $\varphi^{i_a}(x) \ne $ constant, and indeed, $\frac{\partial\phi^a}{\partial\widetilde\phi^b} \;\dphit{\mu}^b = \dphit{\mu}^a \ne \dphi{\mu}^a$.

\subsection{Toward defining a metric}
\label{sec:geo_metric_1}

Now that we have a vector $\dphi{\mu}^a$ on the field configuration manifold, any metric $g_{ab}[\phi]$ must render the combination $\frac12\,g_{ab}[\phi]\,\dphi{\mu}^a\Dphi{\mu}^b$ a scalar under translation-invariant field redefinitions. Meanwhile, we know of a scalar on the field configuration manifold---the action $S[\phi]$. It is therefore tempting to identify:
\begin{equation}
S[\phi] = \frac12 \, g_{ab}[\phi] \, \dphi{\mu}^a \Dphi{\mu}^b\,,
\label{eq:S}
\end{equation}
or, more explicitly:
\begin{align}
S[\phi] &= \,\frac12 \int_{x_a, x_b} g_{(i_a x_a) (i_b x_b)} [\phi] \, \bigl(\partial_\mu\phi^{i_a}(x_a)\bigr) \, \bigl(\partial^\mu\phi^{i_b}(x_b)\bigr) \notag\\
&=  -\frac12 \int_{p_a,p_b} g_{(i_a p_a) (i_b p_b)} [\phi] \, (p_a\cdot p_b) \,\phi^{i_a}(p_a)\, \phi^{i_b}(p_b)\,.
\end{align}
This is the approach taken in Refs.~\cite{Cheung:2022vnd, Cohen:2024bml}. As a technical note, since $\dphi{\mu}^a$ transforms like a vector only under translation-invariant field redefinitions, whereas $S[\phi]$ is a scalar under all field redefinitions, the identification in \cref{eq:S} is consistent only across field bases that are related by translation-invariant field redefinitions. Practically this does not cause any issue because we almost always work with field bases where translation invariance is manifest (e.g., there are no spacetime-dependent couplings),\footnote{An exception is when we expand the fields around a spacetime-dependent background, as discussed at the end of \cref{sec:geo_pre}, in which case one obtains couplings that depend on the nonconstant background fields. So for example, normal coordinates around a point $\phi$ that is not on the constant field submanifold would not be a basis where we can identify $S[\phi]$ with $\frac12 \, g_{ab}[\phi] \, \dphi{\mu}^a \Dphi{\mu}^b$. However, once we have obtained the metric via \cref{eq:S} in a basis where translation invariance is manifest, it is straightforward to obtain the metric in any other bases (including normal coordinates) by simply performing the appropriate field redefinition.} and any two such bases are related by a translation-invariant field redefinition.

As a simple example, consider a one-flavor $\phi^3$ theory:
\begin{align}
S[\phi] &= \int_x\, \biggl[\, \frac12 (\partial_\mu\phi)(\partial^\mu\phi) -\frac12\s m^2\phi^2 -\frac{1}{6}\s\mu\s\phi^3 \biggr] \notag\\[5pt]
&= - \frac12\int_{p_a, p_b} \delta_{p_a p_b} (p_a\cdot p_b + m^2) \, \phi(p_a)\,\phi(p_b)
-\frac16 \,\mu \int_{p_a,p_b,p_c} \delta_{p_ap_bp_c} \,\phi(p_a)\phi(p_b)\phi(p_c) \,.
\end{align}
One can readily identify:
\begin{align}
g_{ab}[\phi] &= \frac{1}{p_a\cdot p_b}\, \biggl[ \,\delta_{p_a p_b} \bigl(p_a\cdot p_b + m^2\bigr) + \frac13\,\mu\int_{p_c} \delta_{p_a p_b p_c} \phi(p_c)\biggr] \notag\\[5pt]
&= \delta_{p_ap_b} \,\biggl(1 -\frac{m^2}{p_a^2}\biggr)
+ \frac13\,\mu\int_{p_c} \delta_{p_a p_b p_c}\frac{1}{p_a\cdot p_b}\,\phi(p_c) \,.
\label{eq:metric_phi3}
\end{align}
This example demonstrates a general property of metrics satisfying \cref{eq:S}: for massive theories, $g_{ab}$ vanishes in the vacuum and on-shell limits~\cite{Cohen:2024bml}. In \cref{eq:metric_phi3}, the (metastable) vacuum limit corresponds to $\overline\phi = 0$, so $\overline g_{ab}$ has a zero at $p_a^2 = m^2$ if $m\ne 0$. As a result, the Levi-Civita connection $\overline\Gamma^a_{bc}$, which involves the inverse metric, becomes singular when $p_a$ goes on-shell. This is not necessarily a problem: as mentioned at the end of \cref{sec:functional_bb} and elaborated in \cref{app:osc}, a simple pole in the connection does not ruin the on-shell covariance of $\V_{1\dots m}$ under nonsingular field redefinitions. 

Another notable feature of \cref{eq:S} is that, while a metric uniquely defines an action, an action does not uniquely define a metric~\cite{Cheung:2022vnd, Cohen:2024bml}. Given an EFT action, if one finds a $g_{ab}[\phi]$ that satisfies \cref{eq:S}, an equally good metric choice would be $g_{ab}[\phi]+h_{ab}[\phi]$, where $h_{ab}[\phi]$ is an arbitrary functional that satisfies $h_{ab}[\phi]\,\dphi{\mu}^a\, \Dphi{\mu}^b = 0$. For example, in the $\phi^3$ theory above, an alternative choice of metric that satisfies \cref{eq:S} would be:
\begin{equation}
g_{ab}'[\phi] = \delta_{p_ap_b} \,\biggl(1 -\frac{m^2}{p_a^2}\biggr) + \mu\int_{p_c} \delta_{p_a p_b p_c}\frac{1}{p_a\cdot p_b + p_b\cdot p_c + p_c \cdot p_a}\,\phi(p_c) \,.
\end{equation}
All we can say is that, if we start from an EFT action written in some basis and identify {\it a metric} that satisfies \cref{eq:S}, its tensor transformation would give {\it a metric} in the new basis, in the sense that $\widetilde S[\widetilde\phi] = \frac12 \,\widetilde g_{ab}[\widetilde\phi] \, \dphit{\mu}^a\, \Dphit{\mu}^b$. However, it might not be easy to identify the same metric $\widetilde g_{ab}[\widetilde\phi]$ starting from $\widetilde S[\widetilde\phi]$. In other words, we do not have a prescription to extract a unique metric from the action that is consistent across all bases. Different metric choices giving the same EFT action define different Riemannian geometries of the field configuration manifold---for example, they give different curvatures. Therefore, the lack of a prescription to unambiguously fix a metric poses a challenge if one wishes to assign physical meaning to geometric objects.

To further illustrate this challenge, let us consider an extreme example. Starting from a general scalar EFT action,
\begin{equation}
S[\phi] = \frac{1}{2}\,\overline\Delta^{\,-1}_{ab}[\phi]\,\phi^a\phi^b +\sum_{n=3}^\infty \frac{1}{n!}\,\overline V_{1\dots n}[\phi] \,\phi^1\dots \phi^n \,,
\label{eq:S_phi}
\end{equation}
where we assume the vacuum is at $\overline\phi = 0$, we can make the following perturbative field redefinition:
\begin{equation}
\widehat\phi^a = \phi^a + (c_2)^a_{12} \,\phi^1\phi^2 + (c_3)^a_{123}\,\phi^1\phi^2\phi^3 + \cdots \,.
\label{eq:redef_hat}
\end{equation}
By making judicious choices of $(c_2)^a_{12}$, $(c_3)^a_{123}$, etc., we can actually arrive at a free theory in the $\widehat\phi$\, basis:
\begin{align}
&(c_2)^a_{12} = \frac{1}{3!}\, \overline\Delta^{ab}\, \overline V_{b12} \,,\quad
(c_3)^a_{123} = \frac{1}{4!} \biggl[ \overline\Delta^{ab}\, \overline V_{b123} -\frac{1}{9}\, \bigl(\overline\Delta^{ab}\, \overline\Delta^{cd}\, \overline V_{bc1} \, \overline V_{d23} \bigr)_\text{3 terms} \biggr]\,, \quad\dots \notag\\[5pt]
&\hspace{80pt}\Rightarrow\qquad \widehat S [\widehat \phi] = \frac{1}{2}\, \overline\Delta^{\,-1}_{ab}\,\widehat\phi^a\widehat\phi^b \,.
\end{align}
While turning an interacting action into a free action might appear unusual, there is really no contradiction here. We have made a nonlocal field redefinition (due to the $\overline{\Delta}^{ab}$ factors) for which modifications to the source term cannot be neglected~\cite{Cohen:2024fak}; in fact, all interactions are encoded in the coupling to the source in the $\widehat\phi$\, basis. Making the obvious choice of metric in the $\widehat\phi$\, basis, 
\begin{align}
\widehat g_{ab}[\widehat\phi] = -\frac{1}{p_a\cdot p_b}\, \overline\Delta^{\,-1}_{ab}\,, 
\end{align}
we then obtain a metric in the original $\phi$ basis via a tensor transformation:
\begin{align}
g_{ab}[\phi] &= \frac{\partial\widehat\phi^c}{\partial\phi^a}\,\frac{\partial\widehat\phi^d}{\partial\phi^b}\;\widehat g_{cd}[\widehat\phi] \notag\\[5pt]
&= -\frac{1}{p_a\cdot p_b}\, \overline\Delta^{\,-1}_{ab} + \frac{1}{3}\, \biggl(\frac{1}{p_a^2} + \frac{1}{p_b^2} \biggr) \, \overline V_{abc} \,\phi^c \notag\\[5pt]
&\quad + \biggl\{\biggl(\frac{1}{p_a^2} + \frac{1}{p_b^2} \biggr) \biggl[ \,\frac{1}{8} \, \overline V_{abcd} -\frac{1}{72} \, \overline\Delta^{ef} (\overline V_{eab}\, \overline V_{fcd} + 2\, \overline V_{eac}\, \overline V_{fbd}) \biggr] \notag\\[5pt]
&\qquad\quad -\frac{1}{9}\, \frac{1}{(p_a + p_c) \cdot (p_b + p_d)} \, \overline\Delta^{ef} \, \overline V_{eac} \, \overline V_{fbd} \biggr\} \, \phi^c \,\phi^d + \cdots \,.
\label{eq:g_flat}
\end{align}
Once we have \cref{eq:g_flat}, we can forget about the $\widehat\phi$\, basis because it is straightforward to verify that \cref{eq:g_flat} is a valid metric choice for the action in \cref{eq:S_phi}, in the sense that \cref{eq:S} is satisfied. This metric is flat on the entire field configuration manifold because it is secretly the tensor transformation of a constant metric. One can also check explicitly, order by order in $\phi$, that the Riemann curvature tensor computed from \cref{eq:g_flat} vanishes identically. Therefore, we have shown that for any action of the form \cref{eq:S_phi}, there exists a metric consistent with \cref{eq:S} for which the curvature of the field configuration manifold is identically zero.\footnote{Note that there is no contradiction between a flat metric and nonvanishing amplitudes, because amplitude expressions contain nontensorial terms which do not vanish in the vacuum and on-shell limits in the present case (cf.\ the discussion at the end of \cref{sec:functional_bb}). There is also no contradiction between nonvanishing $\V_{1\dots m}$ vertices in the $\phi$ basis and vanishing vertices in the $\widehat\phi$\, basis: the two bases are related by a singular field redefinition that introduces single-particle poles, so the proof of on-shell covariance of $\V_{1\dots m}$ in \cref{app:osc} does not go through.}

\subsection{An unambiguous metric}
\label{sec:geo_metric_2}

In light of the challenge discussed in the previous subsection, we would like to consider an alternative approach to identify a metric on the field configuration manifold. Our starting point is again that $\dphi{\mu}^a$ is a vector, so any metric $g_{ab}[\phi]$ must render the expression $\frac12\,g_{ab}[\phi]\,\dphi{\mu}^a\Dphi{\mu}^b$ a scalar under translation-invariant field redefinitions. However, rather than identifying this scalar with the full action $S[\phi]$, we identify it with the two-derivative part of $S[\phi]$ in a special set of bases which we call {\it Warsaw-like bases}:
\begin{equation}
S_{(\partial^2)}[\phi] = \frac12 \, g_{ab}[\phi] \, \dphi{\mu}^a \Dphi{\mu}^b \qquad\qquad\text{(Warsaw-like bases)}\,.
\label{eq:S_2d}
\end{equation}
Concretely, up to any fixed derivative order, we can enumerate the operators and then follow the same procedure that leads to the Warsaw basis for SMEFT dimension-six operators~\cite{Grzadkowski:2010es} to minimize the number of derivatives acting on each field. Up to $\O(\partial^4)$, this procedure yields:
\begin{align}
    S[\phi] = \int_x\, \biggl[&-U(\phi) +\frac12\, g_{ij}(\phi) (\partial_\mu\phi^i) (\partial^\mu\phi^j ) +\frac18\, \lambda_{ijkl}(\phi) (\partial_\mu\phi^i) (\partial^\mu\phi^j ) (\partial_\nu\phi^k) (\partial^\nu\phi^l ) \notag\\[5pt]
    & +\O(\partial^6) \biggr]\hspace{150pt} \text{(Warsaw-like bases)}\,,
    \label{eq:S_Warsaw}
\end{align}
where $U$, $g_{ij} = g_{(ij)}$ and $\lambda_{ijkl} = \lambda_{(ij)(kl)} = \lambda_{(kl)(ij)}$ are real analytic functions of $\phi$; see \cref{app:warsaw} for details. Note that \cref{eq:S_Warsaw} specifies a set of bases rather than a single basis, because there is still the freedom to make nonderivative field redefinitions, which preserve the form of \cref{eq:S_Warsaw}. In other words, a nonderivative field redefinition takes us from one Warsaw-like basis to another Warsaw-like basis. 

We can rewrite \cref{eq:S_Warsaw} as:
\begin{align}
    S[\phi] &= -U[\phi] +\frac12 \,g_{ab}[\phi]\dphi{\mu}^a\Dphi{\mu}^b +\frac18 \,\lambda_{abcd}[\phi]\dphi{\mu}^a\Dphi{\mu}^b\dphi{\nu}^c\Dphi{\nu}^d \notag\\[5pt]
    &\quad\; +\O(\partial^6)\hspace{150pt} \text{(Warsaw-like bases)}\,,
    \label{eq:S_Warsaw_2}
\end{align}
where
\begin{subequations}
\label{eq:g_warsaw}
\begin{align}
    U[\phi] &\equiv \int_x\, U\bigl(\phi(x)\bigr) \,,\\
    g_{ab}[\phi] &\equiv \int_x\, e^{-i(p_a+p_b)\cdot x}\, g_{i_ai_b}\bigl(\phi(x)\bigr) \,,\label{eq:g_def}\\
    \lambda_{abcd}[\phi] &\equiv \int_x \, e^{-i(p_a+p_b+p_c+p_d)\cdot x}\, \lambda_{i_ai_bi_ci_d}\bigl(\phi(x)\bigr) \,.
\end{align}
\end{subequations}
With $g_{ab}[\phi]$ identified as the metric on the field configuration manifold in a Warsaw-like basis, we can follow the standard tensor transformation rule to obtain the metric in any other basis that is related to our Warsaw-like basis by a field redefinition. Similarly, we can demand that $U[\phi]$ and $\lambda_{abcd}[\phi]$ transform as a scalar and a $(0,4)$ tensor, respectively, and obtain their expressions in any other basis. See \cref{app:warsaw} for details. Note that if we have worked with a different Warsaw-like basis, we would obtain the same metric, in the sense that its components are related to $g_{ab}[\phi]$ by the correct tensor transformation that corresponds to the nonderivative field redefinition connecting the two bases; we expand on this point in \cref{app:nd}. The metric defined in this way is nonsingular and free from the ambiguity of the prescription discussed in \cref{sec:geo_metric_1}.\footnote{We note that our metric choice here coincides with that in Ref.~\cite{Finn:2019aip} when the action is truncated at $\O(\partial^2)$, and was also discussed in Ref.~\cite{Cohen:2024bml} in the special case of nonlinear sigma models. Since many phenomenologically interesting EFTs (e.g., SMEFT and HEFT) contain an infinite series of higher-derivative operators, the prescription introduced here is needed to define a metric that has the correct tensorial transformation under derivative field redefinitions.}

A key advantage of the metric definition in this subsection is that the relation to field space geometry becomes totally transparent. If we neglect four- and higher-derivative terms in the action, $g_{i_ai_b}(\phi)$ is nothing but the metric in field space geometry; the metric in functional geometry $g_{ab}[\phi]$ is then obtained via a Fourier transform. Similarly, the (generalized) Christoffel symbols and Riemann curvature are given by:
\begin{subequations}
\begin{align}
\Gamma^a_{1\dots m} [\phi] &= \int_x\, e^{i(p_a-p_1-\cdots -p_m)\cdot x} \, \Gamma^{i_a}_{i_1\dots i_m} \bigl(\phi(x)\bigr) \,, \label{eq:Gamma_fg}\\[5pt]
R^a{}_{bcd}[\phi] &= \int_x \, e^{i(p_a-p_b-p_c-p_d)\cdot x}\, R^{i_a}{}_{i_bi_ci_d}\bigl(\phi(x)\bigr) \,.
\end{align}
\end{subequations}
where $\Gamma^{i_a}_{i_1\dots i_m}$, $R^{i_a}{}_{i_bi_ci_d}$ are the corresponding quantities in field space geometry, as one can verify using:
\begin{equation}
\frac{\partial}{\partial\phi^{(ip)}} = (2\pi)^d\int_x\, \frac{\delta\phi^j(x)}{\delta\phi^i(p)} \,\frac{\delta}{\delta\phi^j(x)} = \int_x\, e^{-ip\cdot x} \, \frac{\delta}{\delta\phi^i(x)}\,.
\label{eq:partial_phi}
\end{equation}
Covariant derivatives of the potential, curvature and coefficient functional $\lambda_{abcd}$ 
on the field configuration manifold are similarly related to those on the target manifold via a Fourier transform:
\begin{subequations}
\begin{align}
U_{;1\dots m}[\phi] &= \int_x e^{-i(p_1+\cdots +p_m)\cdot x}\, U_{;i_1\dots i_m}\bigl(\phi(x)\bigr) \,,\label{eq:U_cd_fg}\\[5pt]
R^a{}_{bcd;1\dots m}[\phi] &= \int_x e^{i(p_a -p_b-p_c-p_d - p_1-\cdots -p_m)\cdot x}\, R^{i_a}{}_{i_bi_ci_d;\,i_1\dots i_m}\bigl(\phi(x)\bigr) 
\,,\\[5pt]
\lambda_{abcd;1\dots m}[\phi] &= \int_x e^{-i(p_a +p_b+p_c+p_d +p_1+\cdots +p_m)\cdot x}\, \lambda_{i_ai_bi_ci_d;\,i_1\dots i_m}\bigl(\phi(x)\bigr) 
\,.
\end{align}
\end{subequations}

A useful feature of the Riemannian geometry on the field configuration manifold defined in this subsection is that $\dphi{\mu}^a$ is a set of Killing vectors associated with spacetime translation symmetry. Indeed, from \cref{eq:T_phi} we see that $\dphi{\mu}^a$ generates a diffeomorphism on the field configuration manifold that is equivalent to a spacetime translation. The metric $g_{ab}[\phi]$ is translation-invariant because it is an integral of a local function (see \cref{eq:g_def}), so the diffeomorphism generated by $\dphi{\mu}^a$ is an isometry. We can also explicitly verify that $\dphi{\mu}^a = -ip_{a\mu}\phi^a$ satisfies Killing's equation:
\begin{align}
2\,\dphi{\mu}_{(a;b)} &= g_{ac}\,\dphi{\mu}^c{}_{; b} + g_{bc}\,\dphi{\mu}^c{}_{; a} \notag\\[5pt]
&= g_{ac}\,\bigl( -ip_{b\mu}\,\delta^c_b +\Gamma^c_{bd} \,\dphi{\mu}^d \bigr) + g_{bc}\,\bigl( -ip_{a\mu}\,\delta^c_a +\Gamma^c_{ad} \,\dphi{\mu}^d \bigr) \notag\\[5pt]
&= -i(p_a+p_b)_\mu \,g_{ab} + g_{ab,d}\,\dphi{\mu}^d \notag\\[5pt]
&= -i(p_a+p_b)_\mu \,g_{ab} + \int_{p_d, x_d} e^{-ip_d\cdot x_d} \frac{\delta g_{ab}}{\delta\phi^{i_d} (x_d)} \, \bigl(-ip_{d\mu}\,\phi^{i_d} (p_d)\bigr)\notag\\[5pt]
&= \int_x \Bigl[\partial_\mu\bigl(e^{-i(p_a+p_b)\cdot x}\bigr)\, g_{i_ai_b}\bigl(\phi(x)\bigr) +e^{-i(p_a+p_b)\cdot x} \,\partial_\mu g_{i_ai_b}\bigl(\phi(x)\bigr)\Bigr] \notag\\[5pt]
&= \int_x \partial_\mu \Bigl[e^{-i(p_a+p_b)\cdot x}\, g_{i_ai_b}\bigl(\phi(x)\bigr)\Bigr] = 0\,,
\end{align}
where we have used \cref{eq:partial_phi} and $g_{ab;d} = g_{ab,d} -g_{bc}\,\Gamma^c_{da} - g_{ac}\,\Gamma^c_{db} = 0$. An especially useful identity for Killing vectors is
\begin{equation}
\dphi{\mu}^a{}_{;\,bc} = R^a{}_{bcd}\,\dphi{\mu}^d\,,
\label{eq:dphi_2d}
\end{equation}
which follows directly from Killing's equation and the definition of Riemann curvature.\footnote{To the best of our knowledge, \cref{eq:dphi_2d} first appeared in Ref.~\cite{Ecker:1972tii} in the context of chiral perturbation theory.}

\subsection{Reduction to field space geometry}
\label{sec:geo_fs}

With the discussion in the previous subsection, we now have a very simple procedure to derive on-shell covariant amplitude expressions in field space geometry starting from the more general framework of functional geometry. Neglecting four- and higher-derivative terms, we can write the action as:
\begin{equation}
S[\phi] = \frac{1}{2}\,g_{ab}[\phi] \, \dphi{\mu}^a \Dphi{\mu}^b  -U[\phi] \,.
\end{equation}
To obtain the on-shell covariant vertex functions $\V_{1\dots m}$ defined in \cref{eq:Vcal_fun}, we need to take covariant derivatives of the action on the field configuration manifold. Covariant derivatives of $U[\phi]$ are given by \cref{eq:U_cd_fg}, while covariant derivatives of the metric with respect to the Levi-Civita connection vanish. Using \cref{eq:dphi_2d}, we can express all covariant derivatives of $\dphi{\mu}^a$ in terms of $\dphi{\mu}^a$, $\dphi{\mu}^a{}_{;b}$ and (covariant derivatives of) the Riemann curvature tensor:
\begin{subequations}
	\label{eq:dphi_cd}
	\begin{align}
	\dphi{\mu}^a{}_{;12} &= R^a{}_{12b}\,\dphi{\mu}^b \,,\\[5pt]
	\dphi{\mu}^a{}_{;123} &= R^a{}_{12b}\,\dphi{\mu}^b{}_{;3} + R^a{}_{12b;3} \,\dphi{\mu}^b \,,\\[5pt]
	\dphi{\mu}^a{}_{;1234} &= R^a{}_{12b;4}\,\dphi{\mu}^b{}_{;3} + R^a{}_{12b;3}\,\dphi{\mu}^b{}_{;4} + R^a{}_{12b}\,R^b{}_{34c}\,\dphi{\mu}^c + R^a{}_{12b;34}\,\dphi{\mu}^b \,,\notag\\[2pt]
	\end{align}
\end{subequations}
and so on. To reproduce results in field space geometry, we are interested in the constant field submanifold, $\phi^i(x) = \phi^i = $ constant. In the constant field limit, we have
\begin{subequations}
	\label{eq:fsl}
	\begin{align}
	g_{ab}[\phi] \bigr|_{\phi(x) =\phi} &= g_{i_ai_b}(\phi)\, \delta_{p_ap_b} \,,\\[5pt]
	\Gamma^a_{1\dots m} [\phi] \bigr|_{\phi(x) =\phi} &= \Gamma^{i_a}_{i_1\dots i_m} (\phi) \, \delta^{p_a}_{p_1\dots p_m}\,, \\[5pt]
	U_{;1\dots n}[\phi] \bigr|_{\phi(x) =\phi} &= U_{;i_1\dots i_n} (\phi) \, \delta_{p_1\dots p_n} \,,\\[5pt]
	R^a{}_{12b;3\dots}[\phi] \bigr|_{\phi(x) =\phi} &= R^{i_a}{}_{i_1i_2i_b;i_3\dots}(\phi)\,\delta^{p_a}_{p_1p_2p_bp_3\dots} \,,\\[5pt]
	\dphi{\mu}^a \bigr|_{\phi(x) =\phi} &= 0 \,,\\[5pt]
	\dphi{\mu}^a{}_{;1} \bigr|_{\phi(x) =\phi} &= -ip_{1\mu} \delta^a_1 = -ip_{1\mu} \delta^{i_a}_{i_1}\,\delta^{p_a}_{p_1} \,,\label{eq:dphi_fs}\\[5pt]
	\dphi{\mu}^a{}_{;12} \bigr|_{\phi(x) =\phi} &= 0 \,,\\[5pt]
	\dphi{\mu}^a{}_{;123} \bigr|_{\phi(x) =\phi} &= -ip_{3\mu} \,R^{i_a}{}_{i_1i_2i_3}(\phi)\, \delta^{p_a}_{p_1p_2p_3} \,,\\[5pt]
	\dphi{\mu}^a{}_{;1234} \bigr|_{\phi(x) =\phi} &= -i \bigl(p_{3\mu} \,R^{i_a}{}_{i_1i_2i_3;i_4}(\phi) + p_{4\mu} \,R^{i_a}{}_{i_1i_2i_4;i_3}(\phi)\bigr)\, \delta^{p_a}_{p_1p_2p_3p_4} \,,
	\end{align}
\end{subequations}
and one can similarly obtain higher covariant derivatives of $\dphi{\mu}^a$. Meanwhile, for the (inverse) propagator, we know from \cref{eq:invprop_fsl} that:
\begin{subequations}
\begin{align}
\Delta^{-1}_{ab} [\phi] \bigr|_{\phi(x) =\phi} &= \Delta^{-1}_{i_ai_b}(\phi\,; p_a)\,\delta_{p_a p_b}\,,\\[5pt]
\Delta^{ab}[\phi] \bigr|_{\phi(x) =\phi} &= \Delta^{i_ai_b}(\phi\,; p_a)\,\delta^{p_a p_b}\,.
\end{align}
\end{subequations}

Combining all the results above, we find that in the constant field limit, the on-shell covariant vertices on the field configuration manifold defined in \cref{eq:Vcal_fun} reduce to their counterparts in field space geometry:
\begin{align}
\V_{1\dots m} \bigr|_{\phi(x) =\phi} &= \Bigl[ \R_{i_1\dots i_m} (p_1, \dots, p_m) -U_{;(i_1\dots i_m)} \notag\\[5pt]
&\qquad - \Gamma^{i_c}_{i_1\dots i_m} \, U_{,i_c} +\sum_{a\in\text{ext}} \Gamma^{i_c}_{i_1\dots\slashed{i_a}\dots i_m} \Delta^{-1}_{i_ci_a}(p_a) \Bigr]\, \delta_{p_1\dots p_m} \notag\\[5pt]
& = \V_{i_1\dots i_m}(p_1,\dots, p_m)\,\delta_{p_1\dots p_m}  \,,
\end{align}
where 
\begin{equation}
\delta_{p_1\dots p_m} \, \R_{i_1\dots i_m} (p_1, \dots, p_m) = \frac{1}{2}\, g_{ab}[\phi] \bigl[\dphi{\mu}^a \,\Dphi{\mu}^b\bigr]_{;(1\dots m)} \Bigr|_{\phi(x) =\phi}
\label{eq:R_def}
\end{equation}
is a function of the curvature and its covariant derivatives that can be derived following \cref{eq:dphi_cd} and using the constant field limit expressions in \cref{eq:fsl}. For $n=3$ and $4$, we find:
\begin{subequations}
	\begin{align}
	\R_{i_1 i_2 i_3} (p_1,p_2,p_3) &= 0 \,,\\[5pt]
	\R_{i_1 i_2 i_3 i_4} (p_1,p_2,p_3,p_4) &= -\frac{2}{3}\,\Bigl[(p_1\cdot p_2) \, R_{i_1(i_3 i_4)i_2}\Bigr]_\text{6 terms} \notag \\[5pt]
    &= -\frac{2}{3}\Bigl[ p_{12}^2\, R_{i_1(i_3i_4)i_2}\Bigr]_\text{3 terms} \,, 
	\end{align}
\end{subequations}
in agreement with our results in \cref{sec:field}. Computing higher-point on-shell covariant vertices is straightforward.

We end this section by outlining how our discussion can be generalized to scalar EFT actions containing four- and higher-derivative terms. We have seen that, up to $\O(\partial^2)$ in a Warsaw-like basis, the action is specified by a scalar $U[\phi]$ and a symmetric $(0,2)$ tensor $g_{ab}[\phi]$ which we identify as the metric. At the next order $\O(\partial^4)$, we have an additional coefficient functional $\lambda_{abcd}[\phi]$, which is a $(0,4)$ tensor; see \cref{eq:S_Warsaw_2}. When we take covariant derivatives of $S[\phi]$ to derive the on-shell covariant vertices, we would obtain additional contributions that depend on (covariant derivatives of) $\lambda_{abcd}$ and the Riemann curvature tensor. Going to higher derivative orders, we will encounter additional operators which can be written in terms of:
\begin{equation}
\dphi{\mu}^a\,,\qquad \ddphi{\mu}{\nu}^a\,,\qquad (\D_{\rho}\D_{\nu}\partial_\mu\phi)^a \,,\qquad \dots 
\end{equation}
These are all vectors defined recursively via~\cite{Alonso:2015fsp,Alonso:2016oah}
\begin{align}
(\D_{\mu_m} \D_{\mu_{m-1}}\dots \D_{\mu_2}\partial_{\mu_1}\phi)^a &\equiv \int_x \,e^{ip_a\cdot x} \, \Bigl[\partial_{\mu_m} (\D_{\mu_{m-1}}\dots \D_{\mu_2}\partial_{\mu_1}\phi)^{(i_a x)} \notag\\[5pt]
&\hspace{-20pt} +\Gamma^{i_a}_{i_bi_c}\bigl(\phi(x)\bigr)\,\bigl(\partial_{\mu_m}\phi^{i_b}(x)\bigr)\,(\D_{\mu_{m-1}}\dots \D_{\mu_2}\partial_{\mu_1}\phi)^{(i_c x)}\bigr) \Bigr]\,,
\end{align}
and they satisfy
\begin{align}
(\D_{\mu_m} \D_{\mu_{m-1}}\dots \D_{\mu_2}\partial_{\mu_1}\phi)^a &= \dphi{\mu_m}^a{}_{;b}\,(\D_{\mu_{m-1}}\dots \D_{\mu_2}\partial_{\mu_1}\phi)^b \notag\\[5pt]
&= \dphi{\mu_m}^{a_m}{}_{;a_{m-1}}\, \dphi{\mu_{m-1}}^{a_{m-1}}{}_{;a_{m-2}} \,\cdots\, \dphi{\mu_1}^{a_1} \,,
\label{eq:dnphi}
\end{align}
as one can show using the torsion-free property of the Levi-Civita connection. As in the $\O(\partial^4)$ case, we can demand that the coefficient functional of each operator transforms as a tensor. As a result, scattering amplitudes can be expressed in a manifestly on-shell covariant manner in terms of a finite number of tensors and their covariant derivatives up to any finite order in the derivative expansion. We leave a detailed exploration of the explicit construction of Warsaw-like bases with higher-derivative tensorial operators and applications of functional geometry to EFTs containing higher-derivative operators to future work.

As a final remark, we would like note a subtle difference in philosophy between our approach and some of the recent attempts to extend field space geometry~\cite{Cheung:2022vnd,Cohen:2024bml,Alminawi:2023qtf}. Extending the construction from the target manifold to a larger manifold (either the field configuration manifold or a jet bundle) makes it possible to encode everything about an EFT into a metric, in the sense that specifying a metric on the larger manifold is sufficient to define an EFT action. While the idea of encoding all the physics of an EFT into the intrinsic geometry of a Riemannian manifold is appealing, one must be careful about assigning physical meaning to geometric objects because of the freedom to define a multitude of metrics from the same EFT action. In this paper, we have taken a different approach which centers around establishing a direct connection to the amplitude expressions in field space geometry. In particular, we have emphasized that to achieve a manifestly on-shell covariant construction of EFT amplitudes, all we need are {\it a}) an action, which is a scalar on the field configuration manifold and invariantly defines the EFT, and {\it b}) a connection, which defines a notion of covariant differentiation. The role of a metric is secondary, in that it provides a natural definition of a connection, but is not central to the construction. With our choice of metric in \cref{sec:geo_metric_2}, the physical content of an EFT is captured by not just the metric, but a tower of tensors $\{U\,,\, g_{ab}\,,\, \lambda_{abcd}\,, \,\dots\}$ which all descend from the action and are simply related to their field space geometry counterparts by a Fourier transform.

\section{Conclusions}
\label{sec:conclusions}

In this work, we have presented a unified perspective on both the geometry of the target manifold (commonly known as ``field space geometry'') and its generalization to the field configuration manifold (which we call ``functional geometry'') that accommodates derivative field redefinitions in scalar EFTs. In both cases, we focused on the amputated correlation functions $\M$, whose vacuum and on-shell limits give physical scattering amplitudes. Working at tree level, we found expressions for $\M$ in terms of building blocks that transform covariantly in the vacuum and on-shell limits under coordinate changes on the respective manifolds. Our main findings are summarized as follows.
\begin{itemize}
    \item In field space geometry, $\M$ carry flavor indices $i_1\dots i_n$ and are functions of momenta $p_1, \dots , p_n$. They satisfy:
    \begin{equation}
       \M_{i_1\dots i_n} (p_1, \dots , p_n) = \Mfun_{i_1\dots i_n} \bigl(\Delta^{ij}, \{V_{j_1\dots j_m}\} \bigr)
       = \Mfun_{i_1\dots i_n} \bigl(\Delta^{ij}, \{\V_{j_1\dots j_m}\} \bigr)\,,
       \label{eq:Mfun_fs_2}
    \end{equation}
    where the middle expression represents the function of propagator $\Delta^{ij}$ and vertices $V_{j_1\dots j_m}$ obtained from standard Feynman rules, while the last expression represents the same function, but with the standard vertices replaced by a new set of vertices $\V_{j_1\dots j_m}$ given in \cref{eq:Vcal_fs_gen} for a scalar EFT action truncated at $\O(\partial^2)$. Under coordinate changes on the target manifold, which are nonderivative field redefinitions, $\Delta^{ij}$ and $\V_{j_1\dots j_m}$ transform as tensors in the vacuum and on-shell limits. Away from the vacuum and on-shell limits, $\Delta^{ij}$ and $\V_{j_1\dots j_m}$ generally depend on nontensorial terms, and so do $\M_{i_1\dots i_n}$. The nontensorial terms do not contribute to on-shell amplitudes, and, as we have shown here for the first time, can be understood as originating from a mismatch between a generic field basis and normal coordinates at any given point on the target manifold. 
    \item In functional geometry, $\M$ carry both flavor and momentum indices, collectively denoted by $1\dots n$. They satisfy:
    \begin{equation}
       \M_{1\dots n} = \Mfun_{1\dots n} \bigl(\Delta^{ab}, \{V_{a_1\dots a_m}\} \bigr)
       = \Mfun_{1\dots n} \bigl(\Delta^{ab}, \{\V_{a_1\dots a_m}\} \bigr)\,,
       \label{eq:Mfun_fg_2}
    \end{equation}
    where the middle expression represents the function of propagator $\Delta^{ab}$ and vertices $V_{a_1\dots a_m}$ obtained from standard Feynman rules, while the last expression represents the same function, but with the standard vertices replaced by a new set of vertices $\V_{a_1\dots a_m}$ given in \cref{eq:Vcal_fun} for a general scalar EFT. Under coordinate changes on the field configuration manifold, which are general field redefinitions that may involve derivatives, $\Delta^{ab}$ and $\V_{a_1\dots a_m}$ transform as tensors in the vacuum and on-shell limits. Away from the vacuum and on-shell limits, $\Delta^{ab}$ and $\V_{a_1\dots a_m}$ generally contain nontensorial terms, and so do $\M_{1\dots n}$. The nontensorial terms can be understood as originating from a mismatch between a generic field basis and normal coordinates at any given point on the field configuration manifold. They do not contribute to on-shell amplitudes for an appropriately chosen nonsingular connection.
    \item In both field space geometry and functional geometry, the on-shell covariant building blocks discussed above can be defined for any connection, independent of the existence of a metric. On the target manifold, it is natural to identify the coefficient function of the two-derivative terms in the action as the metric. On the field configuration manifold, there appeared to be an ambiguity in the definition of metric~\cite{Cheung:2022vnd,Cohen:2024bml}. We proposed to sidestep this ambiguity by identifying the coefficient functional of the two-derivative part of the action in Warsaw-like bases as the metric. With this choice of metric, the relation between geometric objects in field space geometry and functional geometry becomes transparent. We then showed that amplitude expressions in field space geometry can be easily obtained from the constant field limit of the corresponding expressions in functional geometry.
\end{itemize}

By formulating field space geometry and functional geometry in a unified manner, our work sets the stage for revisiting many results that have so far only been established using field space geometry---e.g., classification of EFTs (free vs.\ interacting, renormalizable vs.\ nonrenormalizable, SMEFT vs.\ HEFT), soft theorems, EFT matching and renormalization group evolution equations---through the lens of functional geometry, thereby achieving conclusions that are robust under both nonderivative and derivative field redefinitions. In particular, while four- and higher-derivative terms have been rarely considered in field space geometry, they are straightforward to include in functional geometry, following the procedure outlined at the end of \cref{sec:geo_fs}. A systematic study of their effects in the Higgs sector of the Standard Model from the geometric perspective will shed new light on the interpretation of precision measurements at the LHC and future colliders. Another future direction is to extend functional geometry to accommodate fermion and vector fields, thus generalizing recent efforts to include nonzero-spin fields in field space geometry~\cite{Finn:2019aip,Finn:2020nvn, Helset:2022tlf, Assi:2023zid, Gattus:2023gep, Gattus:2024ird, Assi:2025fsm, Craig:2025uoc}. Finally, it would be interesting to see if similar on-shell covariant building blocks can be identified for loop amplitudes and 1PI effective actions. We are optimistic that we are only beginning to uncover applications to the theory and phenomenology of EFTs that can be understood using the geometry of the field configuration manifold.

\acknowledgments

We are grateful to Xiaochuan Lu for collaboration in the early stages of the project. We thank Andreas Helset and Yu-Tse (Alan) Lee for helpful discussions and feedback on a preliminary draft. T.C.\ is supported by the U.S.~Department of Energy under grant DE-SC0011640. X.-X.L.\ and Z.Z.\ are supported in part by the U.S.~National Science Foundation under grant PHY-2412880. This work was performed in part at the Aspen Center for Physics, which is supported by National Science Foundation grant PHY-1607611.

\vspace{4pt}
\noindent
{\it Note added:} We thank Mohammad Alminawi, Ilaria Brivio and Joe Davighi for sharing their new preprint~\cite{Alminawi:NewPaper}, where they present covariant Feynman rules in the complementary formulation of EFT geometry using jet bundles. They incorporate both the potential and two-derivative interactions into a metric on the 0-jet bundle and derive on-shell amplitude expressions that are manifestly covariant under nonderivative field redefinitions.

\appendix

\phantomsection
\addcontentsline{toc}{section}{Appendices}

\section*{Appendices}
\section{On-shell covariance of $\V_{1\dots m}$}
\label{app:osc}

In this appendix, we show that even for a connection $\Gamma^c_{ab}$ on the field configuration manifold that has a simple pole in $p_c$, the $\V_{1\dots m}$ vertices given in \cref{eq:Vcal_fun} are still on-shell covariant. We first note that, in our definition of on-shell equivalence and on-shell covariance in \cref{sec:functional_osc}, the vacuum limit should be taken before contracting with wave function factors because the wave function factors are only defined at the vacuum. Among the two types of nontensorial terms in \cref{eq:Vcal_fun}, $\Gamma^c_{(1\dots m)} \, S_{,c}$ vanishes as soon as the vacuum limit is taken because $\overline S_{,c} = 0$, provided the connection is not singular at the vacuum, which we assume. Therefore, 
\begin{align}
\Gamma^c_{(1\dots m)} \, S_{,c} \stackrel{\text{OS}}{=} 0\,.
\end{align}
All that remains to prove the on-shell covariance of $\V_{1\dots m}$ is to show that $\Gamma^c_{(1\dots \slashed{a}\dots m)}\, \Delta^{-1}_{ca}$ ($a\in\text{ext}$) is on-shell covariant, even if it is not on-shell equivalent to zero. It turns out that symmetrization of the indices is not important, and we will show that $\Gamma^c_{1\dots \slashed{a}\dots m}\, \Delta^{-1}_{ca}$ is on-shell covariant; in other words, we will show that the inhomogeneous terms in the transformation of $\Gamma^c_{1\dots \slashed{a}\dots m}\, \Delta^{-1}_{ca}$ are on-shell equivalent to zero. 

To simplify notation, let us rewrite $m$ as $m+1$ and, without loss of generality, pick the term with $a=m+1$, assuming $p_a = p_{m+1}$ is an external momentum. Under a coordinate transformation $\phi = F[\widetilde\phi]$ on the field configuration manifold, the inverse propagator becomes:
\begin{equation}
\widetilde\Delta^{-1}_{ca} = \frac{\partial\phi^e}{\partial\widetilde\phi^c}\, \frac{\partial\phi^b}{\partial\widetilde\phi^a}\, \Delta^{-1}_{eb} + \frac{\partial^2\phi^b}{\partial\widetilde\phi^c\, \partial\widetilde\phi^a}\, S_{,b}\,.
\end{equation}
This is the functional analog of \cref{eq:Delta_transform}. The transformation of $\Gamma^c_{1\dots m}$ takes the form:
\begin{equation}
\widetilde\Gamma^c_{1\dots m} = \frac{\partial\widetilde\phi^c}{\partial\phi^d}\, \frac{\partial\phi^{1'}}{\partial\widetilde\phi^1} \cdots \frac{\partial\phi^{m'}}{\partial\widetilde\phi^m} \, \Gamma^d_{1'\dots m'} + \dG^c_{1\dots m} \,,
\label{eq:Gamma_transform}
\end{equation}
where $\dG^c_{1\dots m}$ collects the inhomogeneous terms. Combining the two equations above, we have:
\begin{align}
\widetilde\Gamma^c_{1\dots m}\, \widetilde\Delta^{-1}_{ca}  &= \frac{\partial\phi^b}{\partial\widetilde\phi^a}\,\frac{\partial\phi^{1'}}{\partial\widetilde\phi^1} \cdots \frac{\partial\phi^{m'}}{\partial\widetilde\phi^m} \, \Gamma^c_{1'\dots m'} \,\Delta^{-1}_{cb} \notag\\[5pt]
&\quad +\dG^c_{1\dots m}\,\widetilde\Delta^{-1}_{ca}
+\frac{\partial^2\phi^b}{\partial\widetilde\phi^c\, \partial\widetilde\phi^a}\,\frac{\partial\widetilde\phi^c}{\partial\phi^e}\, \frac{\partial\phi^{1'}}{\partial\widetilde\phi^1} \cdots \frac{\partial\phi^{m'}}{\partial\widetilde\phi^m} \, \Gamma^e_{1'\dots m'}\, S_{,b}\,.
\end{align} 
Among the inhomogeneous terms in the second line, the one proportional to $S_{,b}$ is obviously on-shell equivalent to zero because $\overline S_{,b} = 0$. To show that $\dG^c_{1\dots m}\,\widetilde\Delta^{-1}_{ca}$ is also on-shell equivalent to zero, we need to show that $\dG^c_{1\dots m}$ does not contain poles in $p_c$, since $\widetilde\Delta^{-1}_{ca} \oseq 0$.

The proof proceeds by induction. For $m=2$, we have:
\begin{equation}
\dG^c_{12} = \frac{\partial\widetilde\phi^c}{\partial\phi^b}\, \frac{\partial^2\phi^b}{\partial\widetilde\phi^1\,\partial\widetilde\phi^2} \,,
\end{equation}
which obviously does not have a pole as long as we consider nonsingular field redefinitions, meaning field redefinitions that do not artificially introduce single-particle poles.\footnote{Note, for example, that the nonlocal field redefinition that turns an interacting action into a free action discussed around \cref{eq:redef_hat} is a singular field redefinition. So the discussion below does not apply in that case.} To complete the induction step, we assume that $\dG^c_{1\dots m}$ does not have poles in $p_c$. Using \cref{eq:Gamma_def_fun,eq:Gamma_transform} to derive $\dG^c_{1\dots m+1}$, we find:
\begin{align}
\widetilde\Gamma^c_{1\dots m+1} &= \widetilde\Gamma^c_{1\dots m, m+1} - \sum_{b=1}^m \widetilde\Gamma^d_{(m+1) b} \, \widetilde\Gamma^c_{1\dots \slashed{b} d\dots m} \notag\\[5pt]
&= \frac{\partial}{\partial\widetilde\phi^{m+1}} \biggl(\frac{\partial\widetilde\phi^c}{\partial\phi^d}\, \frac{\partial\phi^{1'}}{\partial\widetilde\phi^1} \cdots \frac{\partial\phi^{m'}}{\partial\widetilde\phi^m} \, \Gamma^d_{1'\dots m'} + \dG^c_{1\dots m} \biggr) \notag\\[5pt]
&\qquad -\sum_{b=1}^m \biggl( \frac{\partial\widetilde\phi^d}{\partial\phi^e}\, \frac{\partial\phi^g}{\partial\widetilde\phi^b}\,\frac{\partial\phi^f}{\partial\widetilde\phi^{m+1}} \, \Gamma^e_{fg} +\frac{\partial\widetilde\phi^d}{\partial\phi^e}\,\frac{\partial^2\phi^e}{\partial\widetilde\phi^b\,\partial\widetilde\phi^{m+1}}\biggr) \notag\\[5pt] &\qquad\qquad\quad \biggl(\frac{\partial\widetilde\phi^c}{\partial\phi^h}\, \frac{\partial\phi^{1'}}{\partial\widetilde\phi^1} \cdots \frac{\partial\phi^{b'}}{\partial\widetilde\phi^d} \cdots\, \frac{\partial\phi^{m'}}{\partial\widetilde\phi^m} \, \Gamma^h_{1'\dots b' \dots m'} + \dG^c_{1\dots\slashed{b} d\dots  m} \biggr) \,.
\end{align}
Collecting the inhomogeneous terms, we find:
\begin{align}
\dG^c_{1\dots m+1} &= \frac{\partial}{\partial\widetilde\phi^{m+1}} \biggl(\frac{\partial\widetilde\phi^c}{\partial\phi^d}\, \frac{\partial\phi^{1'}}{\partial\widetilde\phi^1} \cdots \frac{\partial\phi^{m'}}{\partial\widetilde\phi^m} \biggr)\, \Gamma^d_{1'\dots m'} + \dG^c_{1\dots m,m+1} \notag\\[5pt]
&\qquad -\sum_{b=1}^m \biggl(\widetilde\Gamma^d_{(m+1) b} \, \dG^c_{1\dots \slashed{b} d\dots m} 
+ \frac{\partial\widetilde\phi^c}{\partial\phi^d}\, \frac{\partial\phi^{1'}}{\partial\widetilde\phi^1} \cdots \frac{\partial^2\phi^{b'}}{\partial\widetilde\phi^b\,\partial\widetilde\phi^{m+1}} \cdots\, \frac{\partial\phi^{m'}}{\partial\widetilde\phi^m} \, \Gamma^d_{1'\dots b' \dots m'} \biggr) \notag\\[8pt]
&= \dG^c_{1\dots m,m+1} -\sum_{b=1}^m \widetilde\Gamma^d_{(m+1) b} \, \dG^c_{1\dots \slashed{b} d\dots m}
+\frac{\partial^2\widetilde\phi^c}{\partial\phi^d\,\partial\phi^e}\, \frac{\partial\phi^e}{\partial\widetilde\phi^{m+1}}\,\frac{\partial\phi^{1'}}{\partial\widetilde\phi^1} \cdots \frac{\partial\phi^{m'}}{\partial\widetilde\phi^m} \, \Gamma^d_{1'\dots m'} \,.
\label{eq:dG_recursion}
\end{align}
If $\dG^c_{1\dots m}$ does not have poles in $p_c$, differentiation will not generate a pole, so the first term in the last line of \cref{eq:dG_recursion} is safe. The remaining terms depend on $\widetilde\Gamma^d_{(m+1) b}$ and $\Gamma^d_{1'\dots m'}$, which may have poles when $p_d$ goes on-shell. However, at the vacuum point, each object appearing in \cref{eq:dG_recursion} must be proportional to a momentum-conserving $\delta$-function due to translation invariance, e.g. $\frac{\partial^2\widetilde\phi^c}{\partial\phi^d\,\partial\phi^e}\propto \delta^d(p_c-p_d-p_e)$. The momentum $p_c$ is never set equal to $p_d$ by momentum conservation (it is interesting to note that terms proportional to $\frac{\partial\widetilde\phi^c}{\partial\phi^d} \propto \delta^d(p_c-p_d)$ cancel out). Therefore, $\dG^c_{1\dots m+1}$ does not have poles in $p_c$, so we have completed the proof.

\section{Functional geometry up to $\O(\partial^4)$ in a general basis}
\label{app:warsaw}

In this appendix, we explicitly demonstrate the procedure introduced in \cref{sec:geo_metric_2} to derive the potential $U[\phi]$, metric $g_{ab}[\phi]$, and the coefficient functional $\lambda_{abcd}[\phi]$ of the four-derivative operator starting from an EFT action written in a general basis. Up to $\O(\partial^4)$ and neglecting topological terms (which depend on the spacetime dimension and whose incorporation presents no additional difficulty), the most general form of the action is:
\begin{align}
    S[\phi] = \int_x \,\biggl[&
    -U'(\phi) +\frac12\, g''_{ij}(\phi) (\partial_\mu\phi^i)(\partial^\mu\phi^j) +h_i(\phi) (\partial^2\phi^i) \notag\\[5pt]
    & +\frac18\, \lambda''_{ijkl}(\phi)\,\O_1^{ijkl}(\phi) +\frac12\, \lambda''_{ijk}(\phi) \, \O_2^{ijk}(\phi) + \frac12\, \lambda''_{ij}(\phi)\,\O_3^{ij}(\phi) \notag\\[5pt]
    & +\frac12\, \kappa_{ijk}(\phi)\,\O_4^{ijk}(\phi) +\kappa_{ij}(\phi)\, \O_5^{ij}(\phi) + \frac12\, \kappa_{ij}^{\,\prime}(\phi)\, \O_6^{ij}(\phi) \notag\\[5pt]
    & +\kappa_i(\phi)\, \O_7^i(\phi) 
    +\O(\partial^6)
    \biggr]\,,
    \label{eq:S_general}
\end{align}
where
\begin{subequations}
\begin{align}
    \O_1^{ijkl}(\phi) &= (\partial_\mu\phi^i)(\partial^\mu\phi^j)(\partial_\nu\phi^k)(\partial^\nu\phi^l) \,,\\[5pt]
    \O_2^{ijk}(\phi) &= (\partial_\mu\phi^i)(\partial^\mu\phi^j)(\partial^2\phi^k) \,,\\[5pt]
    \O_3^{ij}(\phi) &= (\partial^2\phi^i)(\partial^2\phi^j)\,,\\[5pt]
    \O_4^{ijk}(\phi) &= (\partial_\mu\phi^i)(\partial_\nu\phi^j) (\partial^\mu\partial^\nu\phi^k)\,,\\[5pt]
    \O_5^{ij}(\phi) &= (\partial_\mu\phi^i)(\partial^\mu\partial^2\phi^j)\,,\\[5pt]
    \O_6^{ij}(\phi) &= (\partial_\mu\partial_\nu\phi^i)(\partial^\mu\partial^\nu\phi^j)\,,\\[5pt]
    \O_7^i(\phi) &= (\partial^4\phi^i)\,,
\end{align}
\end{subequations}
and the coefficient functions have the following symmetry properties:
\begin{align}
    & g_{ij}'' = g_{(ij)}''\,,\qquad \lambda''_{ijkl} = \lambda''_{(ij)(kl)} = \lambda''_{(kl)(ij)}\,,\qquad \lambda''_{ijk} = \lambda''_{(ij)k} \,,\notag\\[5pt]
    & \lambda''_{ij} = \lambda''_{(ij)} \,,\qquad \kappa_{ijk} = \kappa_{(ij)k}\,,\hspace{90pt} \kappa'_{ij} = \kappa'_{(ij)} \,.
\end{align}

To go to a Warsaw-like basis, we first note that five of the operators in \cref{eq:S_general} can be rewritten as a linear combination of the remaining operators via integration by parts:
\begin{subequations}
\label{eq:ibp}
\begin{align}
    \int_x \,h_i(\phi)(\partial^2\phi^i) &= -\int_x\, h_{i,\s j}(\phi) \,(\partial_\mu\phi^i) (\partial^\mu\phi^j) \,,\\[5pt]
    \int_x\, \frac12\, \kappa_{ijk}(\phi) \, \O_4^{ijk}(\phi) 
    &= \int_x\, \biggl[ \biggl( \frac14\, \kappa_{ijk,l}(\phi) -\frac12\, \kappa_{ikj,l}(\phi)\biggr)\, \O_1^{ijkl}(\phi) \notag\\[5pt]
    &\qquad\qquad +\frac14 \bigl( \kappa_{ijk}(\phi) -\kappa_{kij}(\phi) -\kappa_{kji}(\phi) \bigr) \,\O_2^{ijk}(\phi)\biggr] \,,\label{eq:ibp_o4}\\[5pt]
    \int_x \, \kappa_{ij}(\phi)\, \O_5^{ij}(\phi) &= -\int_x\,\bigl[ \kappa_{ij,k}(\phi)\,\O_2^{ikj}(\phi) + \kappa_{ij}(\phi)\,\O_3^{ij}(\phi) \bigr] \,,\\[5pt]
    \int_x\,\frac12\, \kappa'_{ij}(\phi)\,\O_6^{ij}(\phi) &= -\frac12\int_x\,\bigl[ \kappa'_{ij,k}(\phi)\,\O_4^{ikj}(\phi) +\kappa'_{ij}(\phi) \, \O_5^{ij}(\phi) \bigr] \,,\\[5pt]
    \int_x\, \kappa_i(\phi)\, \O_7^i(\phi) &= -\int_x \, \kappa_{i,\s j}(\phi)\, \O_5^{ji}(\phi)\,.
\end{align}
\end{subequations}
Showing Eqs.~\eqref{eq:ibp} is mostly straightforward. The only relation that requires a little bit of work to show is \cref{eq:ibp_o4}:
\begin{align}
    \int_x\, \frac12\, \kappa_{ijk}(\phi) \, \O_4^{ijk}(\phi) &= \int_x \, \frac14\, \kappa_{ijk}(\phi) \bigl((\partial_\mu\phi^i)(\partial_\nu\phi^j) + (\partial_\nu\phi^i)(\partial_\mu\phi^j)\bigr) (\partial^\mu\partial^\nu\phi^k) \notag\\[5pt]
    &= -\frac14\int_x\, \biggl[ \kappa_{ijk,l}(\phi) \bigl((\partial_\mu\phi^i)(\partial_\nu\phi^j) + (\partial_\nu\phi^i)(\partial_\mu\phi^j)\bigr) (\partial^\mu\phi^k)(\partial^\nu\phi^l) \notag\\[5pt]
    &\qquad\qquad +\kappa_{ijk}(\phi) \bigl( (\partial_\mu\phi^i)(\partial^2 \phi^j) +(\partial^2\phi^i)(\partial_\mu\phi^j) \bigr) (\partial^\mu\phi^k) \notag\\[5pt]
    &\qquad\qquad +\kappa_{ijk}(\phi) \,\partial_\mu\bigl( (\partial_\nu\phi^i)(\partial^\nu\phi^j)\bigr) (\partial^\mu\phi^k)\biggr] \notag\\[5pt]
    &= \int_x\, \biggl[ \biggl( \frac14\, \kappa_{ijk,l}(\phi) -\frac12\, \kappa_{ikj,l}(\phi)\biggr)\, \O_1^{ijkl}(\phi) \notag\\[5pt]
    &\qquad\qquad +\frac14 \bigl( \kappa_{ijk}(\phi) -\kappa_{kij}(\phi) -\kappa_{kji}(\phi) \bigr) \,\O_2^{ijk}(\phi)\biggr] \,.
\end{align}
As a result of the integration by parts relation in Eqs.~\eqref{eq:ibp}, we have:
\begin{align}
    S[\phi] = \int_x \,\biggl[&
    -U'(\phi) +\frac12\, g'_{ij}(\phi) (\partial_\mu\phi^i)(\partial^\mu\phi^j) \notag\\[5pt]
    & +\frac18\, \lambda'_{ijkl}(\phi)\,\O_1^{ijkl}(\phi) +\frac12\, \lambda'_{ijk}(\phi) \, \O_2^{ijk}(\phi) + \frac12\, \lambda'_{ij}(\phi)\,\O_3^{ij}(\phi) +\O(\partial^6)
    \biggr]\,,\notag\\
\end{align}
where
\begin{subequations}
\begin{align}
    g'_{ij} &= g''_{ij} - h_{i,\s j} - h_{j,i} \,,\\[5pt]
    \lambda'_{ijkl} &= \lambda''_{ijkl} +\frac12\,\bigl(\kappa_{ijk,l} +\kappa_{ijl,k} + \kappa_{kli,\s j} + \kappa_{klj,i} \notag\\[5pt]
    &\quad\qquad\qquad\;\;  - \kappa_{ikj,l} -\kappa_{ikl,\s j} -\kappa_{ilj,k} -\kappa_{ilk,\s j} -\kappa_{jki,l} -\kappa_{jkl,i} -\kappa_{jli,k} -\kappa_{jlk,i}\bigr) \notag\\[5pt]
    &\quad -\kappa'_{ij,kl} -\kappa'_{kl,ij} + \kappa'_{ik,\s jl} +\kappa'_{jl,ik} +\kappa'_{il,\s jk} +\kappa'_{jk,il} \,,\\[5pt]
    \lambda'_{ijk} &= \lambda''_{ijk} +\frac12\, \kappa_{ijk} -\kappa_{k(ij)} - \kappa_{ik,\s j} -\kappa_{jk,i} +\frac12 \bigl(\kappa'_{ik,\s j} + \kappa'_{jk,i}\bigr) +\kappa_{i,\s jk} +\kappa_{j,ik} \,,\\[5pt]
    \lambda'_{ij} &= \lambda''_{ij} -2\,\kappa_{(ij)} +\kappa'_{ij} +\kappa_{i,\s j} +\kappa_{j,i} \,.
\end{align}
\end{subequations}

Next, we can eliminate $\O_2^{ijk}$ and $\O_3^{ij}$ via a derivative field redefinition:
\begin{equation}
    \phi^i = \widetilde\phi^i + c^i_j(\widetilde\phi) \,(\partial^2\widetilde\phi^j) + \frac12 \, c^i_{jk}(\widetilde\phi)\, (\partial_\mu\widetilde\phi^j)(\partial^\mu\widetilde\phi^k) \,,
    \label{eq:warsaw_redef}
\end{equation}
where $c^i_{jk} = c^i_{kj}$. In the $\widetilde\phi$ basis, the action becomes:
\begin{align}
    \widetilde S[\widetilde\phi] = \int_x \,\biggl[&
    -\widetilde U(\widetilde\phi) +\frac12\, \widetilde g_{ij}(\phi) (\partial_\mu\widetilde\phi^i)(\partial^\mu\widetilde\phi^j) \notag\\[5pt]
    & +\frac18\, \widetilde\lambda_{ijkl}(\widetilde\phi)\,\O_1^{ijkl}(\widetilde\phi) +\frac12\, \widetilde\lambda_{ijk}(\widetilde\phi) \, \O_2^{ijk}(\widetilde\phi) + \frac12\, \widetilde\lambda_{ij}(\widetilde\phi)\,\O_3^{ij}(\widetilde\phi) +\O(\partial^6)
    \biggr]\,,\notag\\
\end{align}
where
\begin{subequations}
\label{eq:warsaw_coef}
\begin{align}
    \widetilde U &= U'\,,\\[5pt]
    \widetilde g_{ij} &= g'_{ij} + U'_{,k} \bigl( c^k_{i,\s j} +c^k_{j,i} -c^k_{ij} \bigr) + U'_{,ik}\, c^k_j + U'_{,\s jk}\, c^k_i\,,\label{eq:warsaw_coef_g}\\[5pt]
    \widetilde\lambda_{ijkl} &= \lambda'_{ijkl} -2\,g'_{mn} \bigl( \Gamma'^m_{ij}\, c^n_{kl} +\Gamma'^m_{kl}\, c^n_{ij} \bigr) -U'_{,mn} \, c^m_{ij}\, c^n_{kl} \,,\\[5pt]
    \widetilde\lambda_{ijk} &= \lambda'_{ijk} -g'_{kl}\, c^l_{ij} -2\, g'_{lm} \, \Gamma'^l_{ij}\, c^m_k -U'_{,lm}\, c^l_{ij} \,c^m_k\,,\\[5pt]
    \widetilde\lambda_{ij} &= \lambda'_{ij} -g'_{ik}\, c^k_j -g'_{jk}\, c^k_i -U'_{,kl}\, c^k_i\, c^l_j\,.
\end{align}
\end{subequations}
In these equations, $\Gamma'^k_{ij} \equiv \frac12 \, g'^{kl} (g'_{li,\s j} +g'_{lj,i} -g'_{ij,l})$ represents the Levi-Civita connection computed as if $g'_{ij}$ is the metric.

If we choose $c^i_j$ and $c^i_{jk}$ such that
\begin{equation}
    \widetilde\lambda_{ijk} = 0 \qquad\text{and}\qquad \widetilde\lambda_{ij} = 0\,,
    \label{eq:warsaw_cond}
\end{equation}
then the $\widetilde\phi$ basis is a Warsaw-like basis. The condition \cref{eq:warsaw_cond} can be solved perturbatively if we assume a power counting $\lambda'_{ijk} \sim\frac{1}{\Lambda^3}$, $\lambda'_{ij} \sim \frac{1}{\Lambda^2}$, while $U'_{ij} \sim m^2 \ll \Lambda^2$, where $\Lambda$ is the EFT cutoff, in which case we obtain:
\begin{subequations}
\begin{align}
    c^i_j &= \frac12 \, g'^{ik}\, \lambda'_{kj} + \O\biggl(\frac{1}{\Lambda^4}\biggr)\,,\\[5pt]
    c^i_{jk} &= g'^{il}\, \bigl( \lambda'_{jkl} -\Gamma'^m_{jk}\, \lambda'_{lm} \bigr) + \O\biggl(\frac{1}{\Lambda^5}\biggr)\,.
\end{align}
\end{subequations}
In the Warsaw-like $\widetilde\phi$ basis, we have, according to \cref{eq:g_warsaw}:
\begin{subequations}
\begin{align}
    \widetilde U[\widetilde\phi] &\equiv \int_x\, \widetilde U\bigl(\widetilde\phi(x)\bigr) \,,\\
    \widetilde g_{ab}[\widetilde\phi] &\equiv \int_x\, e^{-i(p_a+p_b)\cdot x}\, \widetilde g_{i_ai_b}\bigl(\widetilde\phi(x)\bigr) \,,\label{eq:g_def_warsaw}\\
    \widetilde\lambda_{abcd}[\widetilde\phi] &\equiv \int_x \, e^{-i(p_a+p_b+p_c+p_d)\cdot x}\, \widetilde\lambda_{i_ai_bi_ci_d}\bigl(\widetilde\phi(x)\bigr) \,,
\end{align}
\end{subequations}
where the functions in the integrand on the right-hand side of these equations are given by Eqs.~\eqref{eq:warsaw_coef}.

Now that we have the Warsaw-like basis expressions, we can obtain $U[\phi]$, $g_{ab}[\phi]$ and $\lambda_{abcd}[\phi]$ in the original basis following the standard tensor transformation rules. For the scalar potential, we have:
\begin{equation}
    U[\phi] = \widetilde U\bigl[F[\phi]\bigr] = \int_x \,\widetilde U\bigl( F[\phi](x)\bigr) = \int_x \,U'\bigl( F[\phi](x)\bigr) \,,
    \label{eq:warsaw_u}
\end{equation}
where $\widetilde\phi^i = F^i[\phi]$ is the inverse of the field redefinition in \cref{eq:warsaw_redef}, obtained iteratively as:
\begin{align}
    F^i[\phi]
    &= \phi^i -c^i_j(\phi) (\partial^2\phi^j) -\frac12\, c^i_{jk}(\phi) (\partial_\mu\phi^j)(\partial^\mu\phi^k) \notag\\[5pt]
    &\qquad +\biggl[ c^i_{j,l}(\phi) (\partial^2\phi^j) +\frac12\, c^i_{jk,l}(\phi) (\partial_\mu\phi^j)(\partial^\mu\phi^k) +c^i_l(\phi)\,\partial^2 +c^i_{jl}(\phi) (\partial_\mu\phi^j)\, \partial^\mu \biggr] \notag\\[5pt]
    &\qquad\qquad \biggl[ c^l_m(\phi) (\partial^2\phi^m) +\frac12\, f^l_{mn}(\phi) (\partial_\nu\phi^m)(\partial^\nu\phi^n) \biggr] +\O(\partial^6)\,.
    \label{eq:warsaw_redef_inv}
\end{align}
Substituting \cref{eq:warsaw_redef_inv} into \cref{eq:warsaw_u} and expanding around $\phi$, we obtain:
\begin{align}
    U[\phi] &= \int_x\, \biggl\{ U'(\phi) + \frac12\, \bigl( g'_{ij}(\phi) -\widetilde g_{ij}(\phi)\bigr)\, (\partial_\mu\phi^i)(\partial^\mu\phi^j) \notag\\[5pt]
    &\hspace{20pt} +U'_{,i}(\phi)\, \biggl[ c^i_{j,l}(\phi) (\partial^2\phi^j) +\frac12\, c^i_{jk,l}(\phi) (\partial_\mu\phi^j)(\partial^\mu\phi^k) +c^i_l(\phi)\,\partial^2 +c^i_{jl}(\phi) (\partial_\mu\phi^j)\, \partial^\mu \biggr] \notag\\[5pt]
    &\hspace{70pt} \biggl[ c^l_m(\phi) (\partial^2\phi^m) +\frac12\, f^l_{mn}(\phi) (\partial_\nu\phi^m)(\partial^\nu\phi^n) \biggr] \notag\\[5pt]
    &\hspace{20pt} +\frac12\, U'_{,ij}(\phi)\, \biggl[ c^i_k(\phi) (\partial^2\phi^k) +\frac12\, f^i_{kl}(\phi) (\partial_\mu\phi^k)(\partial^\mu\phi^l) \biggr] \notag\\[5pt]
    &\hspace{70pt} \biggl[ c^j_m(\phi) (\partial^2\phi^m) +\frac12\, f^j_{mn}(\phi) (\partial_\nu\phi^m)(\partial^\nu\phi^n) \biggr]
    +\O(\partial^6)\biggr\}\,.
    \label{eq:U}
\end{align}
With some algebra, one can also rewrite the $\O(\partial^4)$ terms as a linear combination of $\O_1^{ijkl}(\phi)$, $\O_2^{ijk}(\phi)$ and $\O_3^{ij}(\phi)$. 
We see from \cref{eq:U} that the potential functional contains not only the spacetime integral of the potential function $U'(\phi)$ in the original basis, but also part of the two- and four-derivative terms.

Next, to derive the metric in the original basis, it is easier to work in position space, where
\begin{equation}
    g_{(i_ax_a)(i_bx_b)} [\phi] = \int_{y_a,y_b} \frac{\delta\widetilde\phi^{(j_ay_a)}}{\delta\phi^{(i_ax_a)}} \frac{\delta\widetilde\phi^{(j_by_b)}}{\delta\phi^{(i_bx_b)}} \,\widetilde g_{(j_ay_a)(j_by_b)}\bigl[F[\phi]\bigr]\,,
    \label{eq:g_trans}
\end{equation}
with
\begin{align}
    \widetilde g_{(j_ay_a)(j_by_b)}\bigl[F[\phi]\bigr] &= \int_{p_a, p_b} e^{i(p_a\cdot y_a +p_b\cdot y_b)} \,\widetilde g_{(j_ap_a)(j_bp_b)}\bigl[F[\phi]\bigr] \notag\\[5pt]
    &= \int_x \,\delta^d(x-y_a)\, \delta^d(x-y_b)\, \widetilde g_{j_aj_b}\bigl(F[\phi](x)\bigr) \notag\\[5pt]
    &= \int_x \,\delta^d(x-y_a)\, \delta^d(x-y_b)\, \biggl[\,\widetilde g_{j_aj_b}\bigl(\phi(x)\bigr) \notag\\[5pt]
    &\qquad -\widetilde g_{j_aj_b,i}(\phi)\, \biggl(c^i_j(\phi) (\partial^2\phi^j) +\frac12\, c^i_{jk}(\phi) (\partial_\mu\phi^j)(\partial^\mu\phi^k)\biggr) +\O(\partial^4)\biggr]\,.\notag\\
    \label{eq:g_pos}
\end{align}
The functional derivatives entering \cref{eq:g_trans} can be obtained as:
\begin{align}
    \frac{\delta\widetilde\phi^{(jy)}}{\delta\phi^{(ix)}} &= \biggl[\delta^j_i - c^j_{k,i}(\phi) (\partial^2\phi^k) -\frac12\, c^j_{kl,i}(\phi) (\partial_\mu\phi^k)(\partial^\mu\phi^l) \notag\\[5pt]
    &\qquad -c^j_i(\phi) \,\partial^2 -c^j_{ik}(\phi) (\partial_\mu\phi^k)\,\partial^\mu +\O(\partial^4)\biggr]_y \, \delta^d(y-x) \,,
    \label{eq:warsaw_deriv}
\end{align}
where $[\,\dots]_y$ means the fields and their derivatives are evaluated at $y$, and open derivatives are taken with respect to $y$. Substituting \cref{eq:g_pos,eq:warsaw_deriv} into \cref{eq:g_trans}, and Fourier transforming back to momentum space, we obtain
\begin{align}
    g_{ab} [\phi] &= \int_{x_a,x_b} e^{-i(p_a\cdot x_a +p_b\cdot x_b)}\, g_{(i_ax_a)(i_bx_b)}[\phi] \notag\\[5pt]
    &= \int_x\, e^{-i(p_a+p_b)\cdot x} \biggl\{ \widetilde g_{i_ai_b}(\phi) \notag\\[5pt]
    &\hspace{40pt}
    -\Bigl[\, \widetilde g_{i_ai_b,i}(\phi)\, c^i_j(\phi) +\widetilde g_{i_ai}(\phi)\, c^i_{j,i_b}(\phi)  +\widetilde g_{i_bi}(\phi)\, c^i_{j,i_a}(\phi)\Bigr] (\partial^2\phi^j) \notag\\[5pt]
    &\hspace{40pt}
    -\frac12\, \Bigl[ \widetilde g_{i_ai_b,i}(\phi)\, c^i_{jk}(\phi) +\widetilde g_{i_ai}(\phi)\, c^i_{jk,i_b}(\phi) +\widetilde g_{i_bi}(\phi)\, c^i_{jk,i_a}(\phi) \Bigr] (\partial_\mu\phi^j)(\partial^\mu\phi^k) \notag\\[5pt]
    &\hspace{40pt}
    +\widetilde g_{i_ai}(\phi) \Bigl[\,p_b^2\, c^i_{i_b}(\phi) +ip_b^\mu\, c^i_{i_bj}(\phi)\, (\partial_\mu\phi^j)\Bigr] \notag\\[5pt]
    &\hspace{40pt}
    +\widetilde g_{i_bi}(\phi) \Bigl[\,p_a^2\, c^i_{i_a}(\phi) +ip_a^\mu\, c^i_{i_aj}(\phi)\, (\partial_\mu\phi^j)\Bigr]
    +\O(\partial^4)
    \biggr\}\,,
\end{align}
where we have counted $p_a, p_b \sim \O(\partial)$. Noting that $\widetilde g_{ij}$ is given by \cref{eq:warsaw_coef_g} and $c^i_j, c^i_{jk}$ solve \cref{eq:warsaw_cond}, we see that the metric $g_{ab} [\phi]$ contains information about zero- and four-derivative terms as well as two-derivative terms in the original basis.

Finally, for $\lambda_{abcd}[\phi]$, since we are only working up to $\O(\partial^4)$ in the action, it is sufficient to keep the leading order term in \cref{eq:warsaw_deriv}, or equivalently, 
\begin{equation}
    \frac{\partial\widetilde\phi^a}{\partial\phi^b} = \delta^a_b +\O(\partial^2) \,.
\end{equation}
We obtain:
\begin{align}
    \lambda_{abcd}[\phi] &=  \frac{\partial\widetilde\phi^e}{\partial\phi^a} \frac{\partial\widetilde\phi^f}{\partial\phi^b} \frac{\partial\widetilde\phi^g}{\partial\phi^c} \frac{\partial\widetilde\phi^h}{\partial\phi^d} \,\widetilde\lambda_{efgh}\bigl[F[\phi]\bigr] \notag\\[5pt]
    &= \int_x \, e^{-i(p_a+p_b+p_c+p_d)\cdot x}\, \Bigl[\,\widetilde\lambda_{i_ai_bi_ci_d}(\phi) +\O(\partial^2)\Bigr]\,,
\end{align}
with $\widetilde\lambda_{ijkl}$ given in \cref{eq:warsaw_coef}.

\section{Nonderivative field redefinitions on the field configuration manifold}
\label{app:nd}

In this appendix, we discuss in more detail nonderivative field redefinitions as a special class of coordinate transformations on the field configuration manifold. Consider $\phi^i = f^i(\widetilde\phi)$, where $f^i$ is a set of real analytic functions. As a coordinate transformation on the field configuration manifold, it can be written as:
\begin{equation}
\phi^{(ix)} = F^{(ix)} [\widetilde\phi] \equiv \int_y \,\delta^d(x-y) \, f^i\bigl(\widetilde\phi(y)\bigr) \,.
\end{equation}
Therefore,
\begin{equation}
\frac{\delta\phi^{(ix)}}{\delta\widetilde\phi^{(jy)}} = \frac{\partial\phi^i}{\partial\widetilde\phi^j}\,\delta^d(x-y) \,,\quad
\frac{\delta^2\phi^{(ix)}}{\delta\widetilde\phi^{(jy)}\delta\widetilde\phi^{(kz)}} = \frac{\partial^2\phi^i}{\partial\widetilde\phi^j\partial\widetilde\phi^k} \,\delta^d(x-y)\,\delta^d(x-z)\,,
\end{equation}
etc., where $\frac{\partial\phi^i}{\partial\widetilde\phi^j}$ and $\frac{\partial^2\phi^i}{\partial\widetilde\phi^j\partial\widetilde\phi^k}$ represent partial derivatives of $\phi^{i}=f^i(\widetilde\phi)$ with respect to $\widetilde\phi$ evaluated at $x$. We can also use \cref{eq:partial_phi} to take functional derivatives in momentum space:
\begin{subequations}
\label{eq:phi_derivatives}
\begin{align}
\frac{\partial\phi^{(ip)}}{\partial\widetilde\phi^{(jq)}} &= \int_y\, e^{-iq\cdot y} \, \frac{\delta}{\delta\widetilde\phi^{(jy)}} \int_x\, e^{ip\cdot x} \,\phi^{(ix)} = \int_x\, e^{i(p-q)\cdot x}\, \frac{\partial\phi^i}{\partial\widetilde\phi^j} \,,\\[5pt]
\frac{\partial^2\phi^{(ip)}}{\partial\widetilde\phi^{(jq)}\partial\widetilde\phi^{(kr)}} &= \int_z \, e^{-ir\cdot z} \,\frac{\delta}{\delta\widetilde\phi^{(kz)}} \int_x e^{i(p-q)\cdot x}\, \frac{\partial\phi^i}{\partial\widetilde\phi^j} = \int_x\, e^{i(p-q-r)\cdot x}\, \frac{\partial^2\phi^i}{\partial\widetilde\phi^j\partial\widetilde\phi^k} \,,
\end{align}
\end{subequations}
etc.

A direct consequence of these equations is that, under nonderivative field redefinitions, any object on the field configuration manifold that is defined as the Fourier transform of an object on the target manifold transforms in the same way as its target manifold counterpart. As a nontrivial example, consider the Levi-Civita connection obtained from the metric defined in \cref{eq:g_def}, which is the Fourier transform of the Levi-Civita connection in field space geometry (see \cref{eq:Gamma_fg}):
\begin{equation}
\Gamma^a_{bc} [\phi] = \Gamma^{(i_ap_a)}_{(i_bp_b)(i_cp_c)} [\phi] \equiv \int_x e^{i(p_a-p_b-p_c)\cdot x} \, \Gamma^{i_a}_{i_bi_c} \bigl(\phi(x)\bigr)\,.
\end{equation}
Under a nonderivative field redefinition, $\Gamma^{i_a}_{i_bi_c}$ transforms as a connection on the target manifold. So if we compute $\Gamma^a_{bc}$ from the transformed target manifold connection, we would obtain:
\begin{align}
\widetilde\Gamma^a_{bc} [\widetilde\phi] &= \int_x \,e^{i(p_a-p_b-p_c)\cdot x} \,\widetilde\Gamma^{i_a}_{i_bi_c} \bigl(\widetilde\phi(x)\bigr) \notag\\[5pt]
&=\int_x \,e^{i(p_a-p_b-p_c)\cdot x}  \Biggl[ \frac{\partial\widetilde\phi^{i_a}}{\partial\phi^{i_d}}\,\frac{\partial\phi^{i_e}}{\partial\widetilde\phi^{i_b}} \,\frac{\partial\phi^{i_f}}{\partial\widetilde\phi^{i_c}} \,\Gamma^{i_d}_{i_ei_f}\bigl(f(\widetilde\phi)\bigr) + \frac{\partial\widetilde\phi^{i_a}}{\partial\phi^{i_d}}\,\frac{\partial^2\phi^{i_d}}{\partial\widetilde\phi^{i_b}\partial\widetilde\phi^{i_c}} \Biggr] \,.
\end{align}
Using \cref{eq:phi_derivatives}, it is straightforward to check that this is equivalent to
\begin{equation}
\widetilde\Gamma^a_{bc} [\widetilde\phi] = \frac{\partial\widetilde\phi^a}{\partial\phi^d}\, \frac{\partial\phi^e}{\partial\widetilde\phi^b} \, \frac{\partial\phi^f}{\partial\widetilde\phi^c} \, \Gamma^d_{ef}\bigl[F(\widetilde\phi)\bigr] +\frac{\partial\widetilde\phi^a}{\partial\phi^d}\, \frac{\partial^2\phi^d}{\partial\widetilde\phi^b\partial\widetilde\phi^c} \,,
\end{equation}
as expected for a connection on the field configuration manifold. We can see the same in position space, where the connection on the field configuration manifold is given by:
\begin{align}
\Gamma^{(i_ax_a)}_{(i_bx_b)(i_cx_c)}[\phi] &= \int_{p_a,p_b,p_c} e^{-i(p_a\cdot x_a-p_b\cdot x_b -p_c\cdot x_c)}\, \Gamma^{(i_ap_a)}_{(i_bp_b)(i_cp_c)} [\phi] \notag\\[5pt]
&= \int_x\, \Gamma^{i_a}_{i_bi_c} \bigl(\phi(x)\bigr) \, \delta^d(x-x_a) \, \delta^d(x-x_b)\, \delta^d(x-x_c)\,.
\end{align}
Under a nonderivative field redefinition,
\begin{align}
&\widetilde\Gamma^{(i_ax_a)}_{(i_bx_b)(i_cx_c)} [\widetilde\phi] = \int_x \, \widetilde\Gamma^{i_a}_{i_bi_c} \bigl(\widetilde\phi(x)\bigr) \,\delta^d(x-x_a) \, \delta^d(x-x_b)\, \delta^d(x-x_c) \notag\\[5pt]
=\;&\int_x \, \Biggl[ \frac{\partial\widetilde\phi^{i_a}}{\partial\phi^{i_d}}\,\frac{\partial\phi^{i_e}}{\partial\widetilde\phi^{i_b}} \,\frac{\partial\phi^{i_f}}{\partial\widetilde\phi^{i_c}} \,\Gamma^{i_d}_{i_ei_f}\bigl(f(\widetilde\phi)\bigr) + \frac{\partial\widetilde\phi^{i_a}}{\partial\phi^{i_d}}\,\frac{\partial^2\phi^{i_d}}{\partial\widetilde\phi^{i_b}\partial\widetilde\phi^{i_c}} \Biggr] \,\delta^d(x-x_a) \, \delta^d(x-x_b)\, \delta^d(x-x_c) \notag\\[5pt]
=\;& \int_{x_d, x_e, x_f}\frac{\delta\widetilde\phi^{(i_ax_a)}}{\delta\phi^{(i_dx_d)}} \, \frac{\delta\phi^{(i_ex_e)}}{\delta\widetilde\phi^{(i_bx_b)}} \, \frac{\delta\phi^{(i_fx_f)}}{\delta\widetilde\phi^{(i_cx_c)}} \, \Gamma^{(i_dx_d)}_{(i_ex_e)(i_fx_f)} \bigl[f(\widetilde\phi)\bigr] + \int_{x_d}\frac{\delta\widetilde\phi^{(i_ax_a)}}{\delta\phi^{(i_dx_d)}} \,\frac{\delta^2\phi^{(i_ex_d)}}{\delta\widetilde\phi^{(i_bx_b)}\delta\widetilde\phi^{(i_cx_c)}}\,,
\end{align}
which is again the desired transformation of a connection on the field configuration manifold.

\phantomsection
\addcontentsline{toc}{section}{References}

\bibliographystyle{JHEP}
\bibliography{ref}

\end{document}